\documentclass[fleqn,usenatbib]{mnras}

\usepackage{newtxtext,newtxmath}

\usepackage[T1]{fontenc}
\usepackage{ae,aecompl}
\usepackage{multirow}

\usepackage{graphicx}	
\usepackage{amsmath}	

\title[Origin of PSBs]{Rise and fall of post-starburst galaxies in \textit{Magneticum Pathfinder}}

\author[Marcel Lotz et al.]{Marcel Lotz$^{1,2}$\thanks{E-mail: mlotz@usm.lmu.de},
Klaus Dolag$^{1,2,3}$,
Rhea-Silvia Remus$^{1}$,
and Andreas Burkert$^{1,2,4}$
\\
$^{1}$Universit\"{a}ts-Sternwarte M\"{u}nchen, Fakult\"{a}t f\"{u}r Physik, LMU Munich, Scheinerstr. 1, 81679 M\"{u}nchen, Germany\\
$^{2}$Excellence Cluster ORIGINS, Boltzmannstra\ss{}e 2, 85748 Garching, Germany\\
$^{3}$Max-Planck Institute for Astrophysics, Karl-Schwarzschild-Stra\ss{}e 1, 85741 Garching, Germany\\
$^{4}$Max-Planck Institute for Extraterrestrial Physics, Giessenbachstra\ss{}e 1, 85748 Garching, Germany
}

\date{Accepted XXX. Received YYY; in original form ZZZ}

\pubyear{2021}

\begin{document}
\label{firstpage}
\pagerange{\pageref{firstpage}--\pageref{lastpage}}
\maketitle

\begin{abstract}
Post-starburst galaxies (PSBs) belong to a short-lived transition population between star-forming (SF) and quiescent galaxies. Deciphering their heavily discussed evolutionary pathways is paramount to understanding galaxy evolution. 
We aim to determine the dominant mechanisms governing PSB evolution in both the field and in galaxy clusters.
Using the cosmological hydrodynamical simulation suite \textit{Magneticum Pathfinder}, we identify $647$ PSBs with $z \sim 0$ stellar mass $M_* \geq 5 \cdot 10^{10} \, \mathrm{M_{\odot}}$. We track their galactic evolution, merger history, and black hole activity over a time-span of $3.6\,$Gyr. Additionally, we study cluster PSBs identified at different redshifts and cluster masses.
Independent of environment and redshift, we find that PSBs, like SF galaxies, have frequent mergers. At $z=0$, $89\%$ of PSBs have experienced mergers and $65\%$ had at least one major merger within the last $2.5\,$Gyr, leading to strong star formation episodes. 
In fact, $23\%$ of $z=0$ PSBs were rejuvenated during their starburst.
Following the mergers, field PSBs are generally shutdown via a strong increase in AGN feedback (power output $P_{AGN,PSB} \geq 10^{56}\,$erg/Myr).
We find agreement with observations for both stellar mass functions and $z = 0.9$ line-of-sight phase space distributions of PSBs in galaxy clusters.
Finally, we find that $z \lesssim 0.5$ cluster PSBs are predominantly infalling, especially in high mass clusters and show no signs of enhanced AGN activity. Thus, we conclude that the majority of cluster PSBs are shutdown via an environmental quenching mechanism such as ram-pressure stripping, while field PSBs are mainly quenched by AGN feedback.

\end{abstract}

\begin{keywords}
galaxies: starburst -- galaxies: kinematics and dynamics -- galaxies: star formation -- galaxies: evolution -- galaxies: clusters: general -- methods: numerical
\end{keywords}



\section{Introduction}
\label{sec:intro}

Post-starburst galaxies (PSBs), also referred to as E+A or k+a galaxies based on spectral properties \citep{1999ApJS..122...51D, 1999ApJ...527...54B}, are characterised by a recent starburst and subsequent fast quenching \citep{1983ApJ...270....7D, 1999AJ....117..140C}. 
As such, they offer a unique opportunity to clarify some of the debated details of both the morphological (late- to early-type) and colour (blue cloud to red sequence) transition, which are fundamental to understanding galaxy evolution \citep{2016MNRAS.456.3032P, 2017MNRAS.472.1401A}. Furthermore, PSBs are found in all environments and at all redshifts, suggesting physical processes of universal importance \citep{1995A&A...297...61B, 2003ApJ...599..865T, 2012ApJ...745..179W, 2013ApJ...770...62D, 2019NatAs...3..440P}.
Their formation mechanism remains a matter of debate: Typically, PSBs are assumed to have undergone a recent quenching event rather than gradual depletion, and thus they belong to a transition population between blue disc-like and red early-type galaxies \citep{2008AJ....135.1636D, 2017MNRAS.472.1447W, 2018MNRAS.477.1921A}.

Different pathways have been proposed to explain the observed strong increase of PSBs with redshift \citep{2018MNRAS.480..381M, 2019ApJ...874...17B} and the varying environmental abundance \citep{1999ApJ...518..576P, 2017MNRAS.472..419L, 2019MNRAS.482..881P} of PSBs. 
\cite{2016MNRAS.463..832W} propose two PSB pathways: First, at $z \gtrsim 2$ PSBs are exclusively massive galaxies which formed the majority of their stars within a rapid assembly period, followed by a complete shutdown in star formation. Second, at $z \lesssim 1$ PSBs are the result of rapid quenching of gas-rich star-forming galaxies, independent of stellar mass.
Possible candidates for this rapid quenching at $z \lesssim 1$ include the environment and/or gas-rich major mergers \citep{2016MNRAS.463..832W}. 
More recent work by \cite{2018MNRAS.480..381M} suggests that the $z > 1$ PSB population is the result of a violent event, leading to a compact object, whereas the $z < 1$ population is able to preserve its typically disc-dominated structure, suggesting an environmental mechanism. 
At redshift $z \sim 0.8$, \cite{2020MNRAS.497..389D} find evidence for a fast pathway associated with a centrally concentrated starburst.
Galaxies at redshifts $z < 0.05$ appear to show evidence for three different pathways through the post-starburst phase, mostly occurring in intermediate density environments \citep{2018MNRAS.477.1708P}: First, a strong disruptive event (e.g. major merger) triggering a starburst and subsequently quenching the galaxy. Second, random star formation in the mass range $9.5 < log(M_*/\mathrm{M_{\odot}}) < 10.5$ causing weak starbursts and, third, weak starburst in quiescent galaxies, resulting in a gradual climb towards the high mass end of the red sequence \citep{2018MNRAS.477.1708P}. 
In general, it is clear that different PSB evolutionary pathways exist, however, to date the number of and relevant characteristics of these pathways is not well determined. 

Despite ongoing arguments, simulations \citep{2018MNRAS.479..758W, 2019MNRAS.484.2447D} and observations \citep{1996AJ....111..109S, 2005MNRAS.359..949B} of PSBs in the local low density Universe generally show signs of, or are consistent with, recent galaxy-galaxy interactions and galaxy mergers.
This is not surprising, as galaxy mergers can impact the star formation rate (SFR) of galaxies in opposing ways:
Mergers have been found to increase \citep{2019MNRAS.490.2139R, 2020MNRAS.494.5396B}, not impact \citep{2019A&A...631A..51P}, and decrease \citep{2020ApJ...888...77W} the SFR on varying timescales, depending on the details of the specific merger.
Mechanisms that directly impact the SFR and have been associated with mergers include: introducing turbulence \citep{2018MNRAS.478.3447E}, increasing disc instabilities \citep{2019MNRAS.489.4196L}, triggering nuclear inflows via tidal interactions \citep{2005MNRAS.361..776S, 2008MNRAS.386.1355G}, and gravitational heating, i.e. the process whereby gas is heated and kept hot via the release of potential energy from infalling substructure \citep{2009ApJ...697L..38J}.
Mergers may also influence the SFR indirectly: Potential mechanisms include facilitating the central galactic black hole (BH) growth \citep{2014MNRAS.437.1456B}, thus potentially leading to strong AGN feedback \citep{2013MNRAS.430.1901H} which may lead to galactic gas removal \citep{2014MNRAS.437.1456B}, ultimately causing long term star formation suppression \citep{2014ApJ...792...84Y}.

Several works highlight the relevance of active galactic nucleus (AGN) feedback in explaining the sharp decline in the SFR found in (PSB) galaxies \citep{2010ApJ...709..884Y, 2018MNRAS.480.3993B, 2019A&A...623A..64C, 2020AAS...23520719L}.
Nonetheless, the details of the mechanism(s) driving the nuclear activity in the centres of galaxies remains a major unsettled question \citep{2018MNRAS.481..341S}, especially in PSBs. 
Some studies find no dominant AGN role in shutting down star formation on short timescale:
zCOSMOS survey observations conclude that several mechanisms, both related and unrelated to the environment, are more relevant to the quenching of star formation on short timescales ($<1\,$Gyr) \citep{2010A&A...509A..42V}. Due to the time delay between the starburst phase and AGN activity, \cite{2014ApJ...792...84Y} go a step further and suggest that the AGN does not play a primary role in the initial quenching of starbursts, but rather is responsible for maintaining the post-starburst phase. 
On the other hand, \cite{2005MNRAS.361..776S} find a complex interplay between starbursts and AGN activity when tidal interactions trigger intense nuclear gas inflows. 
SPH simulations find that the BH accretion rate is especially relevant to the inner SFR, but the correlation is less pronounced when using the galactic SFR \citep{2010MNRAS.407.1529H}. 
\citep{2013MNRAS.430.1901H} suggest that strong AGN feedback is required to explain the observed star formation shutdown in post-merger galaxies. 

Generally, galactic quenching mechanisms, i.e. processes whereby star formation is reduced, can be divided into two categories: \textit{mass quenching}, i.e. stellar mass dependent mechanisms which are mostly independent of the environment (e.g. SNe and AGN feedback), and \textit{environmental quenching}, i.e. mechanisms governed by the surroundings (e.g. ram-pressure stripping and galaxy-galaxy interactions) rather than the galaxy itself \citep{2006PASP..118..517B, 2010ApJ...721..193P, 2017MNRAS.469.3670S}.

The specific environment has a strong influence on the abundances of different types of galaxies, as evidenced by the morphology-density relation \citep{1980ApJ...236..351D, 2003MNRAS.346..601G}:
Cluster galaxies are more likely to be characterised by reduced star formation than galaxies in less dense environments \citep{1974ApJ...194....1O, 1978ApJ...219...18B, 1980ApJ...236..351D, 1997ApJ...488L..75B}.  
As a result, the evolution of PSBs in cluster environments differs significantly from the evolution in lower density environments: A study of PSBs in galaxy clusters at $0.04 < z < 0.07$ indicates that the short-timescale star formation quenching channel (e.g. ram-pressure stripping within and galaxy-galaxy interactions outside of the virial radius) contributes two times more than the long timescale one (e.g. strangulation) to the growth of the quiescent cluster population \citep{2017ApJ...838..148P}. 

PSBs and star-forming cluster galaxies share many properties and are characterised by similar distributions of environment \citep{2006ApJ...650..763H}.
Long-slit spectra observations of Coma cluster galaxies suggest close kinematic similarities between star-forming and PSB galaxies, i.e. both appear to be rotating systems and have exponential light profiles \citep{1996AJ....111...78C}.
In a follow-up study of five low redshift galaxy clusters, \cite{1997AJ....113..492C} also find that most of the recent PSBs are in fact disc galaxies. Similarly, kinematic classifications at $z<0.04$ show that PSBs are typically fast rotators \citep{2013MNRAS.432.3131P}.
Meanwhile, cluster PSBs at $z \sim 0.55$ \citep{2010PASA...27..360P} and $z \sim 1$ \citep{2020MNRAS.493.6011M} appear to have more similarities with early-type morphologies.
Depending on how recently and by what mechanism(s) PSBs have undergone their starburst and subsequent shutdown, the morphological classification likely varies from late- to early-type.

Observations of cluster galaxies suggest that interactions with the intra-cluster medium (ICM) rather than mergers (as is the case in the field) are highly relevant for the evolution of cluster PSBs \citep{2009ApJ...693..112P, 2017ApJ...838..148P}.
Coma cluster observations conclude that the starbursts found in Coma PSBs were not the result of major mergers \citep{1996AJ....111...78C}. Similarly, observations of a rich $z \sim 0.55$ cluster argue against a merger origin, favouring a PSB shutdown scenario involving the ICM \citep{2010PASA...27..360P}. This is further supported by the correlation between quenching efficiency and cluster velocity dispersion, which implies that the star formation shutdown is related to the ICM, specifically that more massive clusters quench more efficiently \citep{2009ApJ...693..112P}.
SAMI and GASP observations find additional evidence for ram-pressure stripping being the dominant quenching mechanism in galaxy clusters: Most cluster PSBs have been quenched outside-in, i.e. the outskirts reach undetectable levels of star formation prior the inner regions \citep{2017ApJ...846...27G, 2019ApJ...873...52O, 2020ApJ...892..146V}. 
In summary, the outside-in quenching, the lack of signs of interaction, the fast rotation, and the dense environment, all favour a scenario in which ram-pressure stripping shuts down star formation in cluster PSBs \citep{2020ApJ...892..146V}.

Ram-pressure stripping also appears to influence galaxy evolution beyond direct quenching:
Studies of cluster galaxies undergoing ram-pressure stripping find that ram-pressure stripping may enhance star formation prior to gas removal \citep{2018ApJ...866L..25V, 2020ApJ...899...98V, 2020MNRAS.495..554R}. 
Furthermore, there is evidence that an AGN is hosted by six out of seven GASP jellyfish galaxies \citep{2021IAUS..359..108P}, i.e. a potential cluster PSB progenitor \citep{2016AJ....151...78P} which is associated with long tentacles of (star-forming) material extending far beyond the galactic disc \citep{2017Natur.548..304P}. This surprisingly high incidence, compared to the general cluster and field population, suggests that ram-pressure stripping may trigger AGN activity via nuclear gas inflow \citep{2017Natur.548..304P, 2021IAUS..359..108P}. Similarly, the Romulus C simulation finds ram-pressure stripping triggered enhanced black hole accretion prior to quenching \citep{2020ApJ...895L...8R}, implying that the AGN feedback may aid in the quenching of star formation during ram-pressure stripping. Meanwhile, \cite{2019MNRAS.486..486R} find that photo-ionisation by the AGN in GASP jellyfish galaxies is the dominant ionisation mechanism. A detailed analysis of an individual GASP jellyfish galaxy supports the scenario in which the suppression of star formation in the central region of the disc is most likely due to the feedback from the AGN \citep{2019MNRAS.487.3102G}.

We aim to clarify the importance of different mechanisms to the onset of the starburst phase, as well as the reasons for the subsequent rapid shutdown in star formation observed in PSBs. We discuss the Magneticum Pathfinder simulations and our PSB selection process in Section \ref{sec:data}.
In Section \ref{sec:environment}, we study the environment, relevant distributions and the overall evolution of PSBs.
The role of mergers is investigated in Section \ref{sec:mergers}.
In Section \ref{sec:shutdown} we analyse the importance of the AGN and SNe feedback.
Finally, we study PSB evolution within galaxy clusters in Section \ref{sec:clusters}.
We discuss our results in Section \ref{sec:discussion} and present our conclusions in Section \ref{sec:conc}.

\section{Data sample}
\label{sec:data}

\subsection{Magneticum Pathfinder simulations}
\label{sub:Mag}

\textit{Magneticum Pathfinder}\footnote{\url{www.magneticum.org}} is a set of large scale smoothed-particle hydrodynamic (SPH) simulations that employ a mesh-free Lagrangian method aimed at following structure formation on cosmological scales, with open access to many features \citep{2017A&C....20...52R}. The simulations are executed with the Tree/SPH code GADGET-3, a development based on GADGET-2 \citep{2001NewA....6...79S, 2005MNRAS.364.1105S}. 
In this work we primarily use Box2 ($352 \, \mathrm{(Mpc/h)^3}$), and to a lesser extent Box2b ($640 \, \mathrm{(Mpc/h)^3}$) and Box4 ($48 \, \mathrm{(Mpc/h)^3}$).
Box2 and Box4 have a higher temporal resolution compared to Box2b, i.e. a larger number of individual \texttt{SUBFIND} halo finder outputs \citep{2001NewA....6...79S, 2009MNRAS.399..497D}. This facilitates and enhances the temporal tracking of galaxies.
On the other hand, Box2b is larger and provides a greater statistical sample and is solely used to increase the sample size in Section \ref{sub:vlosObsComp}. 

Our standard resolution (for Box2, Box2b, and one of the Box4 runs) is set to `high resolution': dark matter (dm) and gas particles have masses of $m_{\mathrm{dm}} = 6.9 \cdot 10^8 \, h^{-1}\mathrm{M_{\odot}}$ and $m_{\mathrm{gas}} = 1.4 \cdot 10^8 \, h^{-1}\mathrm{M_{\odot}}$, respectively. Stellar particles are formed from gas particles and have $\sim 1/4$ of the mass of their parent gas particle. At this resolution level the softening of the dark matter, gas and stars is $\epsilon_{\mathrm{dm}} = 3.75 \, \mathrm{h^{-1} kpc}$, $\epsilon_{\mathrm{gas}} = 3.75 \, \mathrm{h^{-1} kpc}$ and $\epsilon_{\mathrm{stars}} = 2 \, \mathrm{h^{-1} kpc}$, respectively.
Box2 is comprised of $2 \cdot 1584^3$ particles, while Box2b is comprised of $2 \cdot 2880^3$ particles.
The `ultra-high resolution' Box4 run has a $\sim 20$ higher mass resolution (compared to our standard 'high resolution') \citep{2018MNRAS.480.4636S} and is only used to test the numerical convergence of our results.
Throughout this paper the following cosmology is adopted \citep{2011ApJS..192...18K}: $h = 0.704$, $\Omega_M = 0.272$, $\Omega_{\Lambda} = 0.728$ and $\Omega_b = 0.0451$.

The astrophysical processes modelled within the Magneticum simulations include, but are not limited to: gas cooling and star formation \citep{2003MNRAS.339..289S}, metal and chemical enrichment \citep{2003MNRAS.342.1025T, 2007MNRAS.382.1050T, 2017Galax...5...35D}, black holes and AGN feedback \citep{2005MNRAS.361..776S, 2014MNRAS.442.2304H, 2015MNRAS.448.1504S}, thermal conduction \citep{2004ApJ...606L..97D}, low viscosity scheme to track turbulence \citep{2005MNRAS.364..753D,2016MNRAS.455.2110B}, higher order SPH kernels \citep{2012MNRAS.425.1068D} and magnetic fields (passive) \citep{2009MNRAS.398.1678D}.
For a more details on the precise physical processes refer to \cite{2015ApJ...812...29T, 2014MNRAS.442.2304H, 2017Galax...5...35D}.

The Magneticum simulations have been used in the past to compare and interpret observations, in addition to independently studying various properties.
Galaxy kinematics are in good agreement with observations and may be used to predict the formation pathway \citep{2018MNRAS.480.4636S, 2020MNRAS.493.3778S}. 
The specific angular momentum of disc stars and its relation to the specific angular momentum of the cold gas matches observations, and may be used to (morphologically) classify galaxies \citep{2015ApJ...812...29T}. When comparing with the integral field spectroscopic data from SAMI, Magneticum matches observations well: In particular, it is the only simulation able to reproduce ellipticities typical for disc galaxies \citep{2019MNRAS.484..869V}. 
The mass ratios and orbital parameters of galaxy mergers strongly impact the resulting radial mass distribution: Mini mergers can significantly increase the host disc size, while not changing the global shape \citep{2019MNRAS.487..318K}. 
AGN properties in Magneticum, such as the evolution of the bolometric AGN luminosity function, agree with observations \citep{2014MNRAS.442.2304H}. In fact, merger events, especially minor mergers, do not necessarily drive strong nuclear activity \citep{2016MNRAS.458.1013S}.
Moreover, merger events are not the statistically dominant driver of nuclear activity \citep{2018MNRAS.481..341S}.
Satellite galaxies in galaxy clusters are predominantly quenched by ram-pressure stripping \citep{2019MNRAS.488.5370L}. In general, Magneticum galaxy cluster properties, such as the pressure \citep{2013A&A...550A.131P}, temperature, and entropy profiles \citep{2014ApJ...794...67M}, as well as the distribution of metals \citep{2017Galax...5...35D}, agree with observations.

\subsection{Post-starburst selection}
\label{sub:selection}

Galaxies are selected to have a minimum stellar mass of $M_* \geq 3.5 \cdot 10^{10} \, h^{-1} \mathrm{M_{\odot}}$, corresponding to a minimum of $\sim 1000$ stellar particles for a given galaxy. 
The only exceptions to this stellar mass threshold is found in Section \ref{sec:clusters}, where the threshold is reduced to $M_* \geq 3.5 \cdot 10^{9} \, h^{-1} \mathrm{M_{\odot}}$ to increase the available sample size in cluster environments. The additional use of Box2b and the lowering of the stellar mass threshold is done to increase the abundance of PSBs within galaxy cluster environments.

In order to differentiate between star-forming and quiescent galaxies, the criterion introduced by \cite{2008ApJ...688..770F} is used throughout this paper at all redshifts.
To this end, we use the specific star formation rate (SSFR), i.e. the star formation rate (SFR) divided by the galactic stellar mass $\mathrm{SSFR} = \mathrm{SFR}/M_*$, and the redshift evolving Hubble time $t_{\mathrm{H}} = 1/H(t)$, where $H(t)$ is the Hubble parameter calculated at a given redshift.
Galaxies with a value above $\mathrm{SSFR} \cdot t_{\mathrm{H}} > 0.3$ are classified as star-forming, while galaxies with $\mathrm{SSFR} \cdot t_{\mathrm{H}} < 0.3$ are classified as quiescent. 

Importantly, this co-called `blueness criterion' ($\mathrm{SSFR} \cdot t_{\mathrm{H}} > 0.3$) is time dependent rather than merely being applicable to low redshifts. Hence, this definition encompasses the changing star formation history on a cosmological scale and is well suited to study and compare galaxies at different redshifts. With this criterion, the Milky Way, for example, would have $\mathrm{SSFR} \cdot t_{\mathrm{H}} \sim 0.4$ at $z = 0$ and, hence, be considered star-forming \citep{2015ApJ...806...96L}.

We identify post-starburst galaxies (PSBs) in the Magneticum simulations based on the stellar particle age and the blueness: Of all stellar particles of a galaxy, at least $2\%$ need to be younger than $0.5 \, \mathrm{Gyr}$. In addition, the galaxy's blueness at identification needs to be smaller than $\mathrm{SSFR} \cdot t_\mathrm{H} < 0.3$. These two parameters describe galaxies that have a sufficiently large young stellar population, while also no longer being star-forming, i.e. galaxies that have experienced a recent starburst.
In particular, we choose this criterion as it implies a minimum average SSFR within the past $0.5\,$Gyr of $\mathrm{SSFR} \geq 4 \cdot 10^{-11} \, \mathrm{yr}^{-1}$, similar to the criterion used by \cite{2019MNRAS.484.2447D}.
To verify that our results are robust, we initially varied both the young stellar mass percentage (1, 2, 5 or 10 per cent) and the associated evaluation timescale (0.5, 1 or 2 Gyr). Although the resulting sample size varied, the conclusions and the agreement with observations remained robust.

When considering all Box2 galaxies fulfilling these criteria we obtain a sample of $647$ PSBs at $z \sim 0$. This global sample provides the basis of the majority of our analysis and is complemented by additional specific environmental and redshift selections where necessary.
To understand how PSBs differ from other galaxies, we introduce two stellar mass matched control (SMMC) samples: quenched (QSMMC) and star-forming (SFSMMC) galaxies, using the above blueness criterion for differentiation.
The control samples are constructed by selecting the closest quenched and star-forming stellar mass match for each PSB galaxy at identification redshift.
In terms of the star formation at identification redshift the QSMMC sample is indistinguishable from PSBs.

In order to disentangle the details causing the starburst and the following shutdown in star formation, we consider the temporal evolution of PSBs. 
To this end, we employ two complementary methods to track and trace both PSBs and control galaxies in Box2 of the Magneticum simulations. First, we identify the main galactic black hole particle associated with a galaxy and track this particle and subsequently its host backwards in time. This method provides a temporal resolution of $0.43\,$Gyr, as only every fourth time step has stored particle data. 
Second, we analyse the merger trees of the galaxies in question, yielding a complete merger history with a temporal resolution of $0.11\,$Gyr.

\section{Environment, distribution, and evolution of post-starburst galaxies}
\label{sec:environment}

\subsection{Quenched, PSB-to-quenched, and PSB fractions}
\label{sub:qfrac}

Understanding the abundance of specific galaxy types at different halo masses, i.e. in different environments, is crucial for determining the relevant formation and evolutionary mechanisms of PSBs. Specifically, the environment is key to understanding potential triggers of the starburst phase and, subsequently, the causes of the star formation shutdown. 
\cite{2019MNRAS.488.5370L} already demonstrated good agreement between Box2 and observations of quenched fractions at intermediate stellar masses $\log_{10}(M_*/\mathrm{M_{\odot}}) = [9.7,10.5]$ \citep{2012MNRAS.424..232W}. 
We now extend our investigation to higher stellar mass galaxies $\log_{10}(M_*/\mathrm{M_{\odot}}) = [10.70,12.00]$, as well as presenting predictions of the PSB-to-quenched fraction.

\begin{figure*}
	\includegraphics[width=1.5\columnwidth]{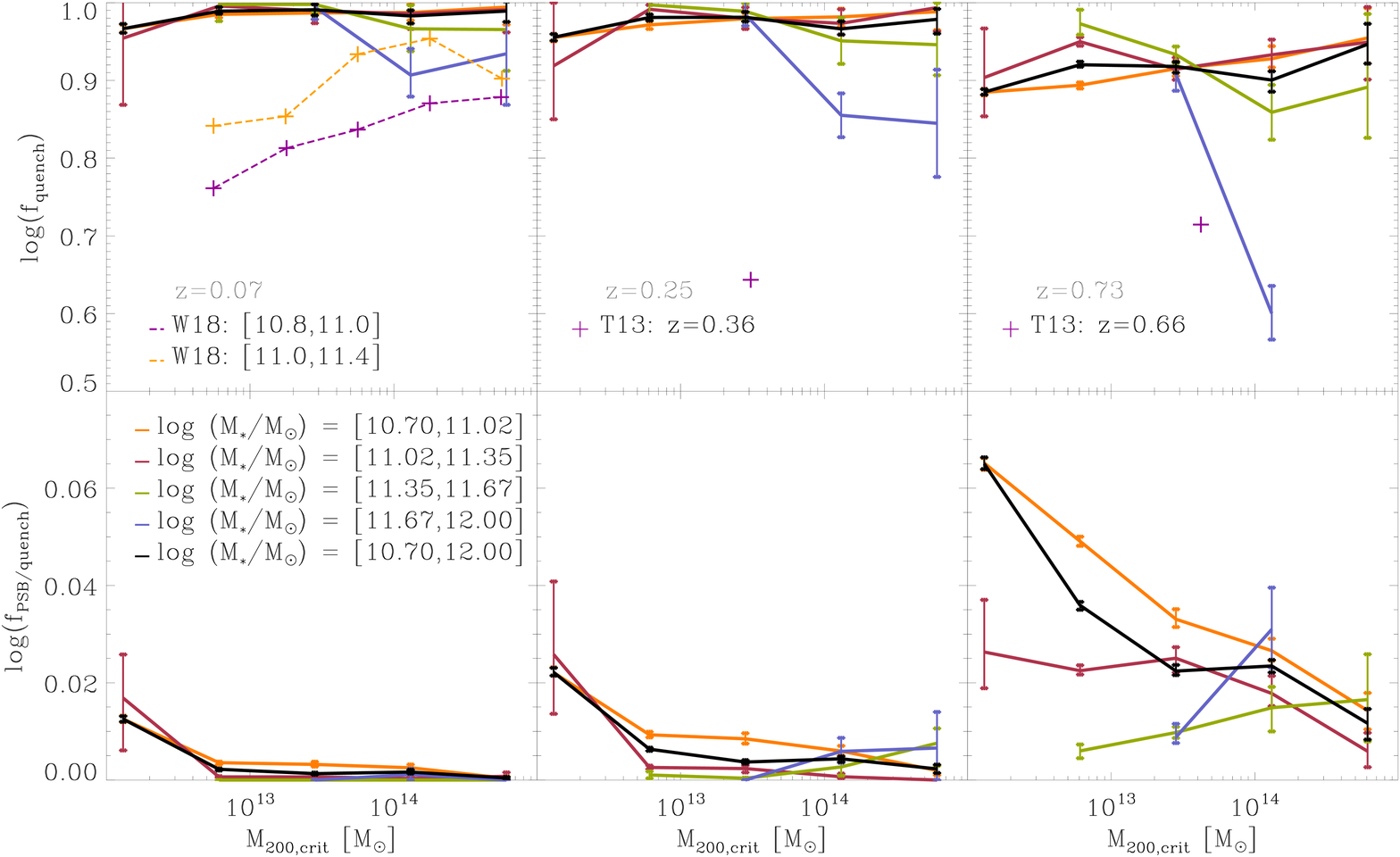}
	\includegraphics[width=1.5\columnwidth]{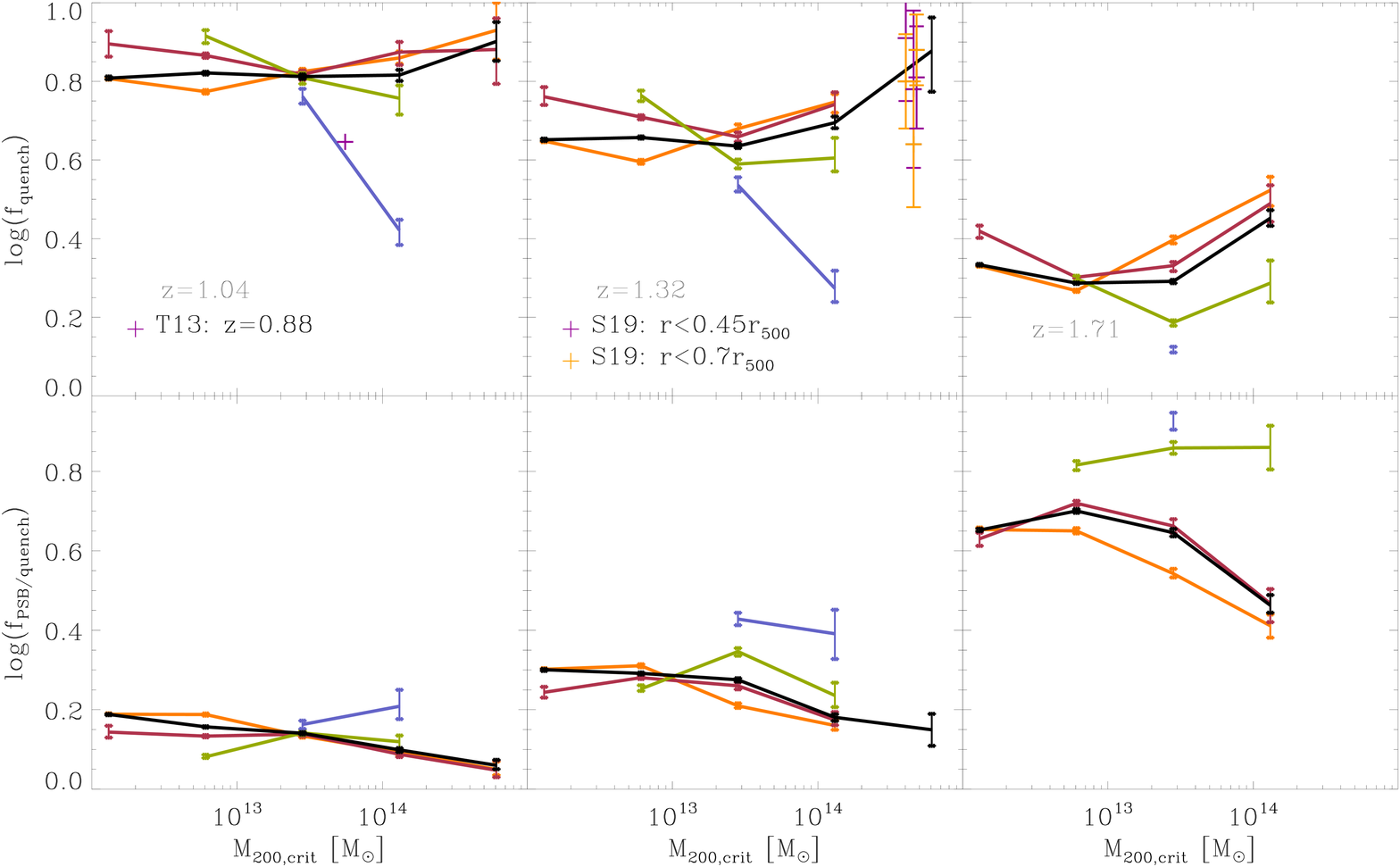}
    \caption{Fraction of quenched galaxies (top) and fraction of PSB-to-quenched galaxies (bottom) as a function of $M_{\mathrm{200,crit}}$ halo mass, at different redshifts $0.07<z<1.71$ (increasing from left to right) for all Box2 galaxies.
    Each panel is subdivided into four unique stellar mass bins (colour coded) and one bin showing the behaviour across the entire evaluated stellar mass range $\log_{10}(M_*/\mathrm{M_{\odot}}) = [10.70,12.00]$ (black). 
    The quenched and PSB-to-quenched fraction is only shown if the denominator in each case is larger than $100$ galaxies. 
    Quenched galaxies are defined as $\mathrm{SSFR} \cdot t_\mathrm{H} < 0.3$ (see Section \ref{sub:selection}).
    Box2 error bars are calculated via bootstrapping. If no observational error bars are shown, then the error is of order the symbol size.
    Note the difference in y-axis range for the PSB-to-quenched fractions between rows.
    At $z=0.07$, we compare the quenched fraction with low redshift observations in the stellar mass range $\log_{10}(M_*/\mathrm{M_{\odot}}) = [[10.8,11.00],[11.0,11.4]]$ \protect\citep{2018ApJ...852...31W}. At intermediate redshifts, we compare the quenched fractions to central galaxies in COSMOS groups \protect\citep{2011ApJ...742..125G, 2013ApJ...778...93T}. At $z=1.32$, we compare our results to $1.38<z<1.45$ cluster galaxies above the common mass completeness limit $\log_{10}(M_*/\mathrm{M_{\odot}}) > 10.85$ within $r < 0.45r_{500}$ and $r < 0.7r_{500}$ of SPT-SZ galaxy clusters \protect\citep{2019A&A...622A.117S}.
    }
    \label{fig:Q_PSBfrac_grid}
\end{figure*}

Figure \ref{fig:Q_PSBfrac_grid} (top panels) shows a number of trends and behaviours relating to the quenched fraction:
First, at redshifts $z \lesssim 1$ the vast majority ($\geq 80\%$) of galaxies in the stellar mass range $\log_{10}(M_*/\mathrm{M_{\odot}}) = [10.70,12.00]$ (black) are quenched, independent of host halo mass. 
The only exception to this is found in the highest stellar mass bin ($\log_{10}(M_*/\mathrm{M_{\odot}}) = [11.67,12.00]$), which shows lower quenched fractions with increasing halo mass, because above halo masses $M_{\mathrm{200,crit}} \geq 10^{14}$ these high mass galaxies are dominated by brightest cluster galaxies (BCGs), which experience episodes of star formation as a result of gas accretion.

Second, Figure \ref{fig:Q_PSBfrac_grid} shows varying agreement with observations: We find broad agreement between our $z=0.07$ quenched fractions and $0.01<z<0.12$ observations by \cite{2018ApJ...852...31W}, which are based on NYU-VAGC \citep{2005AJ....129.2562B} and SDSS DR7 \citep{2009ApJS..182..543A}. Although our Box2 galaxies are characterised by higher quenched fractions and a less distinct split between stellar masses at $z \sim 0$ compared to observations, observations are similarly characterised by high quenched fractions at low redshift, especially towards higher halo mass. 
When comparing our results at $0.25<z<1.04$ to observations of central galaxies in COSMOS groups at median redshift bins $z=[0.36, 0.66, 0.88]$ \citep{2011ApJ...742..125G, 2013ApJ...778...93T}, we find the strongest agreement towards higher redshifts, while the lower redshift comparison lacks good agreement.
At high redshift, we compare our $z=1.32$ results to SPT-SZ cluster galaxies at redshifts $z=[1.38, 1.401, 1.478]$ 
\citep{2019A&A...622A.117S}. The cluster galaxies have stellar masses above the common mass completeness limit $\log_{10}(M_*/\mathrm{M_{\odot}}) > 10.85$, and the quenched fractions are calculated for cluster radii $r < 0.45 \, r_{500}$ and $r < 0.7 \, r_{500}$ \citep{2019A&A...622A.117S}. To better compare with our results, we convert the halo mass from $M_{\mathrm{500}}$ \citep{2019A&A...622A.117S} to $M_{\mathrm{200}}$, assuming an NFW profile with constant concentration ($c=5$) \citep{2003MNRAS.342..163P}. 
The resulting comparison agrees well with our $z=1.32$ results. 
Furthermore, the trend whereby the quenched fraction at constant stellar mass increases towards lower redshift agrees with established models \citep{2008ApJS..175..390H} and simulations \citep{2019MNRAS.488.3143B}. 

Third, towards higher redshifts ($z > 1$) our quenched fraction begins to drop and the differences between the stellar mass bins become larger than the bootstrapped errors associated with the individual bins. At $z=1.71$, we find the highest quenched fraction in the lowest stellar mass bin $\log_{10}(M_*/M_{\odot}) = [10.70,11.02]$. This is likely due to higher stellar mass galaxies at this redshift having undergone more recent mass growth, which is typically associated with star formation, thus leading to lower quenched fractions in high stellar mass compared to low stellar mass galaxies.
In brief, environmental quenching is more effective than mass quenching at high redshift.

Figure \ref{fig:Q_PSBfrac_grid} (bottom panels) shows the PSB-to-quenched fraction. 
The PSB-to-quenched fraction maps the abundance of PSBs relative to the evolving quenched fraction, rather than the total population, thereby avoiding additional systematics associated with the quenched fraction and its evolution.
We find that the qualitative behaviour remains broadly similar at redshifts $z \lesssim 1$: The highest abundance of PSBs is consistently found at low stellar and halo masses. 
Specifically, the PSB-to-quenched fraction is consistently below $7\%$ at redshifts $z \leq 0.73$. Furthermore, the lower the redshift, the lower the PSB-to-quenched fraction. 

At higher redshifts ($z \geq 1.3$) PSBs are no longer most often found at low stellar masses. In particular, the low redshift preference for low stellar masses appears to be inverted at high redshift. High stellar mass galaxies at high redshift belong to the subset of galaxies characterised by the quickest mass assembly. When high stellar mass galaxies become quenched at high redshift, they likely host a significant population of young stars, thus fulfilling our PSB selection criteria. As a result the PSB-to-quenched fraction at high redshift is highest among high stellar mass galaxies.

The PSB-to-quenched fraction as a function of halo mass evolves with redshift:
At low redshifts ($z \leq 0.73$), the PSB-to-quenched fraction exhibits the highest abundances at low halo masses. 
With increasing redshift ($z \geq 1$), the PSB-to-quenched fraction shows less preference for low halo mass. 
Similarly, DEEP2 and SDSS results find that $z \sim 0$ PSBs are found in relatively under-dense environments, while at $z \sim 1$ they are increasingly found in over-dense environments \citep{2009MNRAS.398..735Y}.

The positive correlation between redshift and the PSB-to-quenched fraction is also found for the PSB-to-total fraction:
In the stellar mass range $10.7 < \log_{10}(M_*/\mathrm{M_{\odot}}) < 12.0$ used in Figure \ref{fig:Q_PSBfrac_grid}, we find the following PSB-to-total fractions: $0.45\%$ at $z=0.07$, $0.95\%$ at $z=0.25$, $3.81\%$ at $z=0.73$, $13.4\%$ at $z=1.04$, $19.0\%$ at $z=1.32$, and $20.8\%$ at $z=1.71$.
This behaviour agrees with observations of PSBs with stellar masses $10.0 < \log_{10}(M_*/\mathrm{M_{\odot}}) < 12.5$, which find that the fraction of PSBs declines from $\sim 5\%$ of the total population at $z \sim 2$, to $\sim 1\%$ by $z \sim 0.5$ \citep{2016MNRAS.463..832W}.
At low redshift, the two differing stellar mass ranges yield similar abundances.
However, at high redshift the agreement becomes smaller. This is likely driven by the $\sim 5$ times lower stellar mass threshold used by \cite{2016MNRAS.463..832W}, compared to our threshold. 
To demonstrate, in Box2 at $z=1.71$, this lower stellar mass threshold results in a $\sim 40$ times higher number of total galaxies in the stellar mass range $\log_{10}(M_*/M_{\odot}) = [10.0,12.5]$ compared to $\log_{10}(M_*/M_{\odot}) = [10.7,12.0]$. 
Connecting this with the fact that higher stellar mass galaxies at high redshift are statistically more likely to be classified as PSBs, as illustrated by Figure \ref{fig:Q_PSBfrac_grid}, our higher PSB-to-total fraction at high redshift is expected. Given these considerations, \cite{2016MNRAS.463..832W} and our results agree well.
We conclude, Figure \ref{fig:Q_PSBfrac_grid} (bottom) suggests that both the redshift and environment play an important role in the specific evolution of PSBs.

\subsection{Stellar mass functions of satellite galaxies}
\label{sub:SMF}

\begin{figure*}
	\includegraphics[width=1.99\columnwidth]{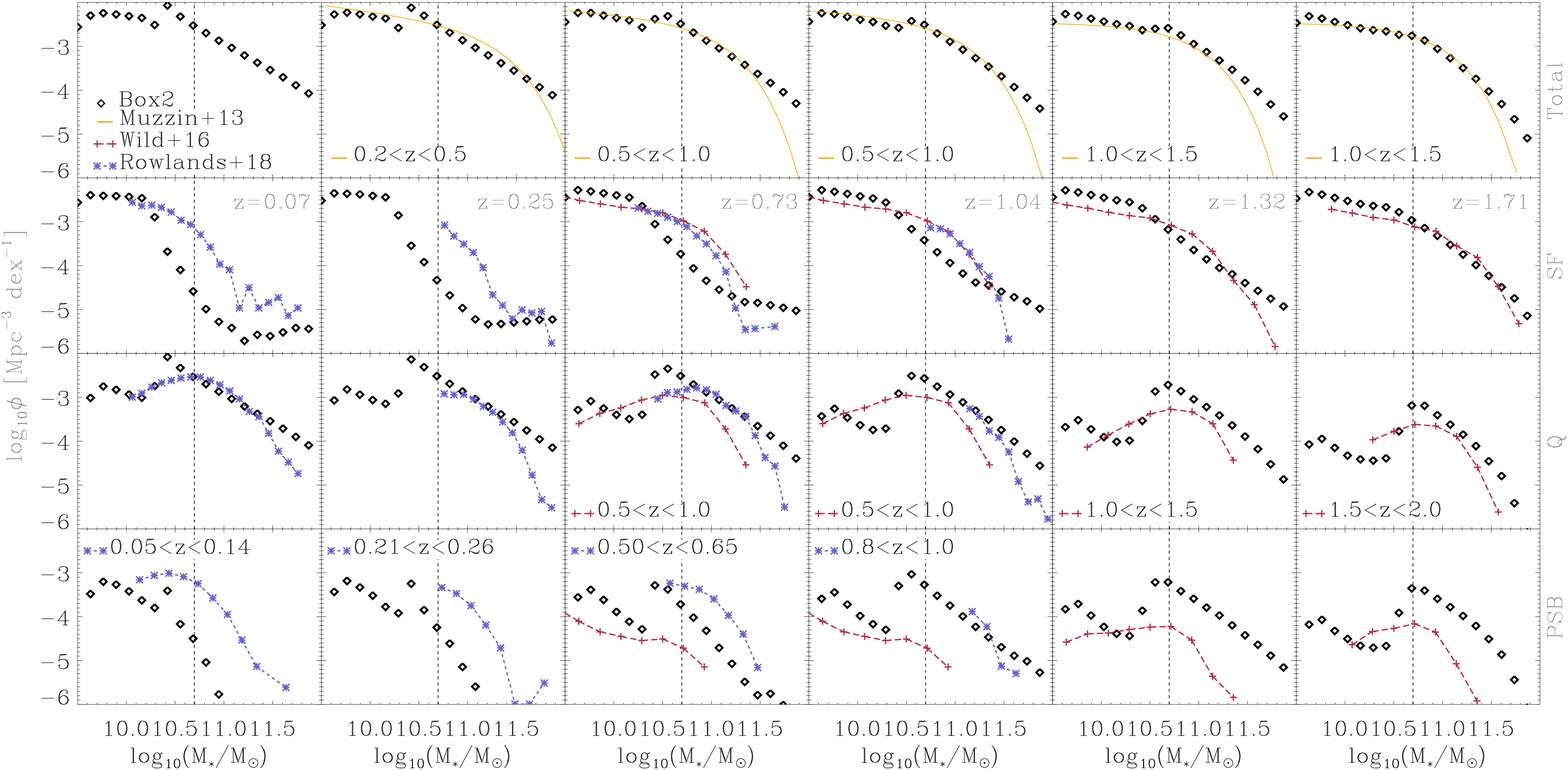}
    \caption{Stellar mass functions of all Magneticum Box2 galaxies, split into total (1st row), star-forming (SF: 2nd row), quenched (Q: 3rd row) and PSB galaxies (4th row) at different redshifts in the range $0.07<z<1.71$ (increasing from left to right). 
    The vertical dashed dotted black line at $\log_{10}(M_*/\mathrm{M_{\odot}}) \sim 10.7$ indicates our standard stellar mass threshold.
    The total stellar mass function (1st row) is compared to $z < 4$ observations based on COSMOS / UltraVISTA \protect\citep{2013ApJ...777...18M}. The SF, Q, and PSB selection is compared to two observational surveys: \protect\cite{2018MNRAS.473.1168R}, based on GAMA ($z < 1$), and \protect\cite{2016MNRAS.463..832W}, based on UKIDSS UDS ($0.5<z<2$).
    }
    \label{fig:SMF_grid}
\end{figure*}

Evaluating the galaxy stellar mass distribution is critical for understanding the relative importance of different evolutionary mechanisms:
Figure \ref{fig:SMF_grid} shows the redshift evolution of the stellar mass function and its various components, as well as comparisons to observations. As such, Figure \ref{fig:SMF_grid} provides a useful extension of Figure \ref{fig:Q_PSBfrac_grid} by displaying the stellar mass distribution and an additional component-wise split into various samples. Although we only consider high mass PSBs ($M_* \geq 4.97 \cdot 10^{10}$) for our analysis, we have extended the stellar mass function below our mass threshold, which is indicated by a vertical dashed dotted black line at $\log_{10}(M_*/\mathrm{M_{\odot}}) \sim 10.7$.

Throughout the studied redshift range ($0.07<z<1.71$) displayed in Figure \ref{fig:SMF_grid}, the total stellar mass function (1st row) shows little evolution.
When comparing the total stellar mass function with observations based on COSMOS / UltraVISTA \citep{2013ApJ...777...18M}, we find agreement at all redshifts, especially towards lower redshifts.
In contrast, the star-forming population (2nd row) shows a significant redshift evolution and only matches observations well at high redshift. The kink in the star-forming stellar mass function at $\log_{10}(M_*/\mathrm{M_{\odot}}) \sim 10.3$, which becomes more evident with decreasing redshift, is the result of our active galactic nucleus (AGN) feedback. Specifically, above these stellar masses the AGN begins to continuously quench galaxies, leading to a relative under-abundance of star-forming galaxies in the stellar mass range $\log_{10}(M_*/\mathrm{M_{\odot}}) \sim [10.3,11.5]$ \citep{2015MNRAS.448.1504S}. This difference becomes most evident when comparing our results to observational surveys based on GAMA ($z < 1$) \citep{2018MNRAS.473.1168R} and on UKIDSS UDS ($0.5<z<2$) \citep{2016MNRAS.463..832W}. 
This relative lack of star-forming galaxies, compared to observations, becomes stronger towards lower redshifts, as more galaxies host AGNs. 
This effect also influences the total and quenched stellar mass functions, as evidenced by the perturbation found at $\log_{10}(M_*/\mathrm{M_{\odot}}) \sim 10.4$ in an otherwise fairly smooth distribution. 

When viewing the evolution of the PSB stellar mass function with redshift in Figure \ref{fig:SMF_grid}, we find a significant evolution: At low redshifts PSBs are primarily found below our stellar mass cut (vertical dashed dotted black line), while they are typically found above our stellar mass cut at high redshifts. 
In other words, the abundance of PSBs above our stellar mass threshold increases significantly with increasing redshift. This strong redshift evolution agrees with VVDS observations, which find that the mass density of strong PSB galaxies is $230$ times lower at $z \sim 0.07$ than at $z \sim 0.7$ \citep{2009MNRAS.395..144W}.
When comparing the shape of the PSB galaxy stellar mass function to observations \citep{2016MNRAS.463..832W, 2018MNRAS.473.1168R}, we do not find close agreement. However, we note that observations at similar redshifts, as indicated by the legend in the bottom row of Figure \ref{fig:SMF_grid}, do not appear to show agreement either. This may be due to different selection mechanisms: While \cite{2016MNRAS.463..832W} derive three eigenvectors, termed super-colours, via a principal component analysis (PCA) of the spectral energy distribution (SED) \citep{2014MNRAS.440.1880W}, \cite{2018MNRAS.473.1168R} use two spectral indices based on a PCA to distinguish different galaxy types. 
In contrast, we determine the percentage of young stars formed within the last $0.5\,$Gyr and the current star formation rate (SFR) (see Section \ref{sub:selection}). Evaluating the SED compared to the numerical star formation may lead to discrepancies.
In short, the PSB stellar mass function appears quite sensitive to the exact selection criteria, both in our simulation and in observations.

\subsection{Galaxy distribution within halos}
\label{sub:FoFNsub}

\begin{figure}
	\includegraphics[width=0.95\columnwidth]{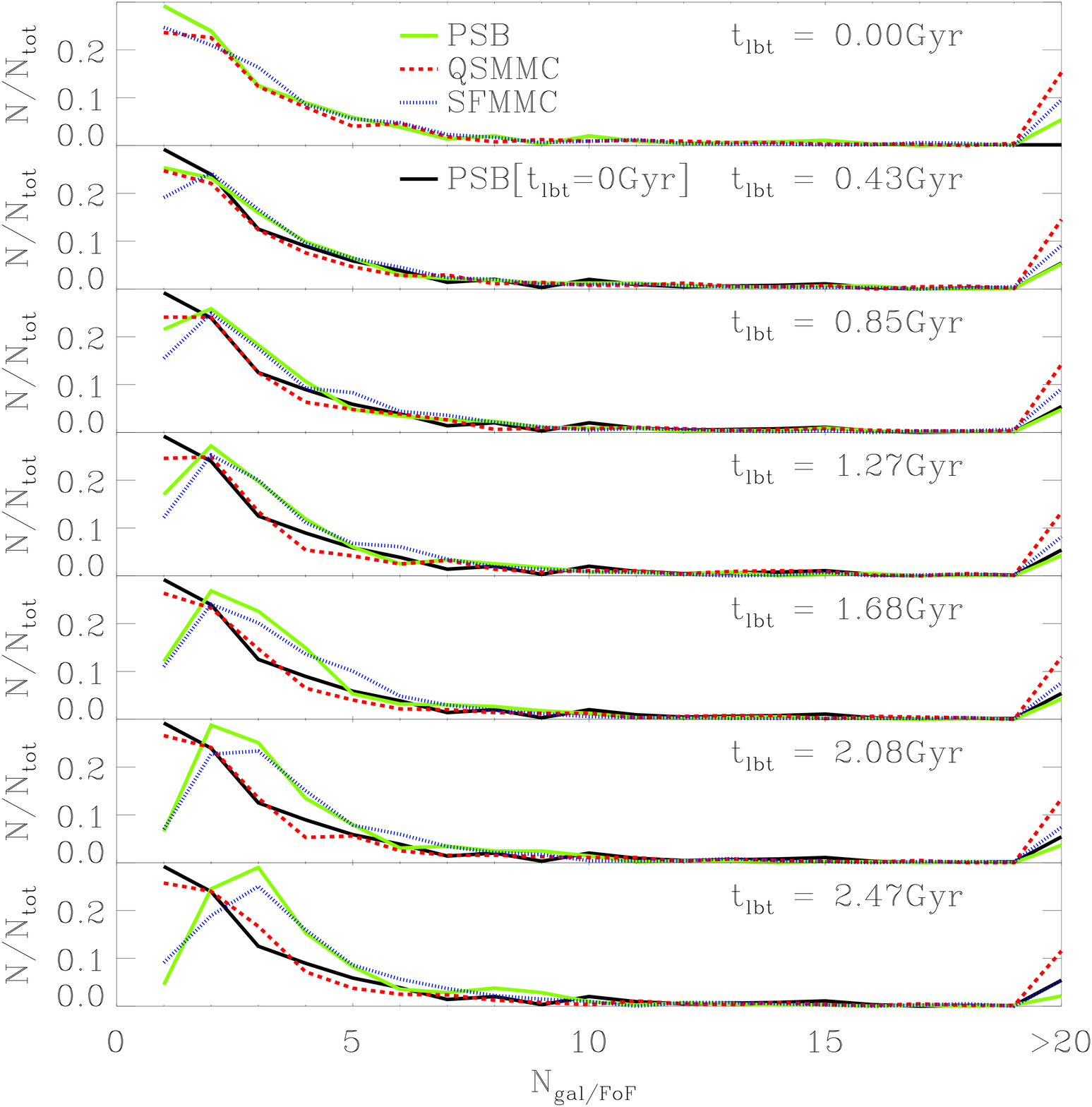}
    \caption{Distribution of $N_{gal/FoF}$, the number of galaxies per Friends-of-Friends (FoF) halo, of PSBs (green), quenched (QSMMC, red), and star-forming stellar mass matched control sample (SFSMMC, blue) galaxies as a function of look-back-time $t_{\mathrm{lbt}}$. Galaxies are identified at $t_{\mathrm{lbt}}=0\,$Gyr  (top panel), thereafter their progenitors are tracked back to $t_{\mathrm{lbt}} \sim 2.5\,$Gyr (bottom panel).
    For a comparison, the PSB $t_{\mathrm{lbt}}=0\,$Gyr distribution is included in each panel as a solid black line. All FoFs with more than $20$ galaxies are grouped together in the last bin. 
    }
    \label{fig:FoFNsubs}
\end{figure}

Figure \ref{fig:FoFNsubs} shows the distribution of the number of galaxies per Friends-of-Friends (FoF\footnote{A FoF linking length of $0.16$ times the mean DM particle separation is used \citep{2009MNRAS.399..497D}. Thereafter, each stellar and gas particle is associated with the nearest DM particle and ascribed to the corresponding FoF group, provided one exists, i.e. has at least $32$ DM particles \citep{2009MNRAS.399..497D}.}) 
halo $N_{\mathrm{gal/FoF}}$ of all PSBs identified at $z \sim 0$ in Box2. All PSBs (green solid lines) were tracked from present-day back over the last $2.5\,$Gyr.
To better understand how PSBs differ from other galaxies, quenched (red dashed lines) and star-forming (blue dotted lines) stellar mass matched control samples (QSMMC and SFSMMC, respectively) of galaxies and their evolution are shown in addition. 

We find a significantly stronger evolution of $N_{\mathrm{gal/FoF}}$ in the PSB (green) and SFSMMC (blue) samples compared to the QSMMC (red) sample. 
At $t_{\mathrm{lbt}}=0\,$Gyr (top panel), the PSB, QSMMC, and SFSMMC samples initially share a similar distribution. The only meaningful exception being the largest bin, i.e. $N_{\mathrm{gal/FoF}} > 20$, which is a factor of $\sim 3$ larger for the QSMMC compared to the PSB sample, indicating a preference of quenched galaxies for richer membership FoF halos. In contrast, PSBs are rarely found in rich membership FoF halos.
In high membership FoF halos, star-forming galaxies lie in intermediate ranges, centred between the other two samples.
The varying galaxy abundances in different halo mass ranges are listed in the bottom row of Table \ref{tab:BHgrowthTable}.
As the look-back-time increases, we find that PSBs, and to lesser degree the SFSMMC galaxies, develop a clear peak around $N_{\mathrm{gal/FoF}} \sim 3$, while values of $N_{\mathrm{gal/FoF}} = 1$ experience a strong decrease. In contrast, the QSMMC distribution remains fairly similar over time. This fundamental difference in evolution of PSB and star-forming galaxies compared to quiescent galaxies suggests that the initial environment at $t_{\mathrm{lbt}} = 2.5\,$Gyr plays an important role in influencing star formation, and subsequently PSB galaxy evolution.

At $t_{\mathrm{lbt}}=2.5\,$Gyr the overwhelming majority of halos in which PSBs (and SFSMMC galaxies) are found, host $N_{\mathrm{gal/FoF}} \sim 2-4$ galaxies. This differs significantly from QSMMC galaxies, which are most often found in halos hosting one galaxy. In contrast, PSBs are rarely found with $N_{\mathrm{gal/FoF}}= 1$, indicating that they are usually not found in isolation\footnote{We note that the number of galaxies found in a given halo is a function of resolution and thus the differences in relative abundance between galaxy types is a more robust quantity.}.
The similarity between the PSB and SFSMMC distributions at $t_{\mathrm{lbt}}=2.5\,$Gyr, shown in the bottom panel of Figure \ref{fig:FoFNsubs}, suggests that star formation is associated with the relative abundance of galaxies in the direct environment. When connecting the initial abundance of galaxies within the FoF halo with the decrease in the number of galaxies found at lower look-back-times, a mechanism linked to the interaction with other galaxies appears likely. Specifically, Figure \ref{fig:FoFNsubs} suggests that galaxy-galaxy processes, such as mergers with nearby galaxies, are important in supporting star formation, as well as possibly being linked to the starburst phase and the following star formation shutdown which characterise PSBs.

\subsection{A closer look: Evolution of massive post-starburst galaxies}

In Table \ref{tab:6massivePSBs} we introduce six massive PSBs, which we study in greater detail alongside the total population of $647$ PSBs. These six massive PSBs are chosen based on their high stellar mass, i.e. higher number of stellar particles, which allows a more detailed (spatial) examination of the involved physical processes. Table \ref{tab:6massivePSBs} lists relevant galactic and halo properties of the six PSBs at $t_{\mathrm{lbt}}=0\,$Gyr.
Similar to the vast majority ($89\%$) of the global $647$ PSB sample (see bottom row in Table \ref{tab:BHgrowthTable}), five of our six massive PSBs are found in halos with halo mass $M_{\mathrm{200,crit}} < 10^{13}\, \mathrm{M_{\odot}}$.

\begin{figure}
	\includegraphics[width=0.95\columnwidth]{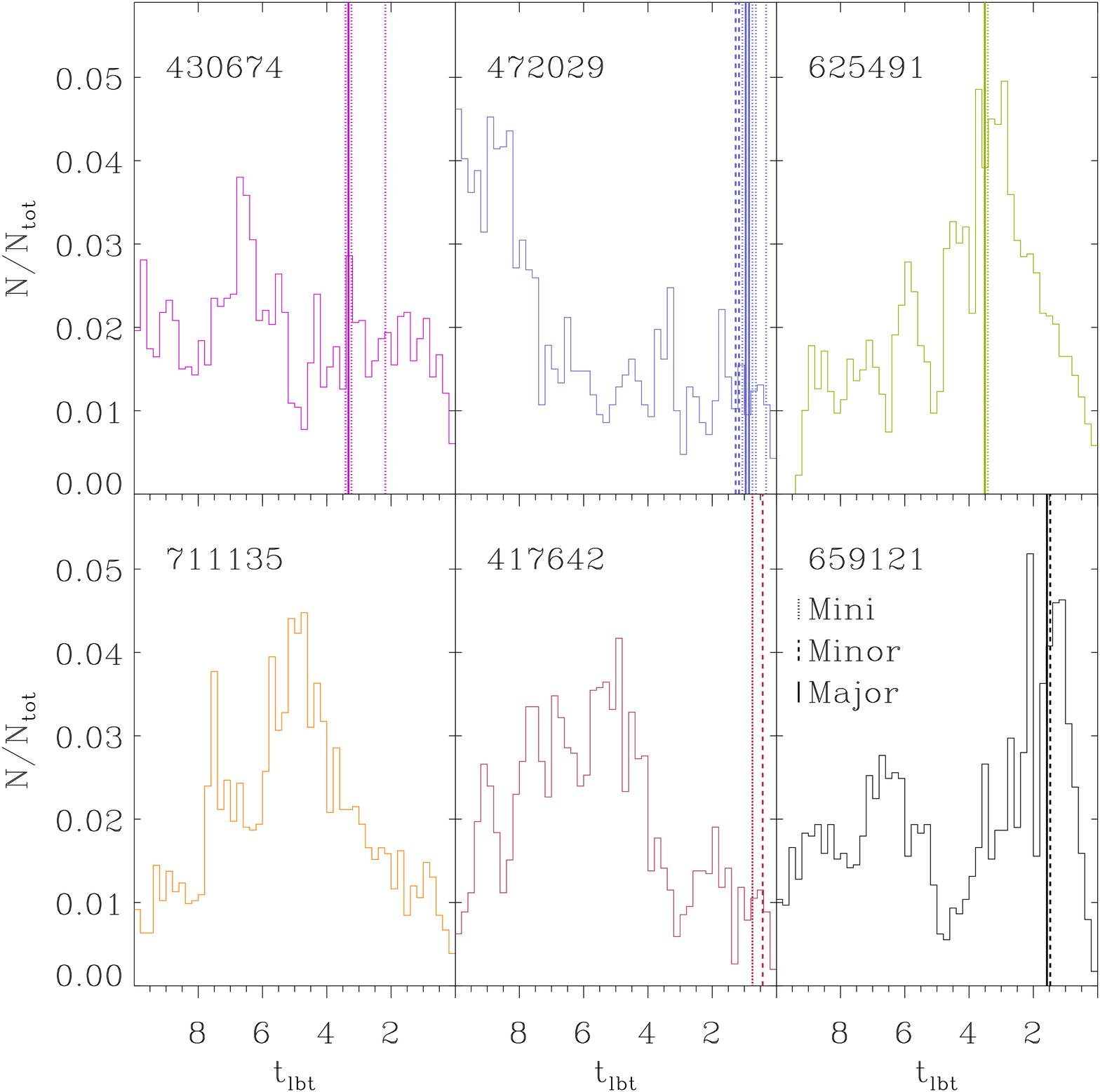}
    \caption{Stellar history, i.e. distribution of newly added stars (in- and ex-situ) as a function of look-back-time for the six massive PSBs introduced in Table \ref{tab:6massivePSBs}. First row IDs from left to right: 430674, 472029, 625491. Second row IDs from left to right: 711135, 417642, 659121. Coloured vertical lines indicate merger events (more easily visible in Figure \ref{fig:AGN_SNe_Energy}): major (solid), minor (dashed), and mini (dotted). 
    Note that only recent mergers with $t_{\mathrm{lbt}} \lesssim 3.6\,$Gyr are shown.
    }
    \label{fig:massiveHistos}
\end{figure}

Figure \ref{fig:massiveHistos} shows the diverse distributions of stellar histories. Specifically, the number of stellar particles added to a given galaxy within a given look-back-time interval for the six massive PSBs is shown. We highlight that this representation includes both internally formed (in-situ) and accreted (ex-situ) stars (whereas the in-situ star formation is shown in Figure \ref{fig:MSpanel}).
The first (last) three galaxies of Table \ref{tab:6massivePSBs} are displayed in the top (bottom) row, as indicated by the IDs.
The stellar histories shown in Figure \ref{fig:massiveHistos} vary: While some massive PSBs are characterised by continuous star formation (and/or accretion) in recent look-back-times (pink, blue), others show strong recent star formation (black). 
Both Table \ref{tab:6massivePSBs} and Figure \ref{fig:massiveHistos} show that massive PSBs with very different properties and stellar histories are captured by our selection criteria.

\begin{table*}
\begin{center}
  \begin{tabular}{| l | c | c | c | c | c | c | c | c |}
    \hline
    ID & $M_*$ [$\mathrm{M_{\odot}}$] & $\mathrm{SSFR} \cdot t_{\mathrm{H}}$ & $M_{\mathrm{gas}}$ [$\mathrm{M_{\odot}}$] & $M_{\mathrm{cgas}}$ [$\mathrm{M_{\odot}}$] & $M_{\mathrm{BH}}$ [$\mathrm{M_{\odot}}$] & $M_{\mathrm{200,crit}}$ [$\mathrm{M_{\odot}}$] & $R_{\mathrm{200}}$ [$\mathrm{kpc}$] & $\#$Galaxies in halo \\ 
    \hline
    430674 & $1.55 \cdot 10^{11}$ & 0.18 & $6.57 \cdot 10^{11}$ & $1.42 \cdot 10^{11}$ & $5.32 \cdot 10^{7}$ & $6.70 \cdot 10^{12}$ & 405 & 13 \\ 
    472029 & $1.52 \cdot 10^{11}$ & 0.00 & $6.42 \cdot 10^{11}$ & $9.58 \cdot 10^{10}$ & $2.25 \cdot 10^{8}$ & $7.67 \cdot 10^{12}$ & 424 & 6 \\ 
    625491 & $1.24 \cdot 10^{11}$ & 0.21 & $2.09 \cdot 10^{11}$ & $9.21 \cdot 10^{10}$ & $8.02 \cdot 10^{7}$ & $2.21 \cdot 10^{12}$ & 280 & 3 \\ 
    711135 & $1.20 \cdot 10^{11}$ & 0.05 & $1.30 \cdot 10^{11}$ & $3.40 \cdot 10^{10}$ & $8.10 \cdot 10^{7}$ & $1.45 \cdot 10^{12}$ & 243 & 1 \\ 
    417642 & $1.11 \cdot 10^{11}$ & 0.00 & $9.04 \cdot 10^{10}$ & $7.43 \cdot 10^{10}$ & $1.76 \cdot 10^{8}$ & $1.09 \cdot 10^{13}$ & 477 & 8 \\ 
    659121 & $1.08 \cdot 10^{11}$ & 0.00 & $1.90 \cdot 10^{11}$ & $6.36 \cdot 10^{10}$ & $1.20 \cdot 10^{8}$ & $1.96 \cdot 10^{12}$ & 269 & 1 \\ 
  \end{tabular}
\end{center} 
\caption{Overview of properties of six massive PSBs at $z \sim 0$ which are studied in greater detail. From left to right: (1) \texttt{SUBFIND} identification (2) Stellar mass, (3) Blueness, (4) Gas mass, (5) Cold gas mass, (6) BH mass, (7) Halo mass, (8) Halo radius, (9) Number of galaxies in halo. All values are given at $t_{\mathrm{lbt}} = 0\,$Gyr.}
\label{tab:6massivePSBs}
\end{table*}

\subsection{Main sequence tracks}
\label{sub:mainseq}

\begin{figure*}
	\includegraphics[width=1.95\columnwidth]{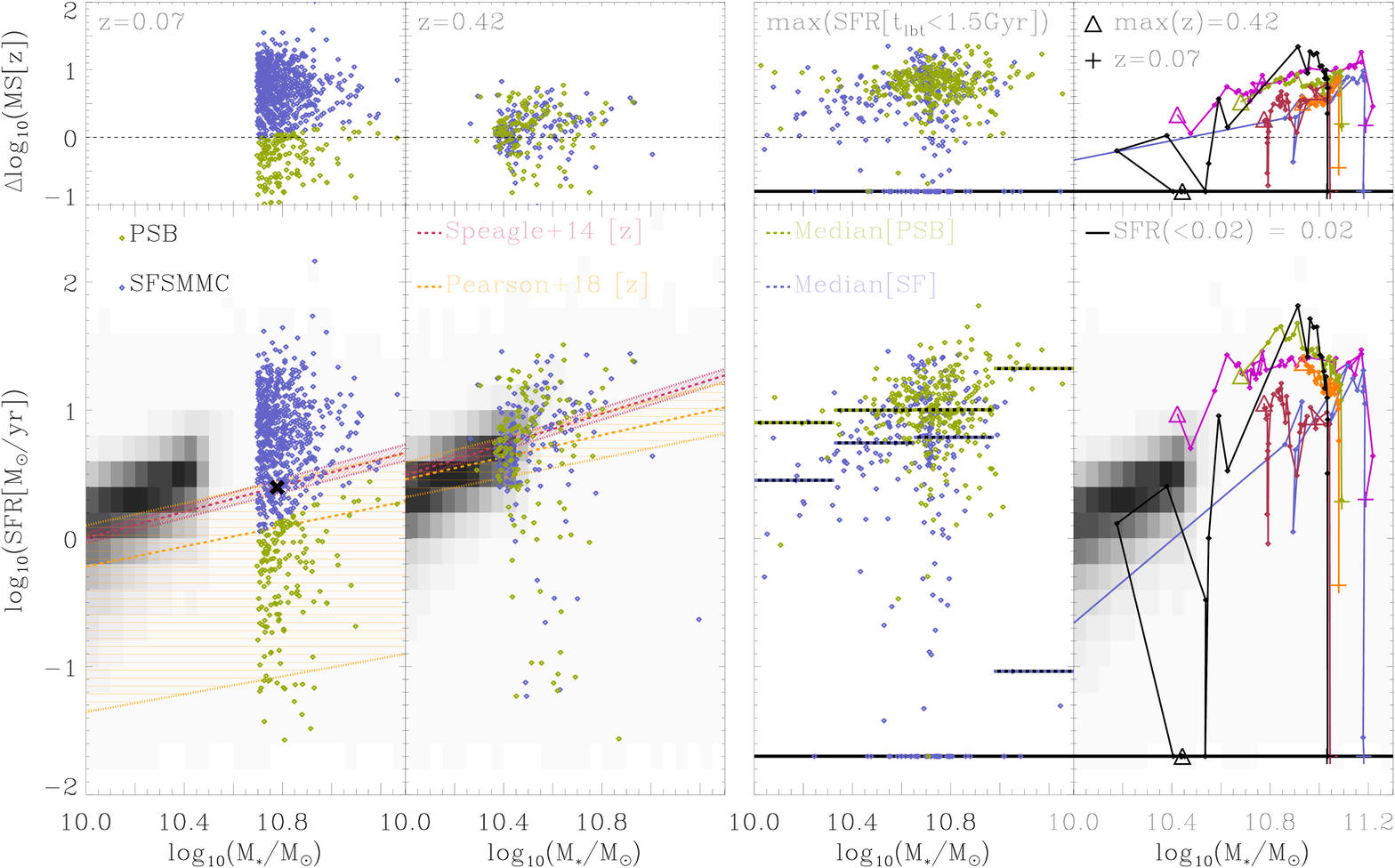}
    \caption{Post-starburst (PSB) progenitor evolution in the stellar mass - star formation rate plane. 1st panel (from left to right): PSB (green) and star-forming stellar mass matched control (SFSMMC: blue) galaxies at $z = 0.07$, i.e. at look-back-time (lbt) $t_{\mathrm{lbt}}=0\,$Gyr. 2nd panel: PSB and SFSMMC progenitors at $z = 0.42$, i.e. $t_{\mathrm{lbt}}=3.6\,$Gyr. 3rd panel: Peak star formation rate (SFR) for each PSB and corresponding SFSMMC progenitor with median SFR bins. 
    4th panel: Evolution of a subset of six massive PSBs (see Table \ref{tab:6massivePSBs}), which are tracked through time, ending at $t_{\mathrm{lbt}}=0\,$Gyr (crosses). Each step (small diamonds), represents an incremental increase of $t_{\mathrm{lbt}} \sim 0.11\,$Gyr, ultimately arriving at the furthest tracked progenitors at a maximum $t_{\mathrm{lbt}}=3.6\,$Gyr (triangles). 
    The shaded regions (1st and 2nd panel) provide redshift dependent observational fits to the main sequence \protect\citep{2014ApJS..214...15S,2018A&A...615A.146P}. 
    The grey density distribution shows the abundance and location of all Magneticum Box2 galaxies at $z = 0.07$ (1st and 4th panel) and $z = 0.42$ (2nd panel).
    For plotting purposes, in the 3rd and 4th panel, low SFR galaxies are artificially set to $SFR(< 0.02) = 0.02$ (and $\Delta[<-0.8] \log_{10}(MS[z]) = -0.8$ in the top panels). 
    For reference, a recent Milky Way stellar mass estimate $M_{*,\mathrm{MW}} \sim 6 \cdot 10^{10}\, \mathrm{M_{\odot}}$ has been included (black X) in the 1st panel \protect\citep{2015ApJ...806...96L}.
    The top panels show the logarithmic relative deviation from the \protect\cite{2018A&A...615A.146P} redshift evolving main sequence, i.e. $\Delta \log_{10}(MS[z]) = \log_{10}(SFR[z]/MS[z])$, for each population in the connected main panel.
    }
    \label{fig:MSpanel}
\end{figure*}

Figure \ref{fig:MSpanel} shows the positions of post-starburst (PSB: green) and star-forming (SF: blue) galaxies and their progenitors in the stellar mass - star formation rate (SFR) plane from left to right at: $z = 0.07$ (1st panel), $z = 0.42$ (2nd panel), peak PSB star formation (3rd panel), and the evolution of the six massive PSBs (4th panel) introduced in Table \ref{tab:6massivePSBs}. To compare the behaviour with observations, we have added main sequence fits (shaded regions) for redshifts $z=0.4$ and $z=0.1$ \citep{2014ApJS..214...15S, 2018A&A...615A.146P}.
The six massive PSBs in the right panel are identified at $z = 0.07$ (crosses), i.e at $t_{\mathrm{lbt}} = 0\,$Gyr. The PSB progenitors are then tracked backwards in intervals of $t_{\mathrm{lbt}} \sim 0.11\,$Gyr, yielding additional data points (small diamonds). PSB progenitors are tracked to a maximum redshift of $z=0.42$ (triangle), i.e. to $t_{\mathrm{lbt}} \sim 3.6\,$Gyr, depending on how recently their black holes have been seeded (see Section \ref{sub:BHgrowthStat} for more details). 
Additionally, the logarithmic relative deviation of galaxies from the observationally based redshift evolving main sequence (MS) fit \citep{2018A&A...615A.146P} is plotted in the top panels, i.e. $\Delta \log_{10}(MS[z]) = \log_{10}(SFR[z]/MS[z])$. In other words, $\Delta \log_{10}(MS[z]) = 0$ galaxies lie on the main sequence and positive (negative) values correspond to the logarithm of the factor they lie above (below) the main sequence. 

The distribution of PSB (green) and SF (blue) galaxies at $z = 0.07$ is shown in the first panel of Figure \ref{fig:MSpanel}. The dichotomy found at $t_{\mathrm{lbt}}=0\,$Gyr, i.e. at our identification time, is the result of our selection criteria (see Section \ref{sub:selection}): By design, PSBs are quenched, while the SFSMMC sample is characterised by star formation. When comparing to observations, we find that this dichotomy is well described by observations \citep{2014ApJS..214...15S, 2018A&A...615A.146P}. Furthermore, we see that the SF galaxies appear as an extension of the grey density distribution describing the abundance of all Box2 Magneticum galaxies, while the PSBs scatter below the main sequence. 
The grey density distribution experiences a strong cut-off at $\log_{10}(\mathrm{M_*} / \mathrm{M_{\odot}}) \gtrsim 10.4$ because the SF population is characterised by a strong decline at this stellar mass (see Figure \ref{fig:SMF_grid}).
We note that the relative deviation from the evolving main sequence shown above the first panel uses a redshift of $z=0.09$, as the observational fit is no longer defined at $z=0.07$ \citep{2018A&A...615A.146P}, the redshift showing Magneticum results.

The $z=0.42$ display of PSB and SF progenitors in the second panel of Figure \ref{fig:MSpanel} shows no meaningful difference between the two populations. 
Both populations match the behaviour of the general distribution of the main sequence of Magneticum galaxies at the same redshift, which is shown in the underlying grey density distribution. Furthermore, the general Box2 Magneticum galaxy distribution (grey), as well as the PSB and SF galaxies are well described by observational fits at $z=0.4$ \citep{2014ApJS..214...15S, 2018A&A...615A.146P}. We note that the least amount of galaxies are found in the second panel, compared to the first and third panel, because not all galaxies can be traced back to higher redshifts. Furthermore, at $z=0.4$ galaxies with $SFR = 0$ are not shown due to the logarithmic scaling.

The third panel in Figure \ref{fig:MSpanel} displays the distribution of PSB and SF progenitors at the height of PSB progenitor star formation within $t_{\mathrm{lbt}} < 1.48\,$Gyr, i.e. since $z = 0.19$. If available, the corresponding SFSMMC galaxy to a given PSB galaxy is displayed, otherwise a random unique SFSMMC at the same redshift is shown for comparison. The median PSB peak star formation occurs at $z=0.13$, i.e. at $t_{\mathrm{lbt}}=0.75\,$Gyr. When comparing the two populations we find that PSBs are characterised by higher SFRs than SFSMMC galaxies, as illustrated by the dashed horizontal lines indicating the median of each population at different stellar mass intervals. 
Interestingly, $\sim 17\%$ of SF progenitors (blue, 3rd panel) at peak PSB progenitor star formation are found on the black horizontal line, i.e. have $SFR < 0.02$. These SFSMMC galaxies were previously quiescent and have become star-forming at $z \sim 0$ via recent mergers, i.e. have been rejuvenated.

In this context, we note that the recent SFR of the most massive progenitors need not always be correlated with a young stellar population at $t_{\mathrm{lbt}}=0\,$Gyr. This is evidenced by the fact that within our PSB sample we have a galaxy which has no in-situ star formation over the evaluated time-span, as illustrated by the green diamond (3rd panel) found on the black horizontal line showing galaxies with $SFR(< 0.02) = 0.02$.
In other words, galaxies need not have formed in-situ stars to host a young stellar population, rather, as is the case for the mentioned PSB galaxy, young ex-situ stars may also be accreted during mergers, leading to a young stellar population in the merger remnant at $t_{\mathrm{lbt}}=0\,$Gyr (see also Section \ref{sub:gasEvol}).

The fourth panel of Figure \ref{fig:MSpanel} shows that massive PSB progenitors are found significantly above the main sequence prior to their quiescent phase at $t_{\mathrm{lbt}}=0\,$Gyr (crosses). PSB progenitors display prolonged strong star formation episodes, with SFRs consistently being significantly larger than the redshift evolving main sequence \citep{2018A&A...615A.146P}. 
Generally, independent of the duration, starbursts of massive PSB progenitors are found in the range $5 \lesssim \Delta MS[z]/MS[z] \lesssim 20$ above the redshift evolving main sequence.

In Figure \ref{fig:MSpanel}, we find both galaxies that continuously remain above the main sequence as well as galaxies that experience rejuvenation, i.e. galaxies which were initially below the main sequence but rise above it during their starburst phase. The starburst timescales ($t_{\mathrm{sb}}$) differ widely and are within the range $t_{\mathrm{sb}} \sim (0.4-3)\,$Gyr.
This spread in timescales is a reflection of the different star formation histories prior to the starburst. As the global $647$ PSB sample is tracked backwards, the sample size is reduced, especially if BHs are recently seeded. This results in a sample size of $455$ tracked PSB progenitors, which reach a $t_{\mathrm{lbt}} \geq 2.5\,$Gyr. Of these $455$ successfully tracked PSBs, $105$ are considered to be rejuvenated galaxies, i.e. $23\%$. 
Independent of whether galaxies are rejuvenated or show sustained star formation, they show a sharp decline in star formation at the end of the starburst phase. Typically, this decline to passive levels of star formation happens within $\lesssim 0.4\,$Gyr. 
This conflicts with our understanding of the typical behaviour of field galaxies, which make up the vast majority of our sample (see last row of Table \ref{tab:BHgrowthTable}), as field galaxies generally experience a gradual decline in average SFR \citep{2007ApJ...660L..43N}. In other words, the (massive) PSBs in Figure \ref{fig:MSpanel} not only show enhanced, often sustained, starbursts, but also experience an abrupt cessation of star formation, the details of which are discussed in Section \ref{sec:shutdown}.

\section{The role of mergers}
\label{sec:mergers}

It is well established that galaxy mergers impact the galactic star formation rate (SFR), both directly \citep{2005MNRAS.361..776S, 2009ApJ...697L..38J, 2018MNRAS.478.3447E, 2019MNRAS.489.4196L} and indirectly \citep{2013MNRAS.430.1901H, 2014MNRAS.437.1456B, 2014ApJ...792...84Y}. 
However, the nature and relevant parameters of the mergers and how they influence the SFR is still debated. For example, while the SIMBA cosmological simulations find an increasing impact \citep{2019MNRAS.490.2139R}, observations based on SDSS, KiDS, and CANDELS find that mergers do not significantly impact the SFR, compared to non-merging systems \citep{2019A&A...631A..51P}, and observations based on $32$ PSBs from LEGA-C suggest that mergers likely trigger the rapid shutdown of star formation found in PSBs \citep{2020ApJ...888...77W}.
To disentangle this complex relationship between mergers and the SFR, we investigate mergers in Box2, both on an individual basis as well as statistically.

\subsection{Case study: Gas evolution}
\label{sub:gasEvol}

\begin{figure*}
	\includegraphics[width=1.95\columnwidth]{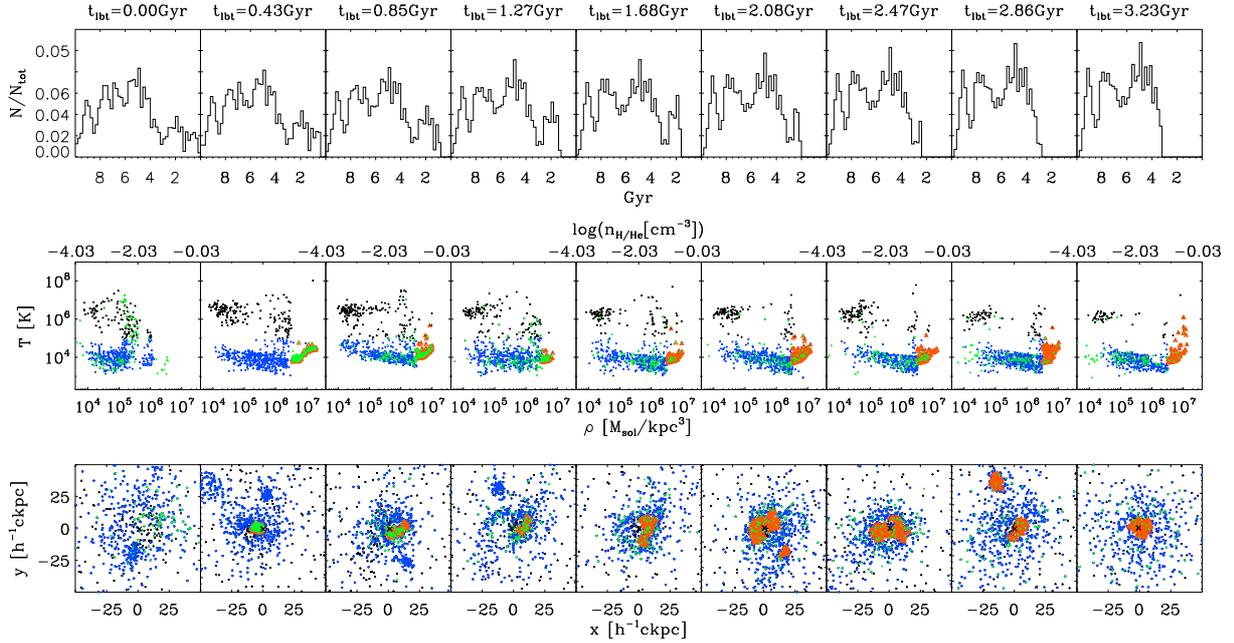}
    \caption{Case study of the gas evolution of one of the PSBs ($417642$), as introduced in Table \ref{tab:6massivePSBs}. The first row shows the stellar history as a function of look-back-time, while the second row shows the gas phase diagram. The third row displays the evolution of the spatial distribution (co-moving side length $100h^{-1}ckpc$) of gas (black), cold gas (blue), i.e. $T < 10^5\,K$, and star-forming gas (orange). To better understand the evolution of the gas involved in star formation, all star-forming gas at $t_{\mathrm{lbt}}=0.43\,$Gyr (green) is identified and tracked through all look-back-times to reveal its origin, as well as its distribution at $t_{\mathrm{lbt}}=0\,$Gyr. The tracked gas (green) uses a smaller symbol size, to show the over-plotted star-forming gas (orange) in the background. The 'x' at the centre of the spatial distribution marks the black hole of the PSB (progenitor).}
    \label{fig:traceGasMassive}
\end{figure*}

The case study of one typical PSB (progenitor), selected from Table \ref{tab:6massivePSBs} (ID=417642), is shown in Figure \ref{fig:traceGasMassive}. The goal is to map the (cold) gas evolution as a means of investigating the initial triggering of the starburst and the following starburst phase. To uncover the mechanisms involved, Figure \ref{fig:traceGasMassive} shows the evolution of the stellar history (1st row), the gas phase (2nd row), and a projection of the spatial gas distribution (3rd row) as a function of look-back-time for the selected PSB (progenitor). To better visualise the evolution of the gas involved in the recent starburst phase, all star-forming gas at $t_{\mathrm{lbt}}=0.43\,$Gyr, i.e. one time-step before the shutdown, is identified and subsequently coloured green. These identified gas particles maintain their green colouring both prior to and after this look-back-time. 

When considering higher look-back-times in Figure \ref{fig:traceGasMassive}, we find that the recent increase in star formation at $t_{\mathrm{lbt}} \sim 2 \,$Gyr coincides with a close galaxy-galaxy interaction, followed by a major merger event at $t_{\mathrm{lbt}} = 0.75 \,$Gyr (see solid red vertical line in Figure \ref{fig:massiveHistos}). The period between $t_{\mathrm{lbt}} \sim (0-2)\,$Gyr is characterised by prolonged star formation.
This is not an exception, but rather most PSB progenitors experience recent merger events (see Table \ref{tab:MergerTable}). It appears that the initial close galaxy-galaxy interactions and the subsequent mergers provide a mechanism by which gas is transported inwards, increasing the number of gas particles above the density threshold (2nd row) required for star formation. 
The increase in the supply of cold, dense gas within the PSB progenitor then enables the starburst.

Similarly to the vast majority of PSBs surveyed in this manner, a strong diffusion of gas is registered in Figure \ref{fig:traceGasMassive} at $t_{\mathrm{lbt}}=0\,$Gyr, following the starburst phase. In the second row, we find a strong decrease in gas density, accompanied by an overall increase in temperature within a timescale of $t \sim 0.4\,$Gyr, as evidenced by the distribution of previously star-forming gas (green) over the entire density and temperature regime. This behaviour at low look-back-times is mirrored in the spatial domain (3rd row), which also provides evidence for a strong redistribution of previously star-forming gas (green). Although the spatial distribution widens, large cold gas reservoirs remain within the PSB galaxy at $t_{\mathrm{lbt}}=0\,$Gyr, agreeing with recent observations \citep{2020ApJ...900..107Y}.
We reviewed the gas evolution of multiple different PSBs and verified that the behaviour shown in Figure \ref{fig:traceGasMassive} is not an exception, but rather typical for our (massive) PSB sample. 

When viewing the stellar history in the first row of Figure \ref{fig:traceGasMassive}, the relative weighting of different components appears to change as time progresses. For example, at $t_{\mathrm{lbt}}=1.68\,$Gyr the onset of the starburst appears to be significantly stronger (compared to the older stars) than at lower look-back-times. Investigating this behaviour, we found that a significant population of older stars are accreted onto the PSB progenitor, thus impacting the relative abundance of different components of the stellar history. In other words, during the presented merging process, more ex-situ old stars are accreted than young in-situ stars are formed (see also Section \ref{sub:mainseq}).

\subsection{Merger statistics}
\label{sub:mergerStat}

To extend the case study conducted in Section \ref{sub:gasEvol} by a statistical analysis, we begin by evaluating the merger history of the $z \sim 0$ global $647$ PSB sample. In addition, we also analyse the two (quiescent and star-forming, respectively) stellar mass matched control samples QSMMC and SFSMMC. The results of the merger tree evaluation for these samples are listed in Table \ref{tab:MergerTable}.

Mergers are defined by their progenitor peak stellar mass ratio within the past four snapshots prior to merger identification, i.e. $t_{\mathrm{lbt}} \leq 0.43\,$Gyr: mini mergers 1:10 - 1:100, minor mergers 1:3 - 1:10, and major mergers 1:1 - 1:3.
The first data row lists the sample size of successfully constructed merger trees. This value is less than the total sample size ($647$), as merger trees only exist over the entire evaluated time-span of $t_{\mathrm{lbt}} = 2.5\,$Gyr if the main progenitor was formed prior to this time-span. The next three rows list the total number of identified mergers (galaxies can have multiple mergers of the same type) for each type. The next three rows in Table \ref{tab:MergerTable} display the percentage of the analysed merger trees which identify at least one merger event of the respective type. The last row lists the percentage of galaxies with at least one merger event, independent of the type.

Table \ref{tab:MergerTable} shows that the PSB sample is characterised by an abundance of merger events. This agrees with low redshift observations, which find that PSBs are associated with interactions and/or mergers \citep{2004ApJ...607..258Y, 2008ApJ...688..945Y, 2009MNRAS.396.1349P, 2017A&A...597A.134M}. 
Specifically, $64.7\%$ of PSBs experience a major merger within the last $2.5\,$Gyr. 
In contrast, only $9.4\%$ of QSMMC galaxies experience a major merger within the same time-span, while this percentage rises to $58.1\%$ for SFSMMC galaxies. 
Compared to the QSMMC, the PSB sample experiences a factor of $\sim 7$ more major merger events.
When comparing the samples, we find close similarities between the PSB and SFSMMC sample, i.e. both show an abundance of mergers. In contrast, the QSMMC sample is characterised by a low abundance of mergers and differs significantly from the other two samples. However, it is not clear that this is typical for PSBs identified at higher redshifts.

\begin{table}
\begin{center}
  \begin{tabular}{| l | c | c | c |}
    \hline
    Criterion & PSBs & QSMMC & SFSMMC \\
    \hline
    Analysed trees  & 632 & 646 & 630 \\ \hline
    $\Sigma(N_{\mathrm{mini}})$  & 343 & 114 & 285 \\ \hline 
    $\Sigma(N_{\mathrm{minor}})$ & 295 & 53 & 260 \\ \hline 
    $\Sigma(N_{\mathrm{major}})$ & 465 & 65 & 415 \\ \hline 
    $N_{\geq 1 \mathrm{ mini}}$      & $40.7\%$ & $14.1\%$ & $33.7\%$  \\ \hline 
    $N_{\geq 1 \mathrm{ minor}}$     & $37.3\%$ & $ 7.4\%$ & $33.8\%$  \\ \hline 
    $N_{\geq 1 \mathrm{ major}}$     & $64.7\%$ & $ 9.4\%$ & $58.1\%$  \\ \hline 
    $N_{\geq 1 \mathrm{ merger}}$    & $88.9\%$ & $23.4\%$ & $79.7\%$  \\ 
  \end{tabular}
\end{center} 
\caption{Overview of different merger abundances of our global $z \sim 0$ identified PSB sample and its stellar mass matched control (SMMC) samples, subdivided into quiescent (QSMMC) and star-forming (SFSMMC) samples. The first data row displays the number of successfully analysed merger trees out of the $647$ galaxies traced for each sample over the time-span $t_{\mathrm{lbt}} = (0.0-2.5)\,$Gyr. The next three rows list the total number of mergers $\Sigma(N)$ encountered over the evaluated time-span, subdivided into the following classes and stellar mass ratios: Mini 1:10 - 1:100, Minor 1:10 - 1:3, Major 1:3 - 1:1. The subsequent three rows list the percentage of galaxies with respect to the analysed merger trees, which encountered at least one merger event of the respective type ($N_{\geq 1}$). The last row shows the percentage of galaxies which encountered at least one merger, independent of type.}
\label{tab:MergerTable}
\end{table}

\begin{table} 
\begin{center}
  \begin{tabular}{| l | c | c | c |}
    \hline
    Criterion & PSBs & QSMMC & SFSMMC \\ \hline
    Analysed trees               & 10520 & 10596 & 10479 \\ \hline
    $\Sigma(N_{\mathrm{mini}})$   & 8559 & 4899 & 8692 \\ \hline 
    $\Sigma(N_{\mathrm{minor}})$  & 6747 & 2439 & 6832 \\ \hline 
    $\Sigma(N_{\mathrm{major}})$  & 6014 & 1638 & 6822 \\ \hline 
    $N_{\geq 1 \mathrm{ mini}}$   & $50.7\%$ & $33.6\%$ & $51.3\%$ \\ \hline 
    $N_{\geq 1 \mathrm{ minor}}$  & $50.5\%$ & $20.6\%$ & $50.1\%$ \\ \hline 
    $N_{\geq 1 \mathrm{ major}}$  & $47.3\%$ & $14.3\%$ & $52.7\%$ \\ \hline 
    $N_{\geq 1 \mathrm{ merger}}$ & $92.6\%$ & $51.9\%$ & $92.3\%$ \\ 
  \end{tabular}
\end{center} 
\caption{Same as Table \ref{tab:MergerTable} but showing an overview of different merger abundances of $z \sim 0.9$ identified PSBs (initial PSB sample size of $10624$) and their control galaxies (QSMMC and SFSMMC). 
The galaxies were traced for each sample over the time-span $t_{\mathrm{lbt}} = (6.5-9.0)\,$Gyr. 
}
\label{tab:Nbeg064MergerTable}
\end{table}

In Section \ref{sec:environment}, we showed the redshift evolution of both the PSB-to-quenched fraction and the PSB stellar mass function. In this context, we investigate the abundance of mergers at redshift $z=0.9$, in the same manner as outlined for our global $z \sim 0$ PSB sample. This is motivated by the desire to separate the redshift evolution of identically selected samples from differences resulting from different (later) environmental selections. We choose redshift $z=0.9$ because we also study the merger abundance in the cluster environment (see Section \ref{sub:clusterMergers}) and compare it to observations (see Section \ref{sub:vlosObsComp}) at this redshift.

As established by Figures \ref{fig:Q_PSBfrac_grid} and \ref{fig:SMF_grid}, the abundance of PSBs increases with increasing redshift. Table \ref{tab:Nbeg064MergerTable} reflects this too, as significantly more PSBs are identified at $z=0.9$ (10624 galaxies), compared to $z\sim 0$ (647 galaxies). 
Beyond this, we find that: First, the percentage of galaxies which experience more than one merger (last row) increases, especially for the QSMMC sample (factor $\sim 2$), less so for the SFSMMC sample (increase by $\sim 12\%$), and least for the PSB sample (increase by $\sim 3\%$). Second, the similarity between the PSB and the SFSMMC sample remains, as both continue to show similar (high) merger abundances compared to the QSMMC sample. Third, the overall increase in the abundance of mergers is especially driven by more mini and minor mergers at $z=0.9$.
This behaviour at $z \sim 0.9$ agrees with LEGA-C observations at $z \sim 0.8$, which find that central starbursts are often the result of gas-rich mergers, as evidenced by the high fraction of PSB galaxies with disturbed morphologies and tidal features ($40\%$) \citep{2020MNRAS.497..389D}.

Albeit differences existing between Tables \ref{tab:MergerTable} ($z \sim 0$) and \ref{tab:Nbeg064MergerTable} ($z = 0.9$), the link between recent (in relation to the identification redshift) star formation and the abundance of mergers appears strong. To summarise, although PSBs are quiescent at identification redshift, they are characterised by recent (strong) star formation.  
The similarity with respect to merger abundances between star-forming and PSB galaxies is likely driven by the ability of mergers to trigger starbursts on short timescales and to provide cold gas on longer timescales to otherwise exhausted galaxies \citep{2010MNRAS.407.2091G, 2012MNRAS.419.3200H}. 
In short, we find strong evidence that mergers are linked to increased star formation, while their absence is linked to quiescent levels of star formation. 
Consequently, the high abundance of mergers appears to be central to the evolution of PSB galaxies, while likely also playing an important role in the subsequent shutdown.

\subsection{Cold gas fractions}
\label{sub:cgas_frac}

\begin{figure*}
	\includegraphics[width=0.95\columnwidth]{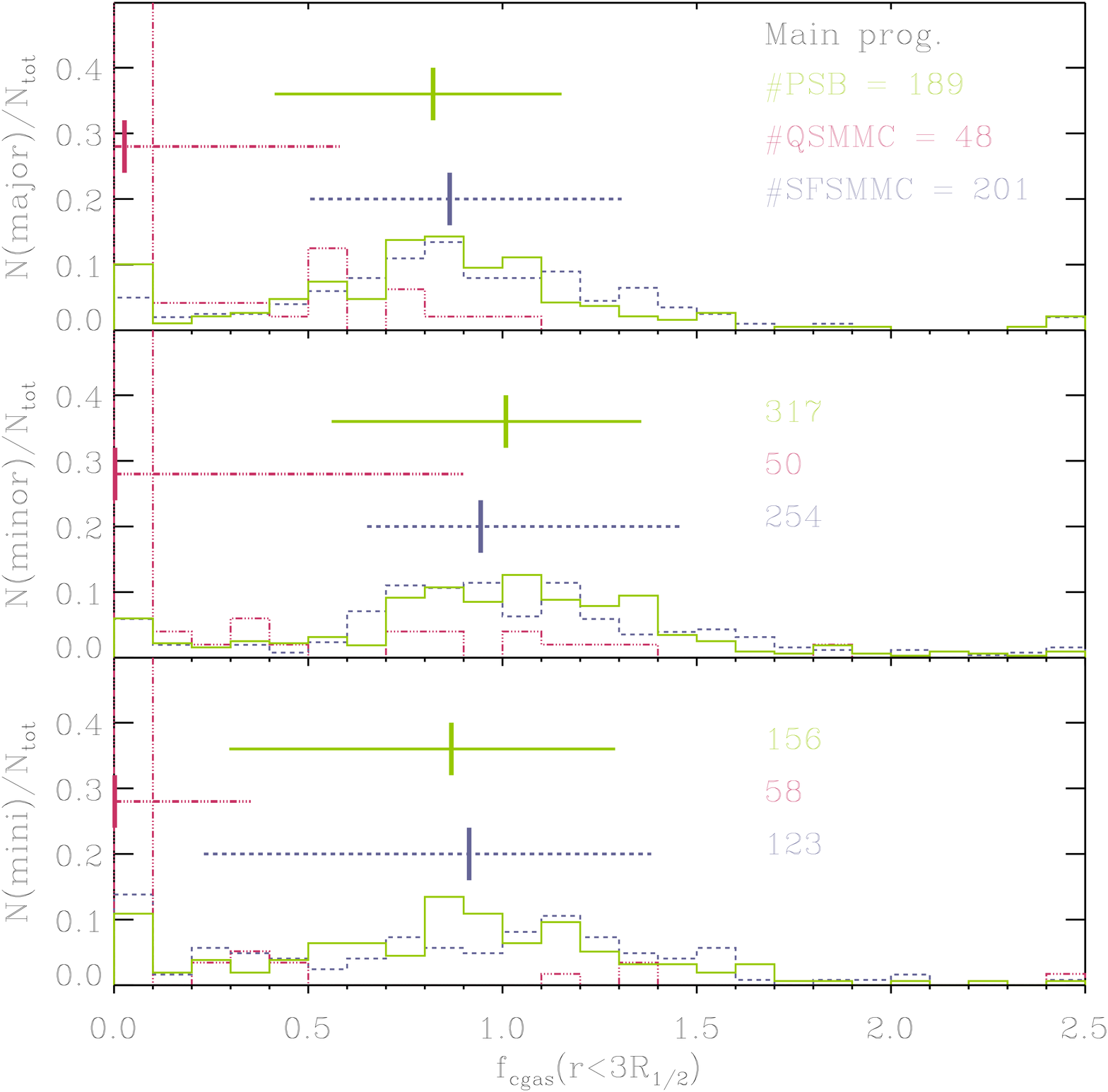}
	\includegraphics[width=0.95\columnwidth]{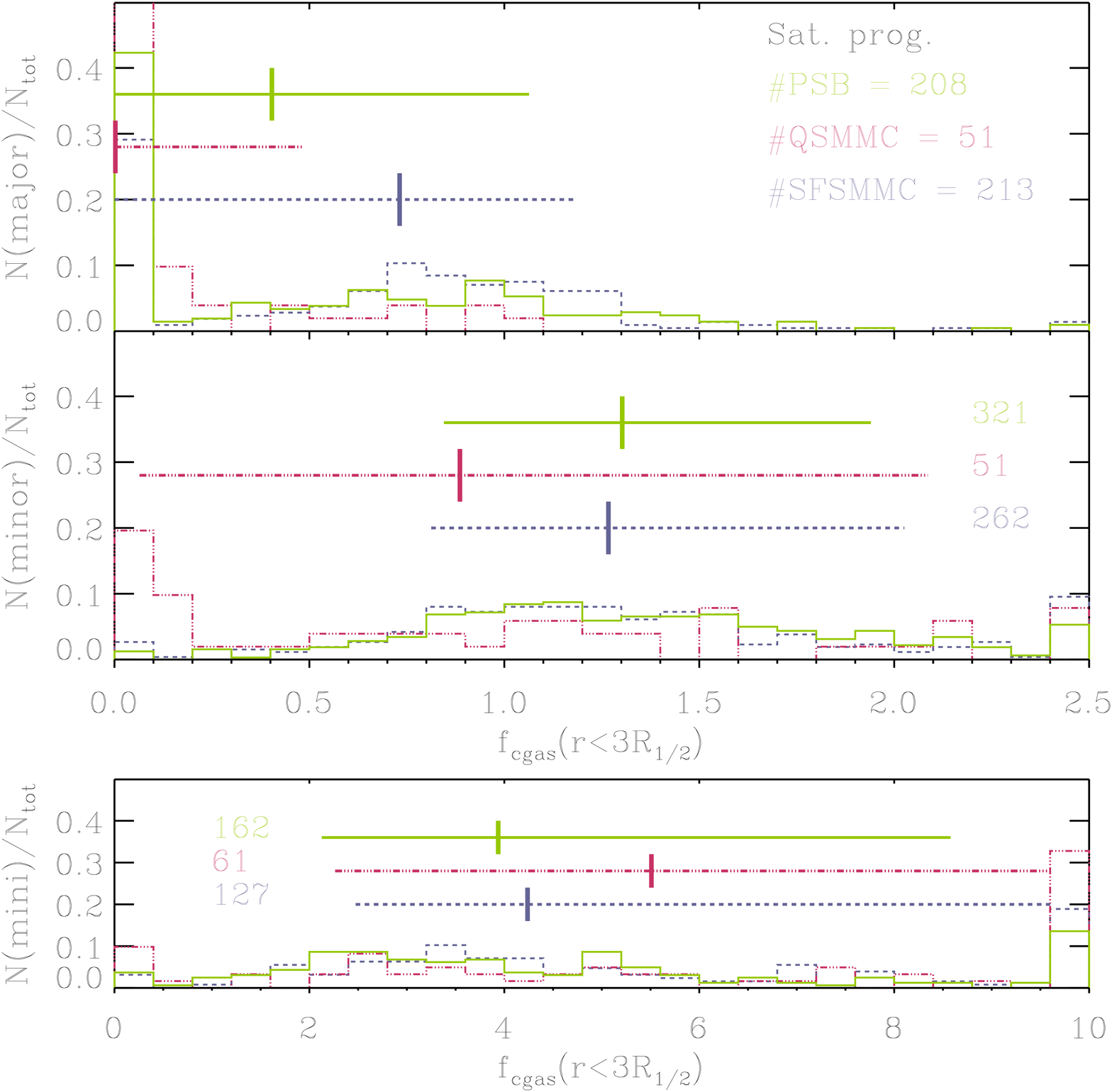}
    \caption{Distribution of cold gas fraction $f_{cgas} = M_{cold,gas}/M_*$ within three half-mass radii $r < 3 \, R_{1/2}$ for main (left) and satellite progenitors (right). The distributions are further split into major (top), minor (middle), and mini (bottom) mergers and show the behaviour of the PSB (green), QSMMC (red), and SFSMMC (blue) samples.
    The short solid vertical lines indicate the median values for each population, while the horizontal lines indicate the $1\,\sigma$ region, i.e. the range between the $15.9\%$ and $84.1\%$ percentile.
    In contrast to all other panels, the panel on the bottom right shows a four times larger $f_{cgas}$ domain. 
    }
    \label{fig:merger_wetness} 
\end{figure*} 

As the timescales of the galaxy-galaxy interactions prior to the detection of a merger event vary widely, depending on the specific geometry of the encounter, we do not individually correlate merger events with the onset of the starburst phase.
Rather, to more closely evaluate the properties of the detected mergers and to investigate their differences, we determine the cold gas fractions $f_{\mathrm{cgas}} = M_{\mathrm{cold,gas}}/M_*$ prior to mergers for the $z \sim 0$ PSB, QSMMC, and SFSMMC samples. 
The cold gas fraction is calculated within three half-mass radii $r < 3 \, R_{1/2}$, where the half-mass radius is defined as the radius of a three dimensional sphere containing half of the total galactic stellar mass. We choose $R_{1/2}$, as its use is well established within our simulations and it is often considered equal to the observationally attained effective radius $R_e$ \citep{2015ApJ...812...29T, 2017MNRAS.464.3742R, 2018ApJ...854L..28T, 2020MNRAS.493.3778S}.
We tested the impact of choosing different half-mass radii ($r/R_{1/2} = [0.5,5]$) on $f_{\mathrm{cgas}}$ and found consistent behaviour for varying half-mass radii.

Figure \ref{fig:merger_wetness} shows the distribution of cold gas fractions $f_{\mathrm{cgas}}$ within three half-mass radii $r < 3 \, R_{1/2}$, split up into main (left) and satellite progenitors (right). We further split the sample into major (top), minor (middle), and mini (bottom) mergers. When a merger event is registered, we determine the cold gas fraction prior to the merger event, i.e. we identify the progenitors peak stellar mass in the $\leq 0.43\,$Gyr before the event is registered and determine the cold gas fraction at this time-step. 
Each progenitor is then assigned to the respective merger type distribution. This is done separately for the PSB (green), QSMMC (red), and SFSMMC (blue) sample. The solid vertical lines indicate the median values of each population, while the horizontal lines are bounded by the percentiles $15.9\%$ and $84.1\%$ respectively, i.e. the equivalent $1\,\sigma$ region of a Gaussian distribution.

All panels showing the individual main progenitor distribution (left) in Figure \ref{fig:merger_wetness} display similar distributions for different merger ratios. The reason for this is that independent of the given merger ratio, by definition, the merging main progenitor has the same cold gas fraction. 
Every time a merger occurs, the population of main progenitors is sampled, resulting in a similar cold gas fraction distribution for all main progenitors, independent of the merger ratio.

In contrast, each sample of satellite progenitors (right) shows an evolving behaviour with merger type. Figure \ref{fig:merger_wetness} (right) displays that the $f_{\mathrm{cgas}}$ distribution for each sample migrates towards higher $f_{\mathrm{cgas}}$ values as the stellar mass ratio between main and satellite progenitor decreases, i.e. when moving towards smaller mergers. In a nutshell, less massive merging satellite progenitors have higher relative abundances of cold gas. 

We find that the main progenitor behaviour of the (median) cold gas fraction distribution of the PSB and SFSMMC sample is similar with $f_{\mathrm{cgas}}(r < 3 \, R_{1/2}) \sim (0.8-1.0)$ for the main progenitors. The PSB and SFSMMC satellite progenitors show an expected (see above) stronger variance in cold gas fractions between merger types. In contrast to the PSB and SFSMMC galaxies, Figure \ref{fig:merger_wetness} shows that the QSMMC sample consistently has lower $f_{\mathrm{cgas}}$ values: The quiescent main progenitors (left) have $f_{\mathrm{cgas}}(r < 3 \, R_{1/2}) \sim 0$, i.e. compared to their stellar mass almost no cold gas is present in the galaxies. The satellite progenitors (right) also show that satellites which merge via major or minor mergers into the QSMMC sample typically have lower cold gas fractions, compared to the PSB and SFSMMC sample.

Taking all this into account, it appears that Figure \ref{fig:merger_wetness} provides some evidence for \textit{galactic conformity}, i.e. the effect whereby properties, e.g. the star formation rate, of satellite galaxies appear correlated to the properties of the central galaxy \citep{2016ApJ...817....9K, 2017MNRAS.472.4769T, 2017MNRAS.472.2504T, 2018MNRAS.477..935T}. In other words, star-forming and PSB main progenitors appear more likely to merge with satellite progenitors which have similarly high cold gas fractions, while quiescent main progenitors appear more likely to merge with satellite progenitors which exhibit more cold gas depletion, i.e. lower $f_{\mathrm{cgas}}$.

Interestingly, the strongest difference between the PSB and SFSMMC sample in Figure \ref{fig:merger_wetness} is found for major mergers of satellite progenitors (top right): Statistically, the median cold gas fraction of SFSMMC major merger satellite progenitors ($f_{\mathrm{cgas}}(r < 3 \, R_{1/2}) = 0.73$) is almost twice as large compared to PSBs ($f_{\mathrm{cgas}}(r < 3 \, R_{1/2}) = 0.40$). 
This is further evidenced by the different abundances at small cold gas fractions: $69\%$ of QSMMC, $42\%$ of PSB, and $29\%$ of SFSMMC satellite major merger progenitors are found within $f_{\mathrm{cgas}}(r < 3 \, R_{1/2}) \lesssim 0.1$. 
As $65\%$ of $z \sim 0$ PSB and $58\%$ of SFSMMC galaxies experienced at least one merger within the last $2.5\,$Gyr (Table \ref{tab:MergerTable}), this difference in cold gas supply marks an important distinction between the, otherwise often similar, populations. The implications associated with the difference in cold gas supply during major mergers, especially for the shutdown of star formation, are discussed in Section \ref{sub:disc:mergers}.

The bottom right panel of Figure \ref{fig:merger_wetness} displays a four times larger domain. The mini mergers of satellite progenitors show a significantly flatter distribution of $f_{\mathrm{cgas}}$, while simultaneously having significantly higher $f_{\mathrm{cgas}}$ values. This is likely the result of infalling cold gas over-densities being classified as mini mergers or gas-rich satellites merging with their host. Subsequently, the low number of stellar particles compared to the abundant (cold) gas particles, drives high values of $f_{\mathrm{cgas}}$. Due to the low resolution of mini merger satellite progenitors, this panel is less relevant to understanding mergers, while still showing that (cold) gas inflow is relatively similar ($f_{\mathrm{cgas}}(r < 3 \, R_{1/2}) \sim 4.0-5.5$) for all analysed samples, with the highest values found in the QSMMC sample.

\section{Shutdown of star formation}
\label{sec:shutdown}

\subsection{Active galactic nucleus and supernova feedback}
\label{sub:AGN+SNe}

We investigate both the active galactic nuclei (AGN) as well as the supernovae (SNe) feedback energy output as a means to better understand the processes involved in shutting down star formation. Specifically, we want to shed light on processes which are linked to the short timescale ($t \sim 0.4\,$Gyr) redistribution and heating of previously star-forming gas, as discussed in Section \ref{sub:gasEvol}. 
We choose these mechanisms in particular because they are able to deposit large amounts of energy on short timescales \citep{2005MNRAS.361..776S, 2015ApJ...803L..21V, 2020MNRAS.494..529W}, thereby potentially strongly impacting star formation.

As a precaution, we also investigated the typical depletion timescales of cold gas in PSB progenitors during peak star formation. We find the timescales to be significantly higher ($t_{\mathrm{depl}} \sim (2-5)\,$Gyr) than the short shutdown timescale ($t_{\mathrm{shutdown}} \lesssim 0.4\,$Gyr) found throughout our PSB sample. In other words, PSBs progenitors do not appear to run out of gas, rather the reservoir of cold, dense gas is abruptly heated and/or redistributed, leading to a shutdown in star formation, as demonstrated in Figure \ref{fig:traceGasMassive}.

\begin{figure*}
	\includegraphics[width=0.95\columnwidth]{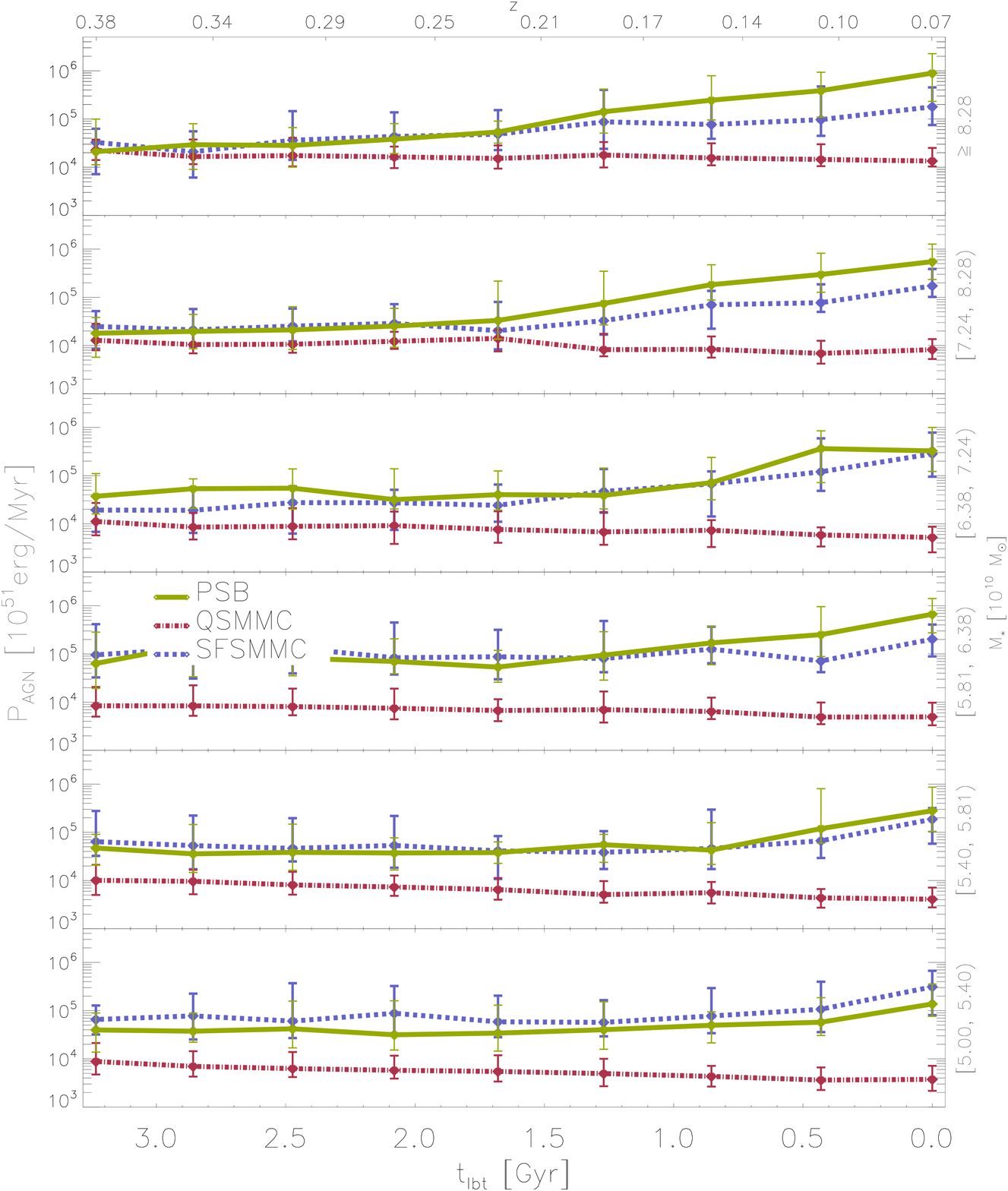}
	\includegraphics[width=0.95\columnwidth]{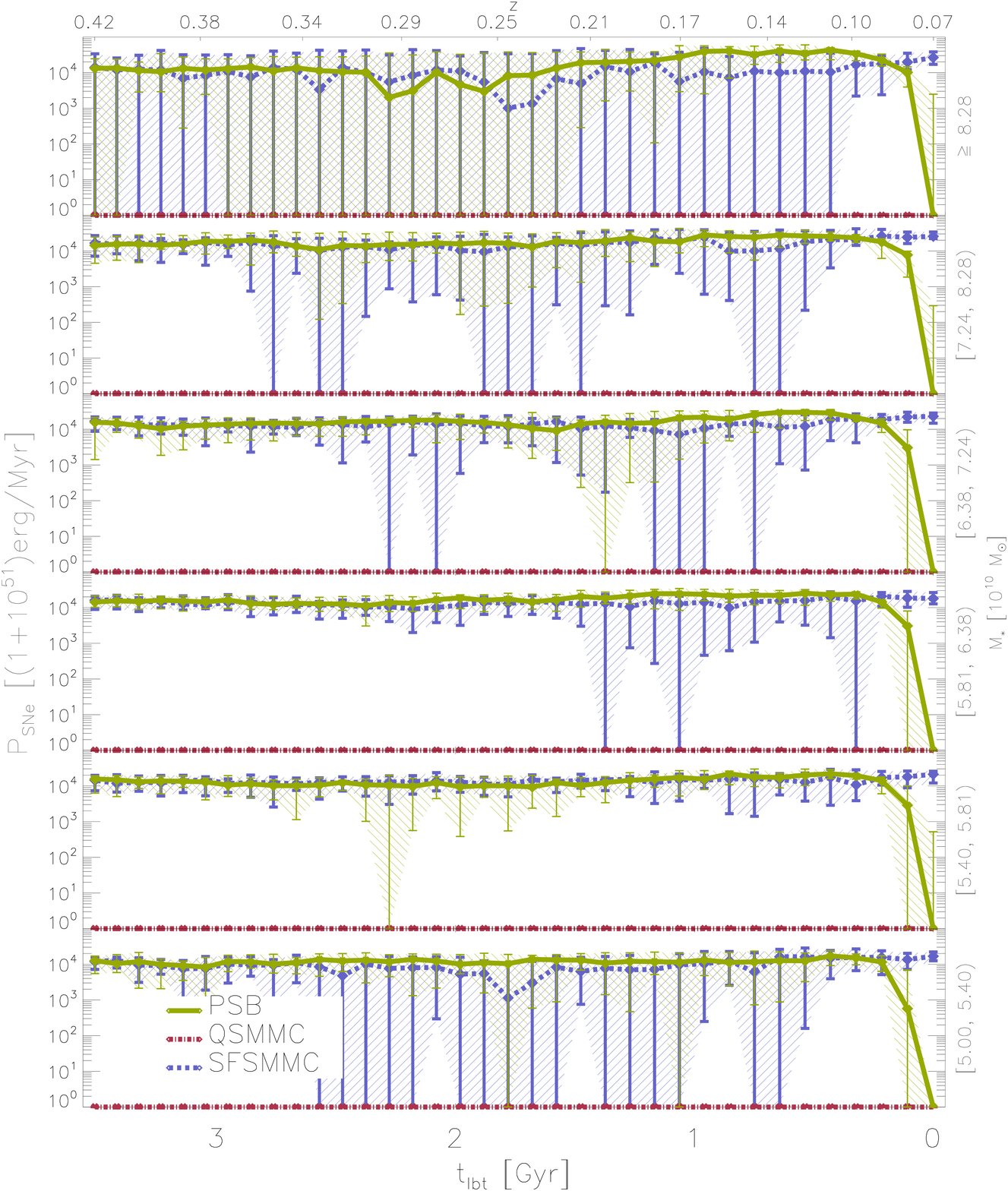}
    \caption{Active galactic nuclei (AGN: left figure) and supernovae (SNe: right figure) power output of the global PSB (green), QSMMC (red), and SFSMMC (blue) samples identified at $z \sim 0$ and evaluated over the past $\sim 3.2\,$Gyr and $\sim 3.5\,$Gyr in units of $10^{51}\,$erg/Myr and $1+10^{51}\,$erg/Myr, respectively. 
    The different panels (from bottom to top) show $t_{\mathrm{lbt}}=0\,$Gyr increasing equal bin size stellar mass intervals: 
    ${M_* \in [[5.00,5.40), [5.40,5.81), [5.81,6.38), [6.38,7.24), [7.24,8.28), \geq 8.28] \cdot 10^{10}\, \mathrm{M_{\odot}}}$. 
    Both figures show the median, as well as the $0.5\,\sigma$ region as error bars for each population. 
    }
    \label{fig:AGN_SNe_Energy_mass_evol} 
\end{figure*}

\begin{figure}
	\includegraphics[width=0.95\columnwidth]{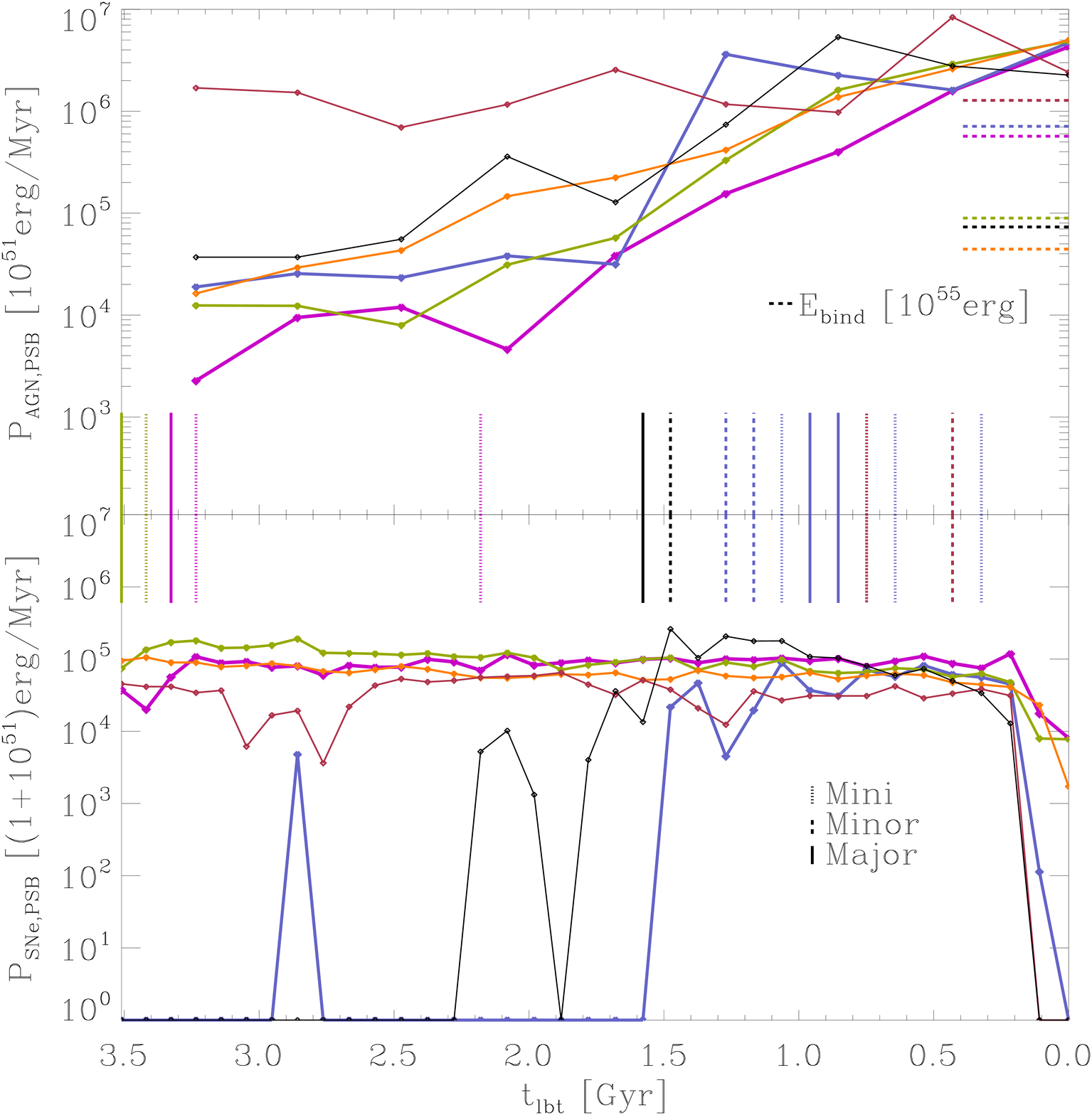}
    \caption{Energy deposited by AGN (top) and SNe (bottom) in units of $10^{51}\,$erg/Myr and $1+10^{51}\,$erg/Myr for the six massive PSBs, introduced in Table \ref{tab:6massivePSBs}, over the past $\sim 3.5\,$Gyr. 
    Vertical lines (following the colour scheme) indicate different merger events: mini (1:10 - 1:100) mergers (dash dotted line), minor (1:3 - 1:10) mergers (dashed line), and major (1:1 - 1:3) mergers (solid lines).
    We note, that the temporal resolution differs by a factor of four between the AGN (top) and SNe (bottom) energy output.
    The horizontal lines in the top panel (right) show an estimation of the spherical binding energy of the massive PSBs in units of $10^{55}\,$erg.
    }
    \label{fig:AGN_SNe_Energy} 
\end{figure}

We calculate the AGN power output $P_{\mathrm{AGN}}$ based on the change in BH mass $\Delta M_{\mathrm{BH}}$ between time steps ($\Delta t = 0.43\,$Gyr) \citep{2014MNRAS.442.2304H}:
\begin{equation}
    P_{\mathrm{AGN}} = (e_r e_f \Delta M_{\mathrm{BH}} \cdot c^2) / \Delta t
    \label{eq:E_AGN}
\end{equation}
where $e_r=0.2$ is the fraction of energy which is thermally coupled to the surrounding gas, and $e_f$ is a free parameter usually set to $e_f=0.15$ (typical for simulations following metal dependent cooling functions \citep{2009MNRAS.398...53B, 2011MNRAS.413.1158B}). 

As we are especially interested in SNe which release their energy on short timescales, our focus is on short lived, i.e. massive, stars. Therefore, supernovae Type II (SNeII), which arise at the end of the lifetime of massive stars, provide the dominant source of supernovae feedback in our analysis \citep{1976ApJ...207..872C}.
Following the star formation model by \cite{2003MNRAS.339..289S}, we expect an average SN energy release per stellar mass of $\epsilon_{\mathrm{SN}} = 4 \cdot 10^{48}\,\mathrm{erg}\mathrm{M_{\odot}}^{-1}$. Combining this with the star formation rate $SFR$ at each time step, (temporal resolution $\Delta t = 0.11\,$Gyr), we receive the following estimation for the SNe power output $P_{\mathrm{SNe}}$: 
\begin{equation}
    P_{\mathrm{SN}} = \epsilon_{\mathrm{SN}} \cdot SFR
    \label{eq:E_SNe}
\end{equation}

The results of these calculations are shown in Figure \ref{fig:AGN_SNe_Energy_mass_evol} for equal bin size stellar mass intervals: 
$M_* \in [[5.00,5.40), [5.40,5.81), [5.81,6.38), [6.38,7.24), [7.24,8.28), \\ \geq 8.28] \cdot 10^{10}\, \mathrm{M_{\odot}}$.
On the left-hand side each data point displays the median AGN power output calculated from the difference in BH mass between time-steps, as indicated in Equation \ref{eq:E_AGN}. 
Following Equation \ref{eq:E_SNe}, data points in the right figure display the SNe power output estimation based on the current star formation rate ($SFR$). 
When the median SFR is zero, which is the case for the entire QSMMC sample, the SNe power output is zero too. Both figures shown in Figure \ref{fig:AGN_SNe_Energy_mass_evol} show the respective median values, as well as the $0.5\,\sigma$ region as error bars (additionally shaded on the right). The different temporal resolution between the two figures is the result of using the BH particle data on the left, which due to storage constraints is only saved every $0.43\,$Gyr, and using \texttt{SUBFIND} data on the right, which is available every $0.11\,$Gyr (see Section \ref{sub:selection}). 

In the stellar mass interval $M_* \in [5.00,5.40) \cdot 10^{10}\, \mathrm{M_{\odot}}$ (bottom panel) characterised by the weakest AGN power output (left), the AGN still strongly outweighs the SNe power output (right):
We find the maximum median SNe power output for PSB galaxies to be $P_{\mathrm{SNe,PSB}} \leq 2 \cdot 10^{55}\,$erg/Myr. In contrast, the maximum median AGN power output is $P_{\mathrm{AGN,PSB}} \geq 10^{56}\,$erg/Myr for the same stellar mass selection.
In other words, Figure \ref{fig:AGN_SNe_Energy_mass_evol} shows that the AGN outweighs the SNe power output by half an order of magnitude, especially at recent look-back-times. 

Figure \ref{fig:AGN_SNe_Energy_mass_evol} (left) shows negligible differences between PSB and SF galaxies at lower stellar masses: Both samples show a recent increase in AGN feedback, which is significantly larger than that of the quenched sample, especially towards more recent look-back-times. However, with increasing stellar mass the difference between PSB and SF galaxies increases. Specifically, in the highest stellar mass interval, i.e. $M_* \geq 8.28 \cdot 10^{10}\, \mathrm{M_{\odot}}$ (top panel), the difference at $t_{\mathrm{lbt}} = 0\,$Gyr is of order half a magnitude between PSBs ($P_{\mathrm{AGN,PSB}} \sim 10^{57}\,$erg/Myr) and SF ($P_{\mathrm{AGN,SF}} \sim 2 \cdot 10^{56}\,$erg/Myr) galaxies. In contrast to the recent elevation in AGN feedback found in PSB and to a lesser degree in SF galaxies (depending on the stellar mass interval), AGN feedback of quiescent galaxies shows no meaningful temporal evolution and only a weak stellar mass evolution ($P_{\mathrm{AGN,Q}} \sim 10^{55}\,$erg/Myr in the highest stellar mass interval).

In Figure \ref{fig:AGN_SNe_Energy_mass_evol} (right) the PSB and SF galaxies show similar median SNe feedback. However, even at $M_* \in [5.00,5.40) \cdot 10^{10}\, \mathrm{M_{\odot}}$ (bottom panel), where PSB and SF galaxies show the most similarities, we see a large spread in SNe feedback in the SF sample, while PSBs show a smaller spread in the distribution of SNe feedback. 
Independent of stellar mass, this is especially the case at recent look-back-times, $t_{\mathrm{lbt}} \sim [0.1,1]\,$Gyr: During this period PSBs are typically experiencing their starburst phase. As a result, the SFR is elevated throughout the entire sample, which due to its linear relation to the SNe feedback (see Equation \ref{eq:E_SNe}) results in a tighter and slightly elevated distribution compared to SF galaxies, as evidenced by smaller error bars. 
Meanwhile, the quiescent galaxy sample is continuously characterised by a lack of SNe feedback, as no meaningful star formation occurs in the sample during the evaluated time span. As dictated by our selection criteria, PSBs show a strong decrease in SNe feedback energy at $t_{\mathrm{lbt}} \sim 0\,$Gyr. 

In addition to our statistical analysis (Figure \ref{fig:AGN_SNe_Energy_mass_evol}), in Figure \ref{fig:AGN_SNe_Energy}, we consider the individual AGN (top) and SNe (bottom) feedback of the six massive PSBs, described in Table \ref{tab:6massivePSBs}. 
We have added vertical lines indicating specific merger events colour coded to match the associated galaxy:
When evaluating the last $3.5\,$Gyr, we find that the six massive PSBs experienced 16 merger events, compared to 10 in the associated SFSMMC, and 2 in the QSMMC sample. 
As previously established in Sections \ref{sub:gasEvol} and \ref{sub:mergerStat}, this further highlights the significance of mergers for the evolution of (massive) PSBs. 

Similarly to the comparison between AGN and SNe feedback shown in Figure \ref{fig:AGN_SNe_Energy_mass_evol}, Figure \ref{fig:AGN_SNe_Energy} also shows that the AGN feedback significantly outweighs the SNe feedback, especially at recent look-back-times. Specifically, within the last $t_{\mathrm{lbt}} \leq 0.5\,$Gyr all six PSBs have an AGN feedback ($P_{\mathrm{AGN,PSB}} \gtrsim 10^{57}\,$erg/Myr) which outweighs the SNe feedback ($P_{\mathrm{SNe,PSB}} \lesssim 10^{56}\,$erg/Myr) by more than an order of magnitude.
Furthermore, Figure \ref{fig:AGN_SNe_Energy} shows that most of the mergers (vertical lines) in the PSB sample occur within the last $\sim 1.5\,$Gyr, i.e. during the same time in which the AGN power output increases by up to $\sim 2$ orders of magnitude.

As a rough comparison, we calculate an estimation of the spherical binding energy ($E_{\mathrm{bind}} = 3GM^2/5R$) of the massive PSBs using the $M_{\mathrm{200,crit}}$ halo mass and $R_{\mathrm{200}}$ radius as displayed in Table \ref{tab:6massivePSBs} for $M$ and $R$, respectively.
The resulting estimation is shown as horizontal lines in the top panel (right) of Figure \ref{fig:AGN_SNe_Energy}. To compare with the power output, the horizontal binding energy lines use a different scale [$10^{55}\,$erg], as indicated by the legend. 
Five out of the six massive PSBs have binding energies with $E_{\mathrm{bind}} \leq 10^{61}\,$erg and the PSB with the most massive halo (shown in Figure \ref{fig:traceGasMassive}) has a binding energy of $E_{\mathrm{bind}} = 1.278 \cdot 10^{61}\,$erg. 
Most binding energies are found within an order of magnitude of the AGN energy released within the last time step $t_{\mathrm{lbt}} \lesssim 0.43\,$Gyr, which further highlights the strong impact the AGN has on (massive) PSBs.
Furthermore, we note that the extensive amount of power deposited by the AGN ($P_{\mathrm{AGN,PSB}} \gtrsim 10^{57}\,$erg/Myr) during $t_{\mathrm{lbt}} \lesssim 0.43\,$Gyr is correlated with the gas temperature increase, gas density decrease, and general redistribution of gas seen in Figure \ref{fig:traceGasMassive} at $t_{\mathrm{lbt}}=0\,$Gyr.
Thus, we find strong evidence that the AGN is connected and probably responsible for the shutdown of the star formation in (massive) PSBs.

\subsection{Black hole growth statistics}
\label{sub:BHgrowthStat}

To complement the analysis in Section \ref{sub:AGN+SNe}, we additionally quantify the black hole (BH) growth for our different samples. 
We calculated both the relative and absolute BH growth:
Indeed, only $7.8\%$ of QSMMC galaxies, compared to $60.2\%$ and $62.7\%$ of the SFSMMC and PSB galaxies, at least double their BH mass over the last $2.5\,$Gyr. In absolute terms, $80.1\%$ of PSB and $73.7\%$ of the SFSMMC galaxies experience a significant mass growth of $\Delta M_{\mathrm{BH}} \geq 10^7\,\mathrm{M_{\odot}}$, while this is only the case for $18.7\%$ of the QSMMC galaxies. 

\begin{figure*}
	\includegraphics[width=0.95\columnwidth]{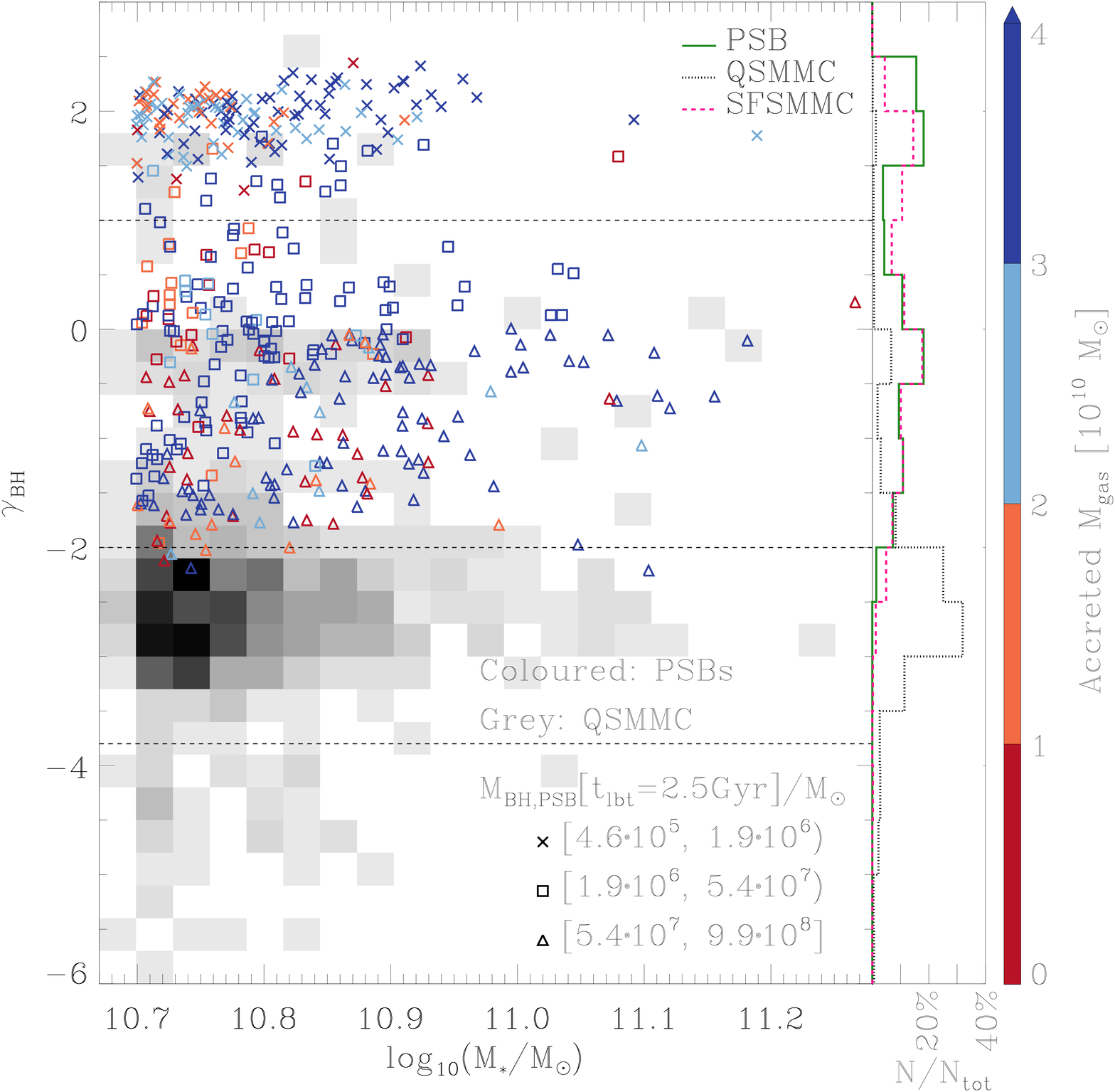}
	\includegraphics[width=0.95\columnwidth]{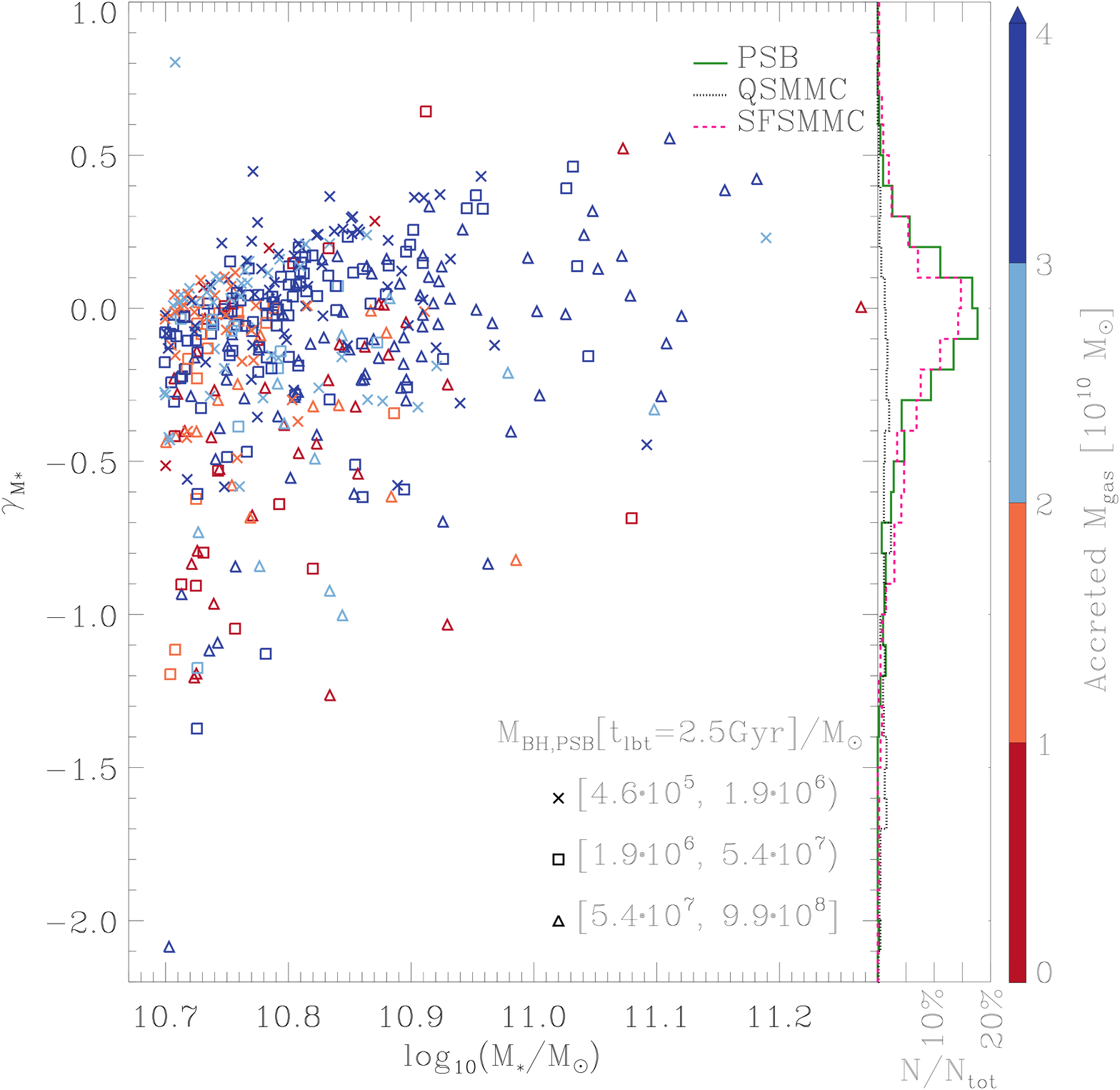}
    \caption{Based on Equation \ref{eq:gammaBH}, the left figure shows the black hole mass growth $\gamma_{BH}$ over a period of $2.5\,$Gyr, as a function of stellar mass for the PSB (coloured points) and the QSMMC sample (grey density). The right figure shows the PSB stellar mass growth $\gamma_{M_*}$, following the same prescription as Equation \ref{eq:gammaBH}, however using stellar mass rather than BH mass. 
    On each side, the colour bar displays the accreted gas mass onto the tracked galaxy due to mergers within the evaluated time-span.
    The symbols (cross, square, triangle) encode the initial BH mass of each PSB galaxy at $t_{\mathrm{lbt}} = 2.5\,$Gyr:
    $M_{\mathrm{BH,PSB}}[t_{\mathrm{lbt}} = 2.5\,\mathrm{Gyr}]/\mathrm{M_{\odot}} \in [[4.6 \cdot 10^5, 1.9 \cdot 10^6), [1.9 \cdot 10^6, 5.4 \cdot 10^7), [5.4 \cdot 10^7, 9.9 \cdot 10^8]]$. 
    The histograms on the right-hand side show the distribution of the PSB (green), QSMMC (black), and SFSMMC (pink) samples along $\gamma_{BH}$ (left) and $\gamma_{M_*}$ (right), respectively.
    To avoid cluttering of points due to an increased stellar mass range, one lone high mass PSB galaxy ($\log_{10}(M_*/M_{_\odot}) = 11.98$) is excluded from the figures. We note that (especially quenched) galaxies exist below the chosen y-range.
    }
    \label{fig:gammaMgas}
\end{figure*} 

To better visualise the scales involved in the BH mass growth over a period of $2.5\,$Gyr, we introduce $\gamma_{\mathrm{BH}}$:

\begin{equation}
\gamma_{\mathrm{BH}} = \log_{10} \left[ \frac{M_{\mathrm{BH}}[t_{\mathrm{lbt}}=0\,\mathrm{Gyr}]-M_{\mathrm{BH}}[t_{\mathrm{lbt}}=2.5\,\mathrm{Gyr}]}{M_{\mathrm{BH}}[t_{\mathrm{lbt}}=2.5\,\mathrm{Gyr}]} \right]
\label{eq:gammaBH}
\end{equation}

Figure \ref{fig:gammaMgas} (left) shows $\gamma_{\mathrm{BH}}$ as a function of stellar mass for QSMMC (grey density) and PSB galaxies, using a colour bar (right-hand side) to encode the galactic accreted gas mass $M_{\mathrm{gas}}$.
As indicated by the legend, different symbols are used to indicate the initial PSB galaxy BH mass at $t_{\mathrm{lbt}} = 2.5\,$Gyr: the least massive BHs are indicated by crosses, i.e. ${M_{\mathrm{BH,PSB}}[t_{\mathrm{lbt}} = 2.5\,\mathrm{Gyr}]/\mathrm{M_{\odot}} \in [4.6 \cdot 10^5, 1.9 \cdot 10^6)}$, intermediate BHs are indicated by squares, i.e. ${M_{\mathrm{BH,PSB}}[t_{\mathrm{lbt}} = 2.5\,\mathrm{Gyr}]/\mathrm{M_{\odot}} \in [1.9 \cdot 10^6, 5.4 \cdot 10^7)}$, and the most massive BHs are indicated by triangles, i.e. ${M_{\mathrm{BH,PSB}}[t_{\mathrm{lbt}} = 2.5\,\mathrm{Gyr}]/\mathrm{M_{\odot}} \in [5.4 \cdot 10^7, 9.9 \cdot 10^8]}$. 

Figure \ref{fig:gammaMgas} (left) clearly shows that, in contrast to QSMMC galaxies, PSBs are consistently found at higher values of $\gamma_{\mathrm{BH}}$. This is in line with previously established behaviour (see Section \ref{sub:AGN+SNe}), where PSBs exhibit a significantly stronger AGN feedback, i.e. BH mass growth, than the QSMMC sample. Interestingly, it appears that the PSB population inhabits distinct regions in the stellar mass - $\gamma_{BH}$ plane. Most noticeably, there appears to be a bimodality, centred around two PSB populations found at $\gamma_{\mathrm{BH}} \sim 2$, i.e. a BH growth by a factor of $\sim 100$, and $\gamma_{BH} \sim 0$, i.e. a doubling of the BH mass over the last $2.5\,$Gyr. 

A strong correlation between decreasing ${M_{\mathrm{BH,PSB}}[t_{\mathrm{lbt}} = 2.5\,\mathrm{Gyr}]/\mathrm{M_{\odot}}}$ and increasing $\gamma_{BH}$ is evident: In Figure \ref{fig:gammaMgas} (left), we see that larger BH growth strongly correlates with smaller $t_{\mathrm{lbt}} = 2.5\,$Gyr BH mass (crosses). Likewise, smaller BH growth correlates with more massive $t_{\mathrm{lbt}} = 2.5\,$Gyr BHs (triangles).
In short, the less massive PSB BHs were at $t_{\mathrm{lbt}} = 2.5\,$Gyr, the more BHs grow in the following $2.5\,$Gyr.

It follows that, the $\gamma_{\mathrm{BH}} \sim 2$ population is characterised by recently seeded BHs at $t_{\mathrm{lbt}}=2.5\,$Gyr. Our BHs are represented by collisionless sink particles, which are seeded with an initial mass of $4.6 \cdot 10^5\, \mathrm{M_{\odot}}$ in galaxies with stellar mass $M_* > 2.3 \cdot 10^{10}\, \mathrm{M_{\odot}}$ \citep{2015MNRAS.448.1504S}. The BHs are seeded below the Magorrian relation, i.e. the relation between BH and bulge mass \citep{1998AJ....115.2285M}. In practice, this means that recently seeded BHs experience an initial rapid BH mass growth at fairly constant stellar mass \citep{2015MNRAS.448.1504S}. 
In contrast, the $\gamma_{\mathrm{BH}} \sim 0$ population is characterised by BHs that have already reached the Magorrian relation at $t_{\mathrm{lbt}}=2.5\,$Gyr. 
Despite the numerical effects associated with seeding BHs, from a physical point of view, the important distinction between PSB and QSMMC galaxies remains: PSBs are characterised by a significantly stronger recent BH mass growth. 
Conversely, the histogram in Figure \ref{fig:gammaMgas} (left) shows a strong overlap between PSB (green) and SFSMMC (pink) galaxies. This further highlights the importance of the specific details of the BH growth, i.e. when, on which timescale, and under which circumstances the growth occurs. 

Figure \ref{fig:gammaMgas} (right) illustrates a number of distinct points:
First, the PSB $\gamma_{\mathrm{BH}}$ bimodality found in the left panel is not reproduced when evaluating the stellar mass growth $\gamma_{\mathrm{M_*}}$ (using Equation \ref{eq:gammaBH}, but substituting $M_{\mathrm{BH}}$ with $M_*$). 
Second, as previously established in Section \ref{sec:environment}, PSB galaxies in Figure \ref{fig:gammaMgas} are overwhelmingly found at lower stellar masses, i.e. close to our mass cut. 
Third, Figure \ref{fig:gammaMgas} shows a weak correlation between stellar mass and gas accretion, as low stellar mass PSBs are more likely to have low gas accretion (red and orange), while gas accretion appears to increase (blue) towards higher stellar masses.
Additionally, we also investigate the correlation between stellar mass and stellar mass (rather than gas) accretion via mergers, finding a stronger correlation than in Figure \ref{fig:gammaMgas}. This is not surprising, as mergers provide a significant pathway for stellar mass growth for massive galaxies \citep{2016MNRAS.458.2371R, 2017MNRAS.464.1659Q, 2021MNRAS.501.3215O}, while in-situ star formation becomes less
\citep{2010ApJ...709.1018V}.

Further investigation shows that the majority ($69\%$) of BHs do not accrete any other BHs, while $21\%$ accrete one other BH within the evaluated time-span.
This reveals that BH growth in our simulation typically happens via smooth accretion, i.e. the process whereby (diffuse) gas is continuously accreted \citep{2009ApJ...694L.158B, 2012A&A...544A..68L}, rather than through the accretion of other BHs.
We find a weak correlation between increasing stellar mass and increasing number of accreted BHs. Again, this is not surprising as larger stellar mass galaxies typically grow their stellar mass via mergers \citep{2010ApJ...709.1018V, 2017MNRAS.464.1659Q}.
As a result more massive galaxies are more likely to merge with satellites which already host a seeded BH, increasing the likelihood of the main BH accreting a satellite BH.
This weak correlation between stellar mass and accreted BHs agrees with our expectations of a hierarchical growth model in a $\Lambda$CDM universe in which large halos are formed late via the coalescence of smaller ones \citep{1997ApJ...490..493N, 2000MNRAS.319..168C, 2006MNRAS.370..645B}.

A closer look at the histogram (right) reveals both a continued similarity between PSB (green) and SFSMMC (pink) galaxies, and a strong difference to QSMMC (black) galaxies:
Compared to PSB and SFSMMC galaxies, far fewer QSMMC galaxies experience a non-negligible stellar mass growth over the considered time-span, $\Delta t \sim 2.5\,$Gyr. As a result, Figure \ref{fig:gammaMgas} (right) is underpopulated with QSMMC galaxies.
This contrast between PSB and SFSMMC on one side, and QSMMC galaxies on the other, further highlights the statistically rich merger history of PSB and SFSMMC galaxies, which are overwhelming located around $\gamma_{\mathrm{M_*}} \sim 0$, i.e. experience a doubling in stellar mass within the past $2.5\,$Gyr.

In short, both the merger history (Table \ref{tab:MergerTable}) and BH growth (Figure \ref{fig:gammaMgas} left) of PSB and SFSMMC galaxies show strong similarities, albeit PSBs are classified as quiescent at $t_{\mathrm{lbt}}=0\,$Gyr. This shows that, when no further stellar mass selection is chosen (as is done in the left panel of Figure \ref{fig:AGN_SNe_Energy_mass_evol}) and the major merger progenitor cold gas content is not taken into consideration (see Figure \ref{fig:merger_wetness} top right), PSBs essentially behave like star-forming galaxies until their recent shutdown in star formation.

\begin{table*}
\begin{center}
  \begin{tabular}{| c | c | c | c | c | c | c | c | c | c |}
    \hline
    & \multicolumn{3}{|c|}{$M_{\mathrm{200,crit}}/\mathrm{M_{\odot}} < 10^{13} [\%]$} & \multicolumn{3}{|c|}{$10^{13} \leq M_{\mathrm{200,crit}}/\mathrm{M_{\odot}} < 10^{14} [\%]$} & \multicolumn{3}{|c|}{$M_{\mathrm{200,crit}}/\mathrm{M_{\odot}} \geq 10^{14} [\%]$} \\ \hline
    Selection & PSBs & QSMMC & SFSMMC & PSBs & QSMMC & SFSMMC & PSBs & QSMMC & SFSMMC \\ \hline
    $\gamma_{BH}  \geq 1.0$ & 95.6 & 91.7 & 87.5 & 3.8 & 8.3 & 10.8 & 0.6 & 0 & 1.7 \\ \hline
    $1.0> \gamma_{BH} \geq -2.0$ & 85.2 & 82.8 & 86.2 & 12.1 & 12.7 & 10.8 & 2.7 & 4.5 & 3.1 \\ \hline
    $-2.0> \gamma_{BH}>-3.8$& 100 & 84.7 & 65.4 & 0 & 12.4 & 26.9 & 0 & 2.9 & 7.7 \\ \hline
    $\gamma_{BH}  \leq -3.8$&  - & 8.9 & 0 & - & 28.9 & 100 & - & 62.2 & 0 \\ \hline 
    \hline
    $3 > \gamma_{BH}>-6$    & 89.4 & 79.1 & 85.0 & 8.7& 13.6 & 12.0 & 1.9 & 7.3 & 2.9 \\ \hline
    \label{tab:gammas}
  \end{tabular}
\end{center} 
\caption{Different subdivisions of $\gamma_{BH}$ (see Equation \ref{eq:gammaBH}), partitioned based on the horizontal lines in Figure \ref{fig:gammaMgas} as a function of different halo masses for the PSB (424), QSMMC (641), and SFSMMC (411) samples.}
\label{tab:BHgrowthTable}
\end{table*} 

Table \ref{tab:BHgrowthTable} displays the BH growth $\gamma_{BH}$ for the PSB, QSMMC, and SFSMMC samples as a function of halo mass, i.e. local environment. The horizontal lines in Figure \ref{fig:gammaMgas} indicate the different $\gamma_{BH}$ subdivisions of Table \ref{tab:BHgrowthTable}. As PSB and SFSMMC galaxies typically experience a more rapid evolution, i.e. their BHs have been more recently seeded, the sample size of galaxies evaluated over $2.5\,$Gyr is smaller for PSBs ($424$) and SFSMMC ($411$), compared to the QSMMC sample ($641$).

The last row of Table \ref{tab:BHgrowthTable} shows that $89.4\%$ of all PSB, $79.1\%$ of QSMMC, and $85.0\%$ of SFSMMC galaxies are found within a field environment ($M_{\mathrm{200,crit}}/\mathrm{M_{\odot}} < 10^{13}$) at $t_{\mathrm{lbt}}=0\,$Gyr. In contrast to $7.3\%$ of QSMMC galaxies, only $1.9\%$ of PSB and $2.9\%$ of SFSMMC galaxies are found in clusters ($M_{\mathrm{200,crit}}/\mathrm{M_{\odot}} > 10^{14}$). This trend reflects the results obtained in Section \ref{sub:FoFNsub}, i.e. that PSBs at $t_{\mathrm{lbt}}=0\,$Gyr are overwhelmingly found in halos with few satellites.

QSMMC galaxies belong to the only sample with a non-negligible population in the $\gamma_{BH} \leq -3.8$ regime (see Figure \ref{fig:gammaMgas} left). Moreover, galaxies found at these low $\gamma_{BH}$ values, i.e. BHs with stagnated growth, are more likely to be found in clusters ($62.2\%$) and groups ($28.9\%$). As galaxy clusters are characterised by an abundance of hot gas and cluster galaxies have high relative velocities, inhibiting galaxy mergers, satellite galaxies have very limited opportunities to replenish their (cold) gas reservoir. Consequently, cluster galaxies have a lower likelihood of gas inflow reaching the galactic centre, resulting in low BH growth, and less numbers of PSBs.

\section{Post-starburst galaxies in galaxy clusters}
\label{sec:clusters}

As observations suggest that the evolution of PSBs differs considerably with environment, we now focus on galaxy clusters \citep{1999ApJ...518..576P, 2005MNRAS.357..937G, 2009MNRAS.395..144W, 2017MNRAS.472..419L, 2019MNRAS.482..881P}. 
Particularly, we want to understand how the environment, specifically galaxy clusters, influence PSB galaxy evolution. To increase our sample size, we lower the mass threshold in this section to include galaxies with at least 100 stellar particles, i.e. $M_* \geq 4.97 \cdot 10^{9} \, \mathrm{M_{\odot}}$. This does not include Table \ref{tab:MuzzinTable} and Figure \ref{fig:AGN_SNe_Energy_mass_evol_cluster}, where the stellar mass threshold ($M_* \geq 4.97 \cdot 10^{10} \, \mathrm{M_{\odot}}$) is kept the same to allow direct comparisons with Table \ref{tab:Nbeg064MergerTable} and Figure \ref{fig:AGN_SNe_Energy_mass_evol}.

\subsection{Galaxy cluster stellar mass function comparison}
\label{sub:soco}

\begin{figure} 
	\includegraphics[width=\columnwidth]{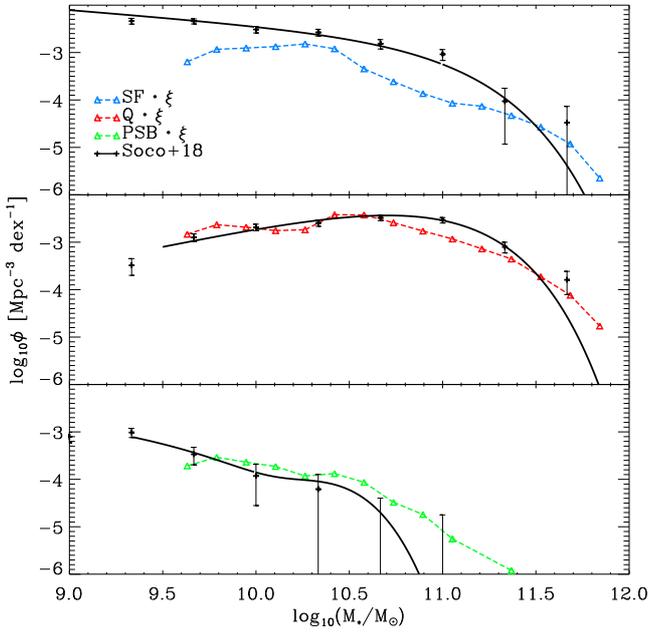} 
    \caption{Stellar mass functions comparing Magneticum Box2 satellite galaxies to \protect\cite{2018MNRAS.476.1242S} at $z = 0.7$ in the group and cluster environments. 
    Following \protect\cite{2018MNRAS.476.1242S}, we select groups and clusters with a member range of $20 \leq N \leq 135$ and stellar mass within radius $R \leq 1 \, \mathrm{Mpc}$ of $10^{11.29} \leq M_{*}/\mathrm{M_{\odot}} \leq 10^{12.45}$. Centred on these groups and clusters, we construct cylinders with height $H_{\mathrm{cyl}} = 5 \, \mathrm{Mpc}$ and evaluate all galaxies contained within the cylindrical volume.
    The black solid line indicates the fit to the stellar mass functions by \protect\cite{2018MNRAS.476.1242S}, while the coloured triangles represent the Magneticum results. 
    The black solid fit line extends to the $90\%$ mass completion limit.
    The different panels show the stellar mass functions of the star-forming (SF: top), the quiescent (Q: middle) and the post-starburst (PSB: bottom) populations. The Magneticum results are normalised by a factor of $\xi = 1/300$ to fit the arbitrarily normalised cluster observations (see \protect\cite{2018MNRAS.476.1242S}).}
    \label{fig:SMF_soco}
\end{figure}

We extend our study of the global stellar mass functions shown in Figure \ref{fig:SMF_grid} by considering the high density environment and comparing to a catalogue of galaxy cluster candidates detected in the Ultra-Deep-Survey (UDS) \citep{2018MNRAS.476.1242S}.
\cite{2018MNRAS.476.1242S} study the environment dependent galaxy evolution in the redshift range $0.5 < z < 1.0$ using the UDS. They identify $37$ clusters, $11$ of which contain more than $45$ members. This results in a sample of $2210$ galaxies, which provide the basis for the stellar mass function calculation \citep{2018MNRAS.476.1242S}.

To compare with the observations, we follow a similar, yet not identical, prescription: Due to redshift uncertainties, \cite{2018MNRAS.476.1242S} sample the volume of cylinders centred on clusters with height $H_{\mathrm{cyl}} = 250 \, \mathrm{Mpc}$. Thereafter, they remove the contaminants by statistically subtracting the field galaxies in each cylinder \citep{2018MNRAS.476.1242S}. To not unnecessarily introduce statistical contamination, we consider smaller cylinders with height $H_{\mathrm{cyl}} = 5 \, \mathrm{Mpc}$. 
In both cases, the cylinder has radius $R_{\mathrm{sph}} = 1 \, \mathrm{Mpc}$ and the stellar mass and the number of satellites $N$ is calculated inside the cylinder.
Following \cite{2018MNRAS.476.1242S}, we select only those clusters with a member range of $20 \leq N \leq 135$ and stellar mass within $1 \, \mathrm{Mpc}$ of $10^{11.29} \leq M_{*}/\mathrm{M_{\odot}} \leq 10^{12.45}$. Subsequently, each cluster is considered along three random yet linearly independent spatial axes, increasing our sampling.

A total of $8406$ Magneticum clusters fulfil the above criteria, with a total of $182213$ member galaxies with stellar mass $M_* \geq 4.97 \cdot 10^{9} \, \mathrm{M_{\odot}}$, of which $43084$ are star-forming, $139129$ are quiescent, and $7704$ are identified as PSBs. The cluster and galaxy counts are the total values across all three spatial axes, i.e. are up to a factor of $\sim 3$ larger than the uniquely identified objects within Box2. 

Figure \ref{fig:SMF_soco} shows the $z = 0.7$ galaxy cluster stellar mass function of star-forming (blue), quenched (red) and PSB (green) galaxies. Similarly to the total sample shown in Figure \ref{fig:SMF_grid} at $z=0.7$, the cluster PSB stellar mass function has two bumps at $\mathrm{log}(M_{*}/\mathrm{M_{\odot}}) \sim 9.7$ and $\mathrm{log}(M_{*}/\mathrm{M_{\odot}}) \sim 10.4$ and is dominated by the low stellar mass end. However, the amplitude of the PSB bumps differs between the total and cluster sample.
Furthermore, we find fewer star-forming and thus more quiescent galaxies in the cluster environment at low stellar mass compared to the total sample (Figure \ref{fig:SMF_grid}).

The observations in Figure \ref{fig:SMF_soco} are fitted by Schechter functions (star-forming and quiescent satellite galaxies) and double Schechter functions (PSBs), respectively \citep{2008MNRAS.388..945B, 2010A&A...523A..13P}. As the stellar mass functions discussed in \cite{2018MNRAS.476.1242S} are arbitrarily normalised, the Magneticum results were also normalised to fit the observational data. Specifically, the Magneticum results (triangles) were multiplied by a factor of $\xi = 1/300$ to vertically adjust them to the observations.
As shown in Figure \ref{fig:SMF_soco}, the shape of the cluster galaxy stellar mass functions from Magneticum are in very good agreement with observations (see also \cite{2015MNRAS.448.1504S}).

There are only two discrepancies: First, the star-forming distribution which, similar to Figure \ref{fig:SMF_grid}, lacks good agreement for masses between $10.5 < \mathrm{log}(M_{*}/\mathrm{M_{\odot}}) < 11.2$. As discussed in Section \ref{sub:SMF}, this is due to the onset of the AGN feedback.
Second, we find evidence for rare massive cluster PSBs which are not found in the significantly smaller observational sample.
Further evidence for good agreement is provided by the replication of the PSB plateau in the mass range $10.0 < \mathrm{log}(M_{*}/\mathrm{M_{\odot}}) < 10.5$, indicating a preferential intermediate mass range.

\subsection{Line-of-sight velocity: Observation and resolution comparison}
\label{sub:vlosObsComp}

\begin{figure}
	\includegraphics[width=\columnwidth]{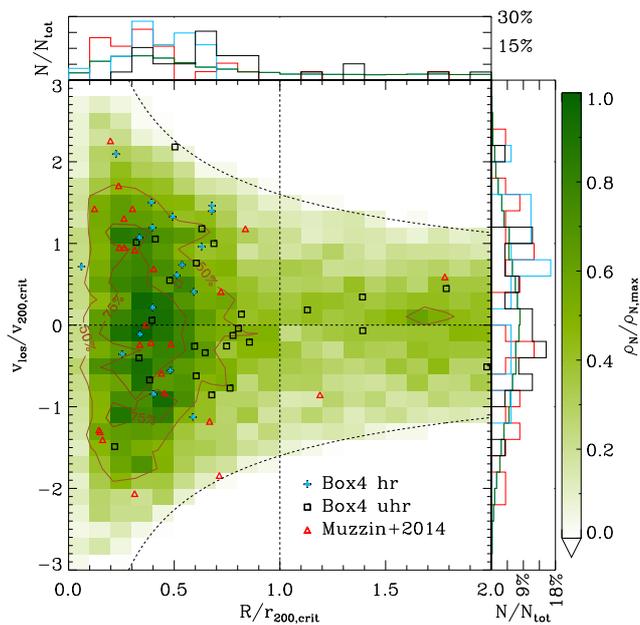}
    \caption{Post-starburst (PSB) galaxy normalised line-of-sight (LOS), $v_{\mathrm{los}}/v_{\mathrm{200,crit}}$, phase space comparison between Box4 at high resolution (hr, blue crosses), Box4 at ultra-high resolution (uhr, black squares), \protect\cite{2014ApJ...796...65M} (red triangles), and Magneticum Box2 and Box2b (green) at $z = 0.9$ in dependence of the cluster-centric 2D projected radial profile, $R/r_{\mathrm{200,crit}}$.
    Following the criteria outlined in \protect\cite{2014ApJ...796...65M}, satellite galaxies, hosted by clusters in the mass range $1 \cdot 10^{14} < M_{\mathrm{200,crit}}/\mathrm{M_{\odot}} < 20 \cdot 10^{14}$, are shown.
    Satellite galaxies are selected above stellar mass $M_* \geq 4.97 \cdot 10^{9} \, \mathrm{M_{\odot}}$.
    The contour lines highlight the regions where the density is $50\%$ and $75\%$ of the maximum density.
    The histograms depict the relative abundance of each population projected onto the respective axes. The enveloping dashed black lines corresponds to $|v_{\mathrm{los}}/v_{\mathrm{200,crit}}| \sim 1.6|({R/r_{\mathrm{200,crit}}})^{-1/2}|$ and is used to exclude interlopers.
    }
    \label{fig:Muzzin} 
\end{figure}

In Figure \ref{fig:Muzzin}, we show the normalised line-of-sight phase space velocity $v_{\mathrm{los}}/v_{\mathrm{200,crit}}$ of PSBs at $z = 0.9$ as a function of the cluster-centric 2D projected radius, $R/r_{\mathrm{200,crit}}$, for both Box2 and Box2b (green density). Box2 and Box2b results are compared to our high (blue crosses) and ultra-high resolution Box4 (black squares), as well as to observations by \cite{2014ApJ...796...65M} (red triangles).
Galaxies with the lower stellar mass threshold of $M_* \geq 4.97 \cdot 10^{9} \, \mathrm{M_{\odot}}$ are shown, previously used in \cite{2019MNRAS.488.5370L}.
We compared this lower stellar mass threshold with our standard stellar mass threshold ($M_* \geq 4.97 \cdot 10^{10} \, \mathrm{M_{\odot}}$) and established the convergence of our results. We choose the lower stellar mass cut, so as to increase our phase space sampling (relevant especially to Figure \ref{fig:PSvradGrid}).

Magneticum PSBs are shown as density maps and are scaled to the maximum density of the PSB galaxy population.
The dashed black lines enveloping the density map in Figure \ref{fig:Muzzin} are based on the virial theorem and are introduced to provide a relationship between the velocity and the radius via ${|v_{\mathrm{los}}/v_{\mathrm{200,crit}}| \sim |({R/r_{\mathrm{200,crit}}})^{-1/2}|}$. 
A proportionality factor of $1.6$ is introduced to scale the enveloping dashed black lines. The factor is motivated by the strongest outlier of the observational data \citep{2014ApJ...796...65M} and is used to filter out interlopers, i.e. galaxies that are only attributed to a cluster due to the line-of-sight projection.

\cite{2014ApJ...796...65M} consider data based on the Gemini Cluster Astrophysics Spectroscopic Survey (GCLASS). They investigate the line-of-sight phase space of these $424$ cluster galaxies at $z \sim 1$, of which $24$ are identified as PSBs according to an absence of $O_{\mathrm{II}}$ emission while also hosting a young stellar population ($D_n(4000) < 1.45$)  \citep{2017ApJ...841...32Z}.

To sample a similar volume as the observations, a cylinder of height $179 \mathrm{Mpc}$ was used.
The cylinder height was calculated by evaluating the scatter around the mean observed redshift, $\sigma_{\mathrm{z}}$, resulting in $\sigma_{\mathrm{z}} = 0.036$ \citep{2013A&A...557A..15V}.
The projections were considered along three linearly independent spatial axes. We identified $20371$ PSBs in $1239$ clusters. 

Figure \ref{fig:Muzzin} shows that both the PSBs identified by \cite{2014ApJ...796...65M} and by Magneticum exhibit a strong preference for the inner region of the clusters. The inner over-density of PSBs found between $R \sim (0.15-0.5) \, r_{\mathrm{200,crit}}$ matches observations well. 
Of the $20371$ identified PSBs, $14790$, i.e. $73\%$, are found inside $r_{\mathrm{200,crit}}$. A subset of $9263$ satellite galaxies, i.e. $45\%$, are even found inside $R < 0.5 \, r_{\mathrm{200,crit}}$. The normalised distributions projected onto each axis further demonstrate the close agreement between observations and our simulation. 
The PSB galaxy preference for a distinct region of phase space, namely $R \sim (0.15-0.5) \, \mathrm{R_{200,crit}}$, suggests a common cause: Most likely an environmental quenching mechanism leads to the shutdown of previously (strongly) star-forming galaxies on a timescale that brings them about halfway through the cluster until star formation is shutdown. This agrees with previous phase space results concerning the fast quenching of star-forming satellite galaxies in clusters \citep{2019MNRAS.488.5370L}.

To test numerical convergence we compare our results with Box4 at high and ultra-high resolution, the latter having a $\sim 20$ times higher mass resolution than the former. 
As Box4 is significantly smaller than Box2 or Box2b (see Section \ref{sub:Mag}), both resolution levels only yield $1$ cluster within the mass range $1 \cdot 10^{14} < M_{\mathrm{200,crit}}/\mathrm{M_{\odot}} < 20 \cdot 10^{14}$.
The clusters host $11$ unique ultra-high and $6$ unique high resolution PSBs, respectively. When considering three linearly independent projections, this yields a sample of $26$ ultra-high and $18$ high resolution PSBs.

Figure \ref{fig:Muzzin} shows agreement between high and ultra-high resolution PSBs, as well as between different boxes. Similarly to Box2 and Box2b, Box4 PSBs (both resolution levels) are also predominantly found within the inner cluster region. Although the 2D projected radial scatter is stronger for the larger ultra-high resolution PSB sample (black squares) compared to the smaller high resolution PSB sample (blue crosses), the difference does not exceed the expected statistical variation.
Meanwhile, the normalised line-of-sight phase space velocity distributions of the two resolution levels, different boxes, and observations shows consistent agreement. In general, the numerical convergence of our results and the associated implications are discussed in Section \ref{sub:disc:numerics}.

\begin{figure}
	\includegraphics[width=\columnwidth]{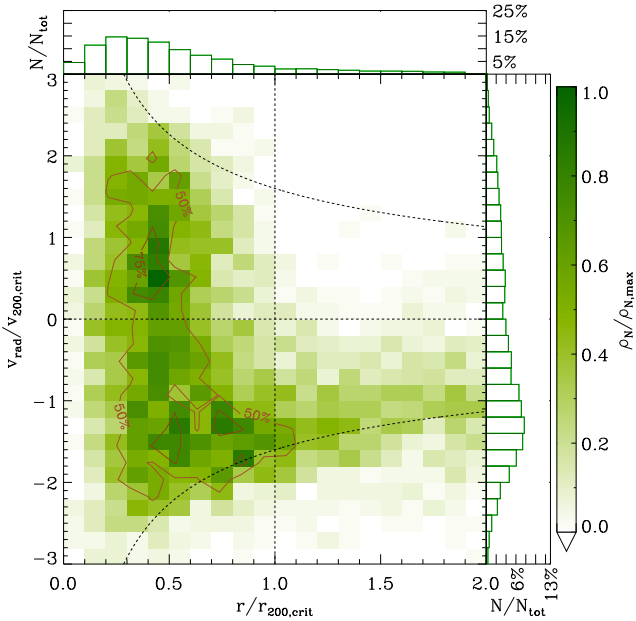} 
    \caption{Same as Fig \ref{fig:Muzzin} but showing the normalised radial velocity $v_{\mathrm{rad}}/v_{\mathrm{200,crit}}$ as a function of the 3D radius $r/r_{\mathrm{200,crit}}$. The enveloping black line is no longer used to filter out galaxies, it only remains to guide the eye.}
    \label{fig:Muzzin_vrad}
\end{figure}

\subsection{Radial velocity as a function of cluster mass and redshift}
\label{sub:PSvradGrid}

To better disentangle the underlying mechanisms potentially involved in triggering the starburst and subsequent shutdown in star formation of PSBs in galaxy cluster environments, we extend our investigation beyond the line-of-sight phase space observational comparison.
Using much of the same nomenclature as Figure \ref{fig:Muzzin}, Figure \ref{fig:Muzzin_vrad} shows the PSBs within a 3D sphere instead of projections. 
Thus, Figure \ref{fig:Muzzin_vrad} no longer shows the PSBs inside a cylindrical volume which is the basis of the projected line-of-sight population but a 3D sphere of radius $2 r_{\mathrm{200,crit}}$, and therefore the sample of PSBs is smaller, i.e. only $5185$ PSBs are plotted. Of this population, $3401$ PSBs (or $66\%$) are infalling, i.e. $v_{\mathrm{rad}}/v_{\mathrm{200,crit}} < 0$. This is similar to the $69\%$ infalling PSBs found in the line-of-sight population.
In addition to PSBs typically being characterised by infall, Figure \ref{fig:Muzzin_vrad} shows an abundance of PSBs in the inner cluster region. Furthermore, it appears that the PSB population, when compared to e.g. the older quiescent cluster population in \cite{2019MNRAS.488.5370L}, is not well mixed within the cluster, clearly indicating a recent infall.

\begin{figure*}
	\includegraphics[width=2\columnwidth]{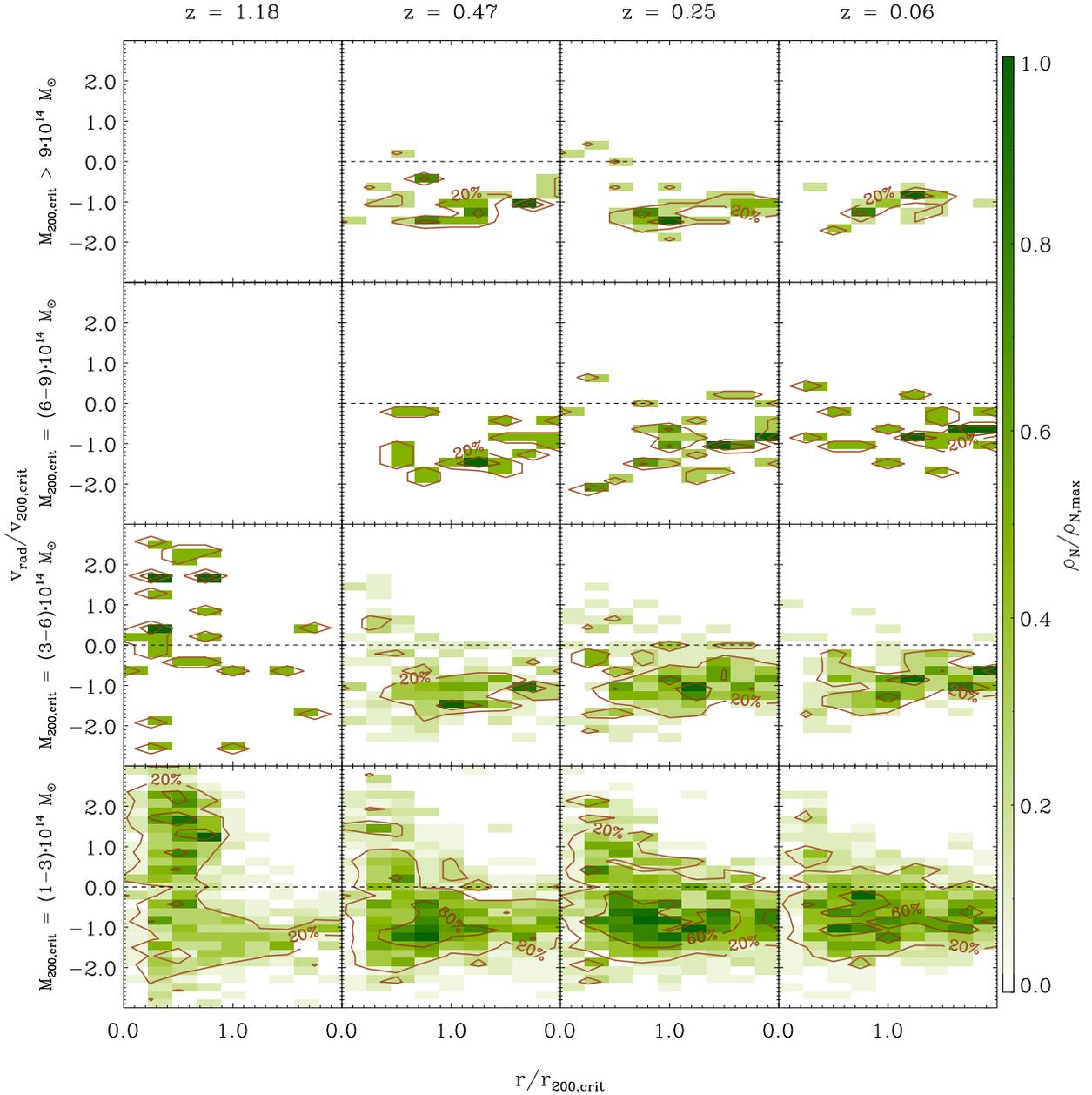}
    \caption{Box2 PSB galaxy ($M_* \geq 4.97 \cdot 10^{9} \, \mathrm{M_{\odot}}$) normalised radial velocity $v_{\mathrm{rad}}/v_{\mathrm{200,crit}}$ as a function of 3D radius $r/r_{\mathrm{200,crit}}$ for different cluster masses and redshifts. 
    Columns from left to right have the following redshifts: $z=1.18$, $z=0.47$, $z=0.25$, and $z=0.06$. Rows from top to bottom have the following cluster masses:$M_{\mathrm{200,crit}} > 9 \cdot 10^{14}\, \mathrm{M_{\odot}}$, $M_{\mathrm{200,crit}} = (6-9) \cdot 10^{14}\, \mathrm{M_{\odot}}$, $M_{\mathrm{200,crit}} = (3-6) \cdot 10^{14}\, \mathrm{M_{\odot}}$, and $M_{\mathrm{200,crit}} = (1-3) \cdot 10^{14}\, \mathrm{M_{\odot}}$. The colour bar displays the relative phase space number density normalised to the maximum value of each individual panel. The contour lines correspond to regions showing $20\%$ and $60\%$ of the maximum density in each panel. The horizontal dashed line marks $v_{\mathrm{rad}}/v_{\mathrm{200,crit}} = 0$: PSBs above this line are moving outwards with respect to the cluster centre, while PSBs below this line are moving into the cluster.
    The empty panels are the result of a lack of clusters in the given redshift and cluster mass range.}
    \label{fig:PSvradGrid}
\end{figure*}

We analyse the normalised radial velocity of cluster PSBs as a function of the 3D cluster-centric radius at different cluster masses and redshifts. Figure \ref{fig:PSvradGrid} shows an overview of four different cluster mass ranges at redshifts $0.06 < z < 1.18$. 
We find that cluster PSBs in all halo mass ranges at redshifts $z \lesssim 0.5$ have unusually negative radial velocities, i.e. they are in the process of infall. 
This means that they are either on their first infall into the cluster or are returning to the cluster after they have left it, typically referred to as backsplash galaxies \citep{2011MNRAS.411.2637P}. 
However, when evaluating cluster PSBs, we find negligible evidence for backsplash orbits, rather the vast majority of cluster PSBs are experiencing their first infall.

In addition to showing that cluster PSBs are overwhelmingly characterised by infall, Figure \ref{fig:PSvradGrid} also reveals two important trends: First, cluster PSBs become increasingly infall dominated towards higher cluster masses, suggesting a density dependent environmental quenching mechanisms, such as ram-pressure stripping. This agrees with observations, which find that processes linked to the termination of star formation in galaxy clusters are more effective in denser environments \citep{2009ApJ...693..112P, 2012A&A...543A..19R}. 
For example, ram-pressure stripping is linearly dependent on the intra-cluster-medium (ICM) density, thus higher mass clusters are more efficient in quenching, i.e. the quenching timescale is shorter \citep{1972ApJ...176....1G, 2019MNRAS.488.5370L}. In this case, higher mass clusters show an increased likelihood of cluster PSBs being fully quenched before they pass their pericenter. 

Second, we find that cluster PSBs become slightly more infall dominated towards lower redshifts, especially visible in the transition from $z \sim 1.2$ to $z \sim 0.5$ in the lowest cluster mass regime in Figure \ref{fig:PSvradGrid}. This is likely driven by the fact that clusters at $z \sim 1.2$, compared to clusters in the same mass range at $z \sim 0.5$, have an increased likelihood of currently undergoing cluster mergers. In other words, clusters at $z \sim 1.2$ are typically not relaxed, while clusters of similar mass at $z \sim 0.5$ have had enough time to at least centrally relax.
Therefore, the same mass clusters at $z \sim 1.2$ are on average more disturbed, which in turn implies a less relaxed and hot ICM. Under these circumstances, the quenching efficiency is inhibited and thus cluster PSBs, on average, are able to penetrate deeper into the galaxy cluster before ram-pressure stripping quenching is efficient.

Considering these findings a picture of cluster PSB galaxy evolution in our simulation emerges, which strongly favours environmental quenching, e.g. ram-pressure stripping, as the responsible shutdown mechanism of cluster PSBs, in contrast to our results found for the field PSBs. 
Specifically, independent of whether an additional starburst is triggered during cluster infall or the PSB progenitors were previously experiencing significant star formation, the cluster environment appears to shut down the star formation during infall. This rapid shutdown increases the likelihood of previously star-forming/star-bursting galaxies to be classified as PSBs. Consequently, it appears that PSBs in galaxy clusters share a similar shutdown mechanism, rather than necessarily sharing the same SFR increasing mechanism.

\subsection{Cluster merger statistics}
\label{sub:clusterMergers}

Following the same approach as in Section \ref{sub:mergerStat}, we analyse merger abundances in Table \ref{tab:MuzzinTable} to understand their statistical relevance to the evolution of cluster PSBs. Of the 411 projection independent PSBs with stellar mass $M_* \geq 4.97 \cdot 10^{10}\, \mathrm{M_{\odot}}$ identified within galaxy clusters in Box2 at $z=0.9$, we successfully trace 410 PSBs to $z=1.7$, i.e. over a time-span of $\sim 2.5\,$Gyr.

\begin{table} 
\begin{center}
  \begin{tabular}{| l | c | c | c |}
    \hline
    Criterion & PSBs & QSMMC & SFSMMC \\ \hline
    Analysed trees              & 410 & 409 & 405 \\ \hline
    $\Sigma(N_{\mathrm{mini}})$  & 370 & 197 & 400 \\ \hline
    $\Sigma(N_{\mathrm{minor}})$ & 290 & 98  & 293 \\ \hline
    $\Sigma(N_{\mathrm{major}})$ & 216 & 94  & 238 \\ \hline
    $N_{\geq 1 \mathrm{ mini}}$       & $57.1\%$ & $32.8\%$ & $53.6\%$ \\ \hline
    $N_{\geq 1 \mathrm{ minor}}$      & $55.6\%$ & $20.3\%$ & $51.4\%$ \\ \hline
    $N_{\geq 1 \mathrm{ major}}$      & $43.2\%$ & $20.8\%$ & $49.9\%$ \\ \hline
    $N_{\geq 1 \mathrm{ merger}}$     & $92.9\%$ & $53.1\%$ & $91.9\%$ \\ 
  \end{tabular}
\end{center} 
\caption{Following the nomenclature as introduced in Table \ref{tab:MergerTable}, but showing results for cluster galaxies in the redshift range $0.9 < z < 1.7$ (compare to Table \ref{tab:Nbeg064MergerTable} in the same redshift range without an environmental selection). Of the 411 uniquely identified cluster PSBs with stellar mass $M_* \geq 4.97 \cdot 10^{10}\, \mathrm{M_{\odot}}$ in Box2 at $z = 0.9$, 410 progenitors were successfully traced over a time-span of $2.5\,$Gyr.
PSB, QSMMC, and SFSMMC galaxies are selected so as to reproduce the cluster criteria outlined in \protect\cite{2014ApJ...796...65M} (see Section \ref{sub:vlosObsComp}).}
\label{tab:MuzzinTable}
\end{table}

When comparing cluster PSBs (Table \ref{tab:MuzzinTable}) with non environmentally selected PSBs (Table \ref{tab:Nbeg064MergerTable}) in the redshift range $0.9 < z < 1.7$, we find very similar merger abundances for all samples. For example, $92.9\%$ of cluster PSBs and $92.6\%$ of non environmentally selected PSBs both at $z=0.9$ have experienced at least one merger event within the last $\sim 2.5\,$Gyr. Additionally the similarity between the PSB and SFSMMC sample appears independent of the environment surveyed, further providing evidence for the importance of mergers for (recently) star-forming galaxies in our simulation. The only difference we find between Tables \ref{tab:MuzzinTable} and \ref{tab:Nbeg064MergerTable} is a slightly higher abundance of mini ($57.1\%$) and minor ($55.6\%$) mergers and a slightly lower abundance of major mergers ($43.2\%$) in cluster PSBs, compared to mini ($50.7\%$), minor ($50.5\%$), and major ($47.3\%$) mergers of non environmentally selected PSBs. Beyond these small differences the abundances found in Table \ref{tab:MuzzinTable} agree with the analysis presented in Section \ref{sub:mergerStat}.

\subsection{Active galactic nuclei and supernovae}
\label{sub:AGN_SNe_clusters}

\begin{figure*}
	\includegraphics[width=0.95\columnwidth]{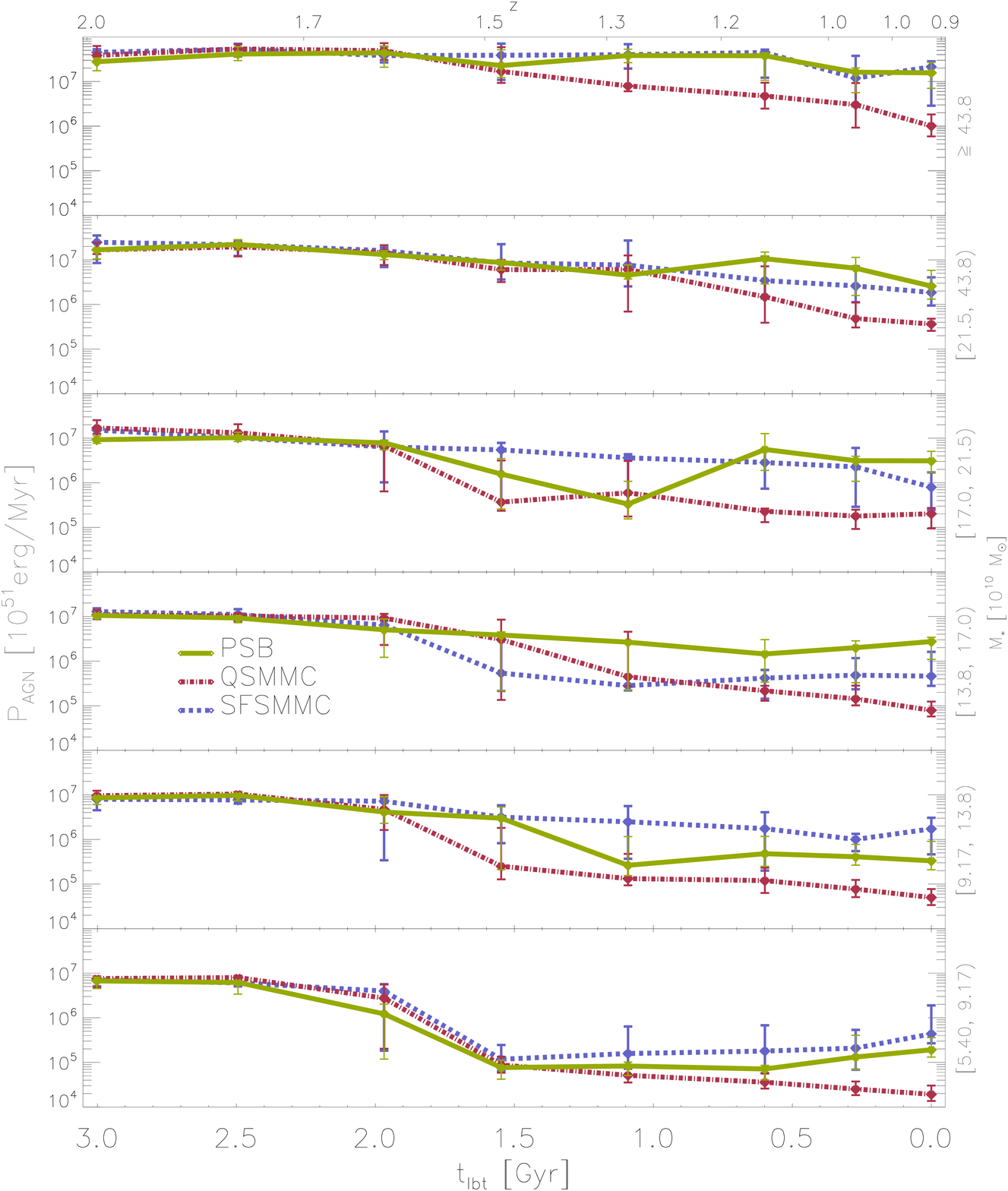}
	\includegraphics[width=0.95\columnwidth]{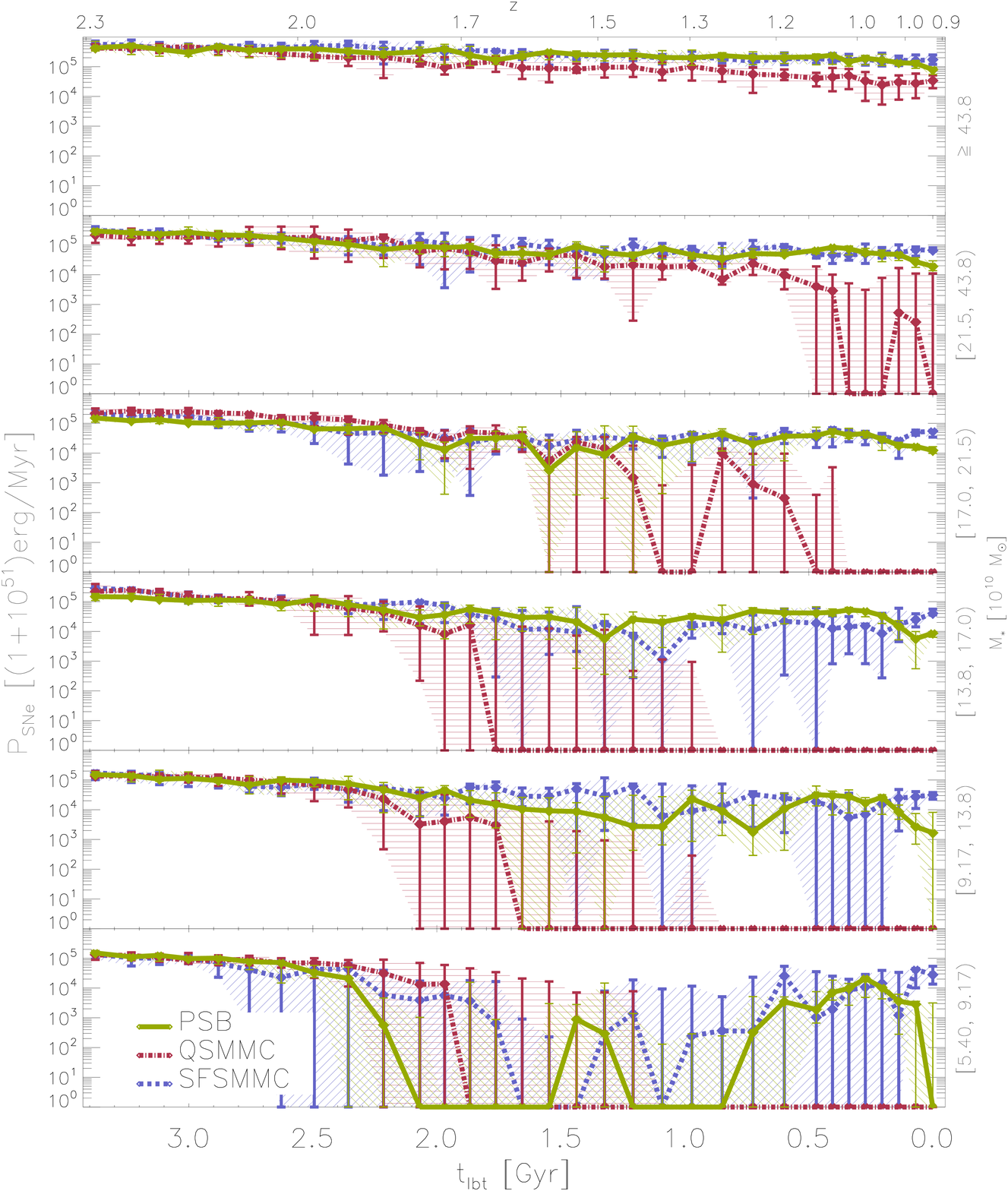}
    \caption{Active galactic nuclei (AGN: left figure) and supernovae (SNe: right figure) power output of cluster PSB (green), QSMMC (red), and SFSMMC (blue) samples identified at $z=0.9$ and evaluated over the past $\sim 3.2\,$Gyr and $\sim 3.5\,$Gyr in units of $10^{51}\,$erg/Myr and $1+10^{51}\,$erg/Myr, respectively. The cluster samples shown are based on the results displayed in Table \ref{tab:Nbeg064MergerTable}.
    The different panels show increasing $z = 0.9$ stellar mass cuts, $M_* > [4.97,6,7,8,9,10] \cdot 10^{10}\, \mathrm{M_{\odot}}$, from the lower to top panel. 
    Both figures show the median, as well as the $0.5\,\sigma$ region as error bars for each population. 
    }
    \label{fig:AGN_SNe_Energy_mass_evol_cluster} 
\end{figure*}

As observations suggest that ram-pressure stripping may trigger AGN activity \citep{2017Natur.548..304P, 2019MNRAS.487.3102G, 2021IAUS..359..108P}, we investigate the AGN and SNe feedback of cluster PSBs at $z=0.9$. We follow the method presented in Section \ref{sub:AGN+SNe} closely, with the exception that, instead of using the global $z \sim 0$ PSB sample, we use the cluster PSB sample presented in Table \ref{tab:MuzzinTable}. Following the same nomenclature as Figure \ref{fig:AGN_SNe_Energy_mass_evol}, Figure \ref{fig:AGN_SNe_Energy_mass_evol_cluster} shows the AGN (left) and SNe (right) power output for the PSB (green), SFSMMC (blue), and QSMMC (red) samples. We note that $t_{\mathrm{lbt}}=0\,$Gyr corresponds to the identification redshift $z=0.9$. Furthermore, we note that at $z \sim 1$ the time-steps in our simulation become larger, which leads the change in abundance of data points in Figure \ref{fig:AGN_SNe_Energy_mass_evol_cluster} at $t_{\mathrm{lbt}} \sim 0.5\,$Gyr. As the feedback energy is normalised per unit time, this does not impact our results. 

Figure \ref{fig:AGN_SNe_Energy_mass_evol_cluster} (left) shows no clear signs of an increase in recent AGN activity. 
As established in Section \ref{sub:PSvradGrid}, cluster PSBs belong to a population of recently in-fallen galaxies. Consequently, if ram-pressure stripping would trigger AGN feedback, we would expect a signal within the last $\sim 1\,$Gyr \citep{2019MNRAS.488.5370L}. 
However, we find no evidence for enhanced AGN activity. 
In fact, it appears that the AGN activity for all samples has been (gradually) declining since $t_{\mathrm{lbt}} \sim 2\,$Gyr. Furthermore, there appears to be no strong stellar mass evolution for the PSB and SFSMMC samples. Only the QSMMC sample shows signs of a stellar mass evolution: As the stellar mass increases, the onset of the decline in AGN activity at $t_{\mathrm{lbt}} \sim 2\,$Gyr begins earlier, while the QSMMC AGN power output at $t_{\mathrm{lbt}} \sim 0\,$Gyr increases by a factor $\sim 2$ between the lowest and highest stellar mass selection. 

In contrast, the SNe power output, i.e. the SFR, shows signs of small increase in activity starting at $t_{\mathrm{lbt}} \sim 0.5\,$Gyr for both the PSB and SFSMMC sample. However, as Figure \ref{fig:AGN_SNe_Energy_mass_evol_cluster} (right) shows, at $t_{\mathrm{lbt}} \sim 0.2\,$Gyr the PSB and SFSMMC samples diverge: The PSBs experience a short timescale decrease in star formation, reaching quiescent levels of star formation according to our blueness criterion, while the SFSMMC galaxies continue to experience an elevated star formation with signs of a small increase. The fact that SFSMMC galaxies are able to sustain star formation despite being located in a high density environment is likely due to their tangential infall orbits, ideally at large cluster-centric radii \citep{2019MNRAS.488.5370L}.
Figure \ref{fig:AGN_SNe_Energy_mass_evol_cluster} (right) also appears to show no indication of a stellar mass evolution: The only exception being the QSMMC sample, which appears to be characterised by more recent quenching with increasing stellar mass.

Generally, integrated over the evaluated time-span, both the AGN and SNe feedback shown in Figure \ref{fig:AGN_SNe_Energy_mass_evol_cluster} is significantly stronger than at $z \sim 0$ in Figure \ref{fig:AGN_SNe_Energy_mass_evol}. As discussed in Section \ref{sub:BHgrowthStat}, recent BH seeding leads to an over-estimation of BH growth. Given the same stellar mass threshold, this becomes more relevant towards higher redshift, as galaxies need to assemble their mass in a shorter time period, i.e. more rapidly. As such, it appears likely that recent BH seeding impacts our $z=0.9$ cluster sample. This likely explains the strong median AGN feedback at high look-back-times. Nonetheless, the fact that the AGN feedback does not increase towards recent look-back-times holds and suggests that AGN feedback is not relevant to shutting down star formation in cluster PSBs. This is further supported by the similarity in AGN feedback between the PSB and SFSMMC sample: While the former population is quenched at $t_{\mathrm{lbt}}=0\,$Gyr, the latter is not.
Compared to low redshift, we also find a higher median SNe feedback at $z=0.9$. This, however, appears purely physical as the star formation, and thus the SNe feedback, was significantly stronger at higher redshift compared to low redshift \citep{2004Natur.428..625H}.
To conclude, we find that cluster PSBs are shutdown via environmental quenching with no evidence that additional galactic feedback is triggered.

\section{Discussion}
\label{sec:discussion}

\subsection{Environment and redshift evolution}
\label{sub:disc:envir}

We find that PSBs at low redshift are more frequently found in low halo mass environments and that the PSB-to-quenched fraction increases with redshift (Figure \ref{fig:Q_PSBfrac_grid}). 
The preference for low halo masses is further strengthened by the fact that $89.4\%$ of $z \sim 0$ PSBs are found in halos with $M_{\mathrm{200,crit}}/\mathrm{M_{\odot}} < 10^{13}$ (Table \ref{tab:BHgrowthTable}). This agrees with DEEP2 and SDSS results which find low redshift PSBs in relatively under-dense environments \citep{2005MNRAS.357..937G} compared to high redshift PSBs \citep{2009MNRAS.398..735Y}. Similarly to the decline of the PSB-to-quenched fraction with decreasing redshift found in our simulation (Figure \ref{fig:Q_PSBfrac_grid}), observations also show that the fraction of PSBs declines from $\sim 5\%$ of the total population at $z \sim 2$, to $\sim 1\%$ by $z \sim 0.5$ \citep{2016MNRAS.463..832W}.
We note that MOSFIRE observations at $z \sim 1$ find a higher number of PSBs, relative to star-forming galaxies, in clusters than in groups or the field \citep{2017MNRAS.472..419L}. Considering that we determine the abundance of PSBs with respect to quiescent galaxies, a comparison is difficult, however, as fewer star-forming than quiescent galaxies are found in high density environments \citep{1980ApJ...236..351D, 2003MNRAS.346..601G}, and our PSB-to-quenched fraction does not show a strong preference for low halo mass at $z \sim 1$ (Figure \ref{fig:Q_PSBfrac_grid}), agreement seems plausible.
Given these points, it appears that PSB galaxy evolution is strongly redshift dependent, favouring decreasing environmental densities towards lower redshifts, supporting the idea that the formation mechanism of PSBs is affected by redshift and environment.

Similarly, we also find a strong stellar mass function evolution (Figure \ref{fig:SMF_grid}): The abundance of PSBs above our stellar mass threshold increases significantly with increasing redshift, matching VVDS observations which find that the stellar mass density ($\log_{10}(M_*/\mathrm{M_{\odot}}) > 9.75$) of strong PSB galaxies is $230$ times higher at $z \sim 0.7$ than at $z \sim 0.07$ \citep{2009MNRAS.395..144W}. 
In contrast, when comparing the PSB galaxy stellar mass function shape to observations at redshifts $0.07<z<1.71$ \citep{2016MNRAS.463..832W, 2018MNRAS.473.1168R} and the observations with each other, we do not find close agreement. These discrepancies are likely due to the stellar mass function sensitivity to the exact selection criteria of PSBs. Interestingly, when comparing the PSB stellar mass function at $z=0.7$ to observations in the group and cluster environment (Figure \ref{fig:SMF_soco}), we find close agreement, including the double Schechter behaviour.

\subsection{The impact of mergers}
\label{sub:disc:mergers}

Evaluating PSBs in relation to the star-forming main sequence (Figure \ref{fig:MSpanel}) shows that during their starburst phase, which is often correlated with recent mergers (Figure \ref{fig:AGN_SNe_Energy} and Table \ref{tab:MergerTable}), massive PSBs are found significantly above the normalised redshift evolving main sequence \citep{2014ApJS..214...15S}. 
Considering our global low redshift PSB sample, of which $89\%$ have experienced a merger in the last $2.5\,$Gyr, we find that during peak star formation PSBs have SFRs which are a few times higher than on the main sequence, with a wide spread in their distribution. 
This behaviour matches observations by \cite{2019A&A...631A..51P}, which, on the one hand, find that mergers have little effect on the SFR for the majority of merging galaxies, but, on the other hand, also find that an increasing merger fraction correlates with the distance above the main sequence, i.e. at sometimes mergers may induce starbursts.
Furthermore, simulations by \cite{2008A&A...492...31D} suggest that strong starbursts, where the SFR is increased by a factor $\geq 5$, are rare and only found in $15\%$ of major galaxy interactions and mergers.
\cite{2020MNRAS.493.3716H} highlights the impact of mergers on the SFR: Star-forming post-merger galaxies, which make up $67\%$ of their post-merger galaxies identified in the IllustrisTNG simulation experience on average a SFR increase by a factor of $\sim 2$. This behaviour is in qualitative agreement with the correlation between mergers and the SFR increase found in our star-forming and, to a stronger extent, PSB galaxies (Figure \ref{fig:MSpanel}).
Additionally, when studying adjacent galaxies in IllustrisTNG, \cite{2020MNRAS.494.4969P} find that the presence of closest companions boost the average specific SFR of massive galaxies by $14.5\%$. This agrees with our study of an individual PSB in Figure \ref{fig:traceGasMassive}, where we find an increase in star formation prior to identifying a merger event while another galaxy is in close proximity.

Figure \ref{fig:MSpanel} also shows that $23\%$ of the tracked PSBs were previously quiescent, i.e. have undergone rejuvenation. 
When comparing quiescently star-forming, quenching, and rejuvenating galaxies in the EAGLE simulation, \cite{2016MNRAS.460.3925T} find that $\sim 1.6\%$ and $\sim 10\%$ of all galaxies can be characterised as fast and slow rejuvenating galaxies, respectively. In other words, although (fast) rejuvenation is generally rare, rejuvenation may well be a relevant pathway for the evolution of PSBs.
Consistent with the high merger abundances throughout our PSB sample, observations find quiescent galaxies may undergo rejuvenation events, e.g. via (gas rich) minor mergers, triggering the required starburst phase found in PSBs \citep{2012ApJ...761...23F, 2014MNRAS.444.3408Y, 2017ApJ...841L...6B, 2020ApJ...900..107Y}.

Even when solely considering isolated merger simulations, much of the ambiguity concerning the quenching impact of mergers remains \citep{2019MNRAS.487..318K}: Different types of mergers have been associated with varying quenching impacts, both directly, e.g. by introducing turbulence \citep{2018MNRAS.478.3447E}, and indirectly, e.g. by facilitating BH growth \citep{2013MNRAS.430.1901H, 2014MNRAS.437.1456B}.
For example, binary galaxy merger simulations find that the termination of star formation by BH feedback in disc galaxies is significantly less important for higher progenitor mass ratios \citep{2008AN....329..956J, 2009ApJ...690..802J}. 
Similar studies find that galaxies, which are dominated by minor merging and smooth accretion in their late formation history ($z \lesssim 2)$, experience an energy release via gravitational heating which is sufficient to form red and dead elliptical galaxies by $z \sim 1$, even in the absence of SNe and AGN feedback \citep{2009ApJ...697L..38J}.
Meanwhile, SPH simulations of major mergers demonstrate that consistency with observations does not require BH feedback to terminate star formation in massive galaxies or unbind large quantities of cold gas \citep{2011MNRAS.412.1341D}.
When linking the BH accretion rate with the galaxy-wide SFR, the disc galaxy mergers in the hydrodynamical simulations by \cite{2015MNRAS.449.1470V}, typically find no temporal correlation and different variability timescales. However, when averaging over time during $\sim (0.2-0.3)\,$Gyr long merger events, they find a typical increase of a factor of a few in the ratio of BH accretion rate to SFR \citep{2015MNRAS.449.1470V}.
This qualitatively agrees with our results shown in Figures \ref{fig:AGN_SNe_Energy_mass_evol} and \ref{fig:AGN_SNe_Energy}, that the recent AGN feedback increase in PSB and star-forming galaxies correlates with high merger abundances.
Note, however, that not all simulations agree in that mergers and AGN feedback are correlated (e.g. \cite{2014MNRAS.442.1992H}). 

The ambiguous nature of merger impacts is also reflected in our results:
Using a statistical approach, i.e. comparing merger abundances at $z \sim 0$ (Table \ref{tab:MergerTable}), we find that $88.9\%$ of PSBs, $23.4\%$ of quenched (QSMMC), and $79.7\%$ of star-forming stellar mass matched control (SFSMMC) galaxies experience at least one merger within the last $2.5\,$Gyr. 
The high merger abundance found in both our PSB and our SFSMMC sample highlights the varying merger impact: While our PSB sample is considered quiescent at $z \sim 0$, the reverse is true for SFSMMC galaxies with similarly rich merger histories, especially compared to the QSMMC sample. A similar behaviour is found when considering merger abundances at $z=0.9$ (Table \ref{tab:Nbeg064MergerTable}).
Our high merger abundance broadly agrees with observations of local PSBs in SDSS, which, in their youngest age bin ($< 0.5\,$Gyr), classify at least $73\%$ of PSBs, far more than their control sample, as distorted or merging galaxies \cite{2017A&A...597A.134M}.
Generally, observations of PSBs in the local low density Universe are associated with galaxy-galaxy interactions and galaxy mergers \citep{1996ApJ...466..104Z, 2001ApJ...547L..17B, 2008ApJ...688..945Y, 2009MNRAS.396.1349P, 2018MNRAS.477.1708P}, in excellent agreement with our results.

When evaluating the cold gas fractions of PSB, QSMMC, and SFSMMC progenitors within three half-mass radii (Figure \ref{fig:merger_wetness}), we find general agreement between PSB and SFSMMC progenitors: Both show a preference for higher cold gas fractions compared to QSMMC progenitors.
This is in line with observations, which find evidence for gas-rich mergers triggering central starbursts \citep{2018MNRAS.477.1708P, 2020MNRAS.497..389D}, fast quenching \citep{2019ApJ...874...17B}, and that recently merged galaxies typically are a factor of $\sim 3$ more atomic hydrogen rich than control galaxies at the same stellar mass \citep{2018MNRAS.478.3447E}.

With regard to cold gas fractions, the only difference between PSB and SFSMMC galaxies is found for the satellite progenitor populations in major mergers (Figure \ref{fig:merger_wetness}): The major merger satellite progenitors of PSBs are characterised by a lower cold gas fraction (almost half) compared to the SFSMMC sample. The subsequent lower cold gas content of post-merger PSBs, compared to SFSMMC galaxies, may be linked to a higher likelihood of a subsequent shutdown in star formation for two reasons: First, less cold gas is available to maintain star formation. Second, given a similar onset of merger triggered AGN feedback, the same amount of energy is distributed across a smaller supply of cold gas, quickening its heating and/or redistribution.

\subsection{Shutting down star formation}
\label{sub:disc:shutdown}

As shown for our global $z \sim 0$ PSB sample (Figure \ref{fig:AGN_SNe_Energy_mass_evol}) and the subset of six massive PSBs (Figure \ref{fig:AGN_SNe_Energy}), we find a significant increase in AGN feedback at recent look-back-times and towards higher stellar masses. The importance of AGN feedback in shutting down PSBs is also evidenced by the decreasing agreement between the otherwise often similarly behaving PSB and SFSMMC sample with increasing stellar mass.
As the fraction of galaxies hosting an AGN is a strong function of stellar mass \citep{2005MNRAS.362...25B}, the apparent lack of strong AGN activity in the SFSMMC sample, compared to the PSB sample at high stellar masses, is a strong indicator of the AGN quenching effectiveness.
In addition, the short shutdown timescale (Figure \ref{fig:MSpanel}), the redistribution and heating of gas (Figure \ref{fig:traceGasMassive}), the correlated BH growth (Figures \ref{fig:gammaMgas}), and the comparatively weak SNe energy (Figure \ref{fig:AGN_SNe_Energy_mass_evol}) all suggest that merger triggered AGN feedback generally is the dominant shutdown mechanism of PSBs at low redshift. However, we can neither fully exclude other causes, nor do all PSBs necessarily experience the same shutdown sequence.
Nonetheless, it appears likely that merger facilitated BH growth, which triggers AGN feedback, plays an important, albeit not necessarily exclusive, role in mediating between the starburst and post-starburst phase within our simulation at low redshifts. 
This agrees with previous Magneticum results, which found that merger events are not statistically dominant in fuelling mechanisms for nuclear activity, while still finding elevated merger fractions in AGN hosting galaxies compared to inactive galaxies, pointing towards an intrinsic connection between AGN and mergers \citep{2018MNRAS.481..341S}.
Generally, the importance of AGN feedback in explaining the sharp decline in the SFR found in (PSB) galaxies is also supported by several other works \citep{2005MNRAS.361..776S, 2013MNRAS.430.1901H, 2019A&A...623A..64C, 2020AAS...23520719L}.

We find evidence for the simultaneous mechanical expulsion and heating of previously star-forming cold gas (Figure \ref{fig:traceGasMassive}). The rapid shutdown in this and similar examples (Figure \ref{fig:MSpanel}) happens on timescales of $t_{\mathrm{shutdown}} \lesssim 0.4\,$Gyr, which, due to the short timescale, generally favours AGN feedback as the expected quenching mechanism \citep{2020MNRAS.494..529W}. 
Although much of the dense cold gas is heated, some cold gas remains in the recently quenched galaxy (Figure \ref{fig:traceGasMassive}). The fact that significant amounts of cold gas are redistributed on short timescales rather than only being directly heated may provide an explanation to observations which find significant non star-forming (molecular) gas reservoirs in PSBs \citep{2013MNRAS.432..492Z, 2015ApJ...801....1F}.
This also agrees with other simulations, suggesting that the SFR is quenched with feedback via gas removal, with little effect on the SFR via gas heating \citep{2014MNRAS.437.1456B}. Furthermore, the large amounts of molecular gas found in PSBs rules out processes such as gas depletion, expulsion, and/or starvation as the dominant shutdown mechanisms \citep{2015ApJ...801....1F}, which is supported by the results presented in this work.

\subsection{Post-starburst galaxies in galaxy clusters}
\label{sub:disc:clusters}

Cluster PSBs are typically infalling, especially towards lower redshift and higher cluster masses (Figure \ref{fig:PSvradGrid}). 
This matches previous results concerning the quenching of satellite galaxies in clusters \citep{2019MNRAS.488.5370L}, showing that star-forming galaxies are more likely to be on their first infall, especially for higher mass clusters, indicating that ram-pressure stripping typically quickly shuts down star formation already during the first infall of satellite galaxies.
This higher quenching effectiveness matches other simulations which find a similar significant enhancement of ram-pressure stripping in massive halos compared to less massive halos \citep{2019MNRAS.484.3968A}. 
In fact, several observations suggest that environmental quenching mechanisms, such as interactions with the ICM \citep{2009ApJ...693..112P, 2010PASA...27..360P} or specifically ram-pressure stripping \citep{2013A&A...553A..90G, 2017ApJ...846...27G, 2019MNRAS.482..881P}, are responsible for the abundance of PSBs in galaxy clusters. 

Generally, different populations of satellite galaxies, e.g. infalling, backsplash and virialised, occupy distinct regions of phase space \citep{2013MNRAS.431.2307O}. Hence, the clear preference of cluster PSBs for infall (Figure \ref{fig:PSvradGrid}) provides a strong indication for environmental quenching such as ram-pressure stripping. 
This is also reflected in the PSB galaxy preference for distinct cluster-centric radii ($R \sim (0.15-0.5)\,r_{\mathrm{200,crit}}$) found in projections (Figure \ref{fig:Muzzin}), showing excellent agreement with observations \citep{2014ApJ...796...65M}.
It also agrees with SAMI observations of recently quenched cluster galaxies, which are exclusively found within $R \leq 0.6\, R_{\mathrm{200}}$ and show a significantly higher velocity dispersion relative to the cluster population \citep{2019ApJ...873...52O}.
Similarly, GASP observations find that PSB galaxies avoid cluster cores and are characterised by a large range in relative velocities \citep{2020ApJ...892..146V}. Furthermore, both the SAMI and GASP phase space behaviour is consistent with recent infall, suggesting that PSBs could be descendants of galaxies which were quenched during first infall via ram-pressure stripping \citep{2019ApJ...873...52O, 2020ApJ...892..146V}. 
Providing multiple lines of evidence, \cite{2020ApJ...892..146V} conclude that the outside-in quenching \citep{2019ApJ...873...52O, 2020MNRAS.493.6011M}, the morphology, and kinematics of the stellar component, along with the position of GASP PSBs within massive clusters point to a scenario in which ram-pressure stripping has shutdown star formation via gas removal. This is in excellent agreement with our findings.

When comparing merger abundances at $z=0.9$ between non environmentally selected PSBs (Table \ref{tab:Nbeg064MergerTable}) and cluster PSBs in (Table \ref{tab:MuzzinTable}), we find broad agreement: Both samples are characterised by high merger abundances, i.e. in both samples $\sim 93\%$ of PSBs experience at least one merger event in the past $2.5\,$Gyr. In contrast to the non environmentally selected sample, cluster galaxies have slightly fewer major mergers ($47.3\%$ vs. $43.2\%$ for PSBs). It appears that mergers are important in enabling the conditions necessary for (strong) star formation in cluster PSB progenitors, while it remains unclear what impact they have on shutting down star formation in cluster PSBs.
In contrast to the majority of cluster PSB galaxy observations, observations of the Cl J1604 supercluster at $z \sim 0.9$ indicate that galaxy mergers are the principal mechanism for producing PSBs in clusters, while both interactions between galaxies and with the ICM also appear effective \citep{2014ApJ...792...16W}.
As found by observations \citep{2019ApJ...873...52O, 2020ApJ...892..146V} and our results (Figure \ref{fig:PSvradGrid}), cluster PSBs belong to a population of recently in-fallen galaxies. Hence, it appears likely that PSB progenitors have had ample opportunity to experience mergers in the outskirts of clusters prior to infall, likely positively impacting recent star formation, thereby building a young stellar population necessary for their later identification as PSBs. As discussed in Section \ref{sub:disc:mergers}, the impact of mergers is varied and need not lead to a subsequent shutdown in star formation, e.g. via triggering AGN feedback. It seems plausible that ram-pressure stripping is more efficient in shutting down star formation than merger triggered mechanisms, and hence is the dominant shutdown mechanism in cluster PSBs.

In a previous paper, we found evidence for a starburst $0.2 \, \mathrm{Gyr}$ after satellite galaxies first fall into their respective clusters, i.e. after crossing the cluster virial radius for the first time \citep{2019MNRAS.488.5370L}. 
Specifically, we found that the average normalised blueness, i.e. $\mathrm{SSFR} \cdot t_{\mathrm{H}}$, of satellite galaxies with stellar masses $M_* > 1.5 \cdot 10^{10}\, \mathrm{M_{\odot}}$ shows a significant starburst lasting $\sim 0.2\,$Gyr.
As discussed by \cite{2019MNRAS.488.5370L}, this is likely driven by the onset of ram-pressure stripping which triggers a short starburst, followed by a complete shutdown in star formation within $< 1\,$Gyr, often on shorter timescales. 
Observations of local cluster galaxies undergoing ram-pressure stripping come to similar conclusions: Ram-pressure likely drives an enhancement in star formation prior to quenching \citep{2018ApJ...866L..25V, 2020MNRAS.495..554R}.
Similarly, cluster galaxies undergoing ram-pressure stripping in the GASP sample show a systematic enhancement of the star formation rate: As the excess is found at all galacto-centric distances within the discs and is independent of both the degree of stripping and star formation in the tails, \cite{2020ApJ...899...98V} suggest that the star formation is most likely induced by compression waves triggered by ram-pressure stripping.
Furthermore, HST observations have found strong evidence of ram-pressure stripping first shock compressing and subsequently expelling large quantities of gas from infalling cluster galaxies, which experience violent starbursts during this intense period \citep{2014ApJ...781L..40E}.
When evaluating the median SNe feedback, i.e. the SFR, at $z=0.9$ for cluster PSBs (Figure \ref{fig:AGN_SNe_Energy_mass_evol_cluster}), we find on average no evidence for a strong starburst at recent look-back-times. However, this signal likely correlates more strongly with cluster-centric radius (e.g. \cite{2019MNRAS.488.5370L}). Hence it does not seem surprising that we find no signal.
To better understand cluster PSB galaxy evolution, we also investigated whether cluster PSBs are found in the vicinity or crossing of cluster shock fronts: We found no evidence for an increased abundance of cluster PSBs near shocks.

Observations of GASP jellyfish galaxies undergoing strong ram-pressure stripping find that the majority host an AGN \citep{2017Natur.548..304P, 2019MNRAS.487.3102G, 2021IAUS..359..108P} and that the suppression of star formation in the central region is driven by AGN feedback \citep{2019MNRAS.486..486R}.
Similarly, the Romulus C simulation finds evidence for ram-pressure stripping triggering AGN feedback, which may aid in the quenching process \citep{2020ApJ...895L...8R}. 
When comparing these results to our study of median AGN feedback in cluster PSBs at $z=0.9$ (Figure \ref{fig:AGN_SNe_Energy_mass_evol_cluster}), we find no evidence for a recent increase in AGN feedback. However, we note that such an increase would likely correlate more strongly when evaluating the AGN feedback as a function of cluster-centric radius, which goes beyond the scope of this paper.
While observations based on the UKIDSS UDS conclude that a combination of environmental and secular processes are most likely to explain the appearance of PSBs in galaxy clusters \citep{2019MNRAS.482.1640S}, all our evidence (Figures \ref{fig:PSvradGrid} and \ref{fig:AGN_SNe_Energy_mass_evol_cluster}) suggests that environmental quenching in the form of ram-pressure stripping leads to the shutdown in star formation found in our cluster PSBs.

\subsection{Numerical considerations}
\label{sub:disc:numerics}

In addition to the ambiguous involvement of various physical mechanisms, the implementation and approximation of known physical mechanisms in simulations comes with its own set of challenges. 
For example, many simulations underestimate the effectiveness of feedback due to excessive radiative losses \citep{2012MNRAS.426..140D}, which, in turn, are the result of a lack of resolution and insufficiently realistic modelling of the interstellar medium \citep{2015MNRAS.446..521S}.
\cite{2020MNRAS.498.1259Z} highlights the difficulty of reproducing very young PSBs in simulations, potentially indicating that new sub-resolution star formation recipes are required to properly model the process of star formation quenching.

To test the numerical convergence of our results, we searched for PSBs at ultra-high resolution in Box4. 
However, due to the low number of PSBs per volume element (Figure \ref{fig:SMF_grid}) and the small volume of Box4 (see Section \ref{sub:Mag}), no $z \sim 0$ PSBs were found with stellar mass $M_* \geq 5 \cdot 10^{10} \, \mathrm{M_{\odot}}$. Similarly, when evaluating Box4 at ultra-high resolution using the same redshift and cluster mass domain as shown in Figure \ref{fig:PSvradGrid} only one PSB galaxy was identified above our stellar mass threshold (at $z=0.47$ and in a cluster with $M_{\mathrm{200,crit}} = (1-3) \cdot 10^{14}\, \mathrm{M_{\odot}}$).
If we lower the stellar mass threshold to $M_* \geq 5 \cdot 10^{9} \, \mathrm{M_{\odot}}$ then $40$ PSBs at $z \sim 0$ are identified, $55\%$ of which have experienced at least one merger in the last $2.5\,$Gyr. We note that fewer recent mergers are expected as stellar mass decreases \citep{2014MNRAS.444.3986R}.

This lower stellar mass threshold was also used in Figure \ref{fig:Muzzin}: Despite the small sample sizes ($\leq 26$ PSBs identified when considering three linearly independent projections sampled from $\leq 11$ unique PSBs), the comparison of individual PSBs at different resolution levels and in different boxes shows agreement.
Additionally, we note that the high resolution run of Box4 employs an updated AGN model \citep{2015MNRAS.448.1504S}.
In short, the comparison between the different boxes and resolutions
has shown that our results do not appear to be driven by the resolution level nor details in the applied AGN model.

Generally, we note that the identification and comparison of Magneticum PSBs to observations may be influenced by a number of effects: As discussed in \cite{2019MNRAS.488.5370L}, we measure the star formation rate rather than the colour of galaxies. In other words, we determine the star formation directly and instantaneously, rather than via the indirect, at times delayed, observation of local and/or global galactic properties.
However, the galaxies selected in Box2 do not reproduce the detailed morphologies, especially concerning the cold thin discs, found in observations. Hence, we cannot capture the details of mechanisms which act on scales similar to our gas softening ($\epsilon_{\mathrm{gas}} = 3.75 \, \mathrm{h^{-1} kpc}$). This, for example, becomes relevant during ram-pressure stripping, where cold thin discs, dependent on infall geometry, provide additional shielding compared to more diffuse galactic configurations, thereby impacting quenching efficiencies and timescales.
These limitations need to be addressed with the next generation of cosmological simulations.

\section{Conclusions}
\label{sec:conc}

In order to understand the physical mechanisms leading to the formation of post-starburst galaxies (PSBs), i.e. the reasons for both the onset of the initial starburst and the abrupt shutdown in star formation, we studied the environment and temporal evolution of PSBs with stellar mass $M_* \geq 5 \cdot 10^{10} \, \mathrm{M_{\odot}}$. To this end, we used Box2 of the hydrodynamical cosmological \textit{Magneticum Pathfinder} simulations to resolve the behaviour of PSBs at varying redshifts $0.07<z<1.71$ both throughout our whole box volume as well as in specific environments such as galaxy clusters.
The principal sample studied consists of $647$ PSBs, identified at $z \sim 0$, i.e. a global sample spanning the whole box volume. Throughout our analysis the behaviour and evolution of PSBs is compared to star-forming (SF) and quiescent (Q) stellar mass matched control (SMMC) galaxy samples at different look-back-times (lbt).
Furthermore, Magneticum PSBs are compared with observed quenched fractions \citep{2011ApJ...742..125G, 2013ApJ...778...93T, 2018ApJ...852...31W, 2019A&A...622A.117S}, stellar mass functions \citep{2013ApJ...777...18M, 2016MNRAS.463..832W, 2018MNRAS.473.1168R}, and the star formation main sequence \citep{2014ApJS..214...15S, 2018A&A...615A.146P} at different redshifts. Especially, we compare Magneticum galaxy cluster PSBs to observed high environmental density stellar mass functions \citep{2018MNRAS.476.1242S} and the cluster phase space behaviour at high redshift \citep{2014ApJ...796...65M}. Our results are summarised as follows:

\begin{itemize}
    \item At $z \sim 0$ PSBs and SF galaxies both are characterised by an abundance of mergers: $89\%$ of PSB and $80\%$ of SF galaxies experience at least one merger event within the last $2.5\,$Gyr, compared to $23\%$ of quiescent galaxies. 
    Over the same time-span, $65\%$ of PSB, $58\%$ of SF, and $9\%$ of quiescent galaxies experience at least one major merger ($M_*$ ratio: > 1:3) event. 
    This established similarity in merger abundances between PSB and SF galaxies is also found at redshift $z \sim 0.9$, both when evaluating the entire box volume as well as when specifically selecting galaxy cluster environments.

    \item Inspecting $z \sim 0$ PSB, quiescent, and SF galaxies with $M_* \in [5.00,5.40) \cdot 10^{10}\, \mathrm{M_{\odot}}$, we find that the AGN feedback, which is associated with recent mergers, consistently outweighs the supernova (SNe) feedback. Within the last $0.5\,$Gyr, the difference between AGN and SNe feedback increases significantly: While the maximum median SNe power output for PSBs is $P_{\mathrm{SNe,PSB}} \leq 2 \cdot 10^{55}\,$erg/Myr, the maximum median AGN power output is $P_{\mathrm{AGN,PSB}} \geq 10^{56}\,$erg/Myr.
    In contrast to the SF galaxies, PSBs are characterised by increasing AGN feedback with increasing stellar mass: At stellar masses $M_* \geq 8.3 \cdot 10^{10} \, \mathrm{M_{\odot}}$, the AGN feedback at $z = 0$ of PSBs ($P_{\mathrm{AGN,PSB}} \sim 10^{57}\,$erg/Myr) is half an order of magnitude larger than that of SF galaxies ($P_{\mathrm{AGN,SF}} \sim 2 \cdot 10^{56}\,$erg/Myr), which, in turn, is significantly larger than that of quenched galaxies ($P_{\mathrm{AGN,Q}} \sim 10^{55}\,$erg/Myr).
    This strongly indicates that the star formation in PSBs generally is shut down by AGN feedback.
    
    \item In our global $z \sim 0$ PSB sample we find that during the star formation shutdown, typically at $t_{\mathrm{lbt}} \lesssim 0.4\,$Gyr, galactic gas, especially previously star-forming gas, is often abruptly heated, while simultaneously being redistributed. This results in a sharp decrease in the (cold) gas density. This is often correlated with a recent strong increase in black hole mass, triggering significant AGN feedback. 
    
    \item In contrast to SF galaxies, PSBs in our global sample, especially at $t_{\mathrm{lbt}} = [0.1,1]\,$Gyr, show less spread, i.e. are more continuous in the distribution of SNe feedback energy.
    As the star formation rate (SFR) linearly impacts the SNe feedback, the smaller spread in the distribution of PSB SNe feedback in combination with slightly elevated median SFRs during recent times, compared to SF galaxies, is a reflection of the recent starburst phase. 
    As the stellar mass increases, the median PSB SNe feedback increases slightly and the difference to SF galaxies, which continue to be associated with a wider distribution in feedback, becomes stronger.
    
    \item When evaluating the cold gas content prior to mergers in our global sample, PSB and SF progenitors show similar cold gas fractions within three half-mass radii ($f_{\mathrm{cgas}} \sim 0.9$) for the main progenitors. However, when considering cold gas abundances of satellite progenitors, i.e. not the most massive progenitors prior to major merger events, PSBs are characterised by lower median cold gas fractions ($f_{\mathrm{cgas}} = 0.40$) compared to SF satellite progenitors ($f_{\mathrm{cgas}} = 0.73$). This is also reflected in the different abundance of satellite major merger progenitors which have low cold gas fractions: $42\%$ of PSBs compared to $29\%$ of SF galaxies have $f_{\mathrm{cgas}} \lesssim 0.1$. 
    This indicates that, statistically, PSBs have less cold gas available following major mergers than SF galaxies, leading to a higher likelihood of a subsequent shutdown in star formation.
    
    \item Prior to the star formation shutdown, PSB progenitors exhibit both sustained long term star formation ($t \sim 3\,$Gyr), as well as short starbursts ($t \sim 0.4\,$Gyr). During the starbursts, independent of the duration, massive PSB progenitors are found at least a factor of $\Delta MS[z]/MS[z] \gtrsim 5$ above the redshift evolving main sequence. Of the tracked PSBs in our global sample, $23\%$ are rejuvenated galaxies, i.e. were considered quiescent before their starburst. 
    At $z \sim 0.4$, Magneticum Box2 main sequence galaxies agree well with observations, while at $z \sim 0.1$ our galaxies lie slightly above observations \citep{2014ApJS..214...15S, 2018A&A...615A.146P}.
    
    \item At $t_{\mathrm{lbt}} \sim 2.5\,$Gyr, PSB and SF progenitors from our global sample are rarely found in isolated halos, whereas quenched progenitors are most often found in isolated halos. This initial difference between the PSB and SF versus quenched distribution of galaxies within a given halo, becomes indistinguishable at $t_{\mathrm{lbt}} \sim 2.5\,$Gyr. This indicates that common initial conditions, i.e. an abundance, albeit not saturation, of galaxies in the immediate vicinity, are shared among SF and PSB galaxies, enabling the rich merger history found in these populations. 
    \item We compared the Box2 total, SF, quenched, and PSB stellar mass functions (SMF) at multiple redshifts $0.07<z<1.71$ to observations, finding broad agreement \citep{2013ApJ...777...18M, 2016MNRAS.463..832W, 2018MNRAS.473.1168R}: While the total and quenched SMF agree well with observations over the evaluated redshift range, the agreement for SF galaxies improves with increasing redshift. Meanwhile, PSB SMFs show that both the agreement between simulation and observations, as well as between observations, is subject to variation.
    When comparing stellar mass functions in group and cluster environments at $z = 0.7$, we are able to closely reproduce the observations \citep{2018MNRAS.476.1242S}. In particular, similarly to observations, we also find evidence for a PSB plateau in the stellar mass range $10.0 < \mathrm{log}(M_{*}/\mathrm{M_{\odot}}) < 10.5$ in group and cluster environments.
    
    \item At redshifts $z \lesssim 1$, PSBs are consistently found close to our stellar mass threshold ($M_* \geq 5 \cdot 10^{10}\, \mathrm{M_{\odot}}$) and at low halo masses. Towards higher redshift the abundance of PSBs increases significantly, especially at higher stellar masses.
    Overall, the PSB-to-quenched fraction increases with redshift, most significantly between $z \sim 1.3$ and $z \sim 1.7$.
    
    \item To compare with line-of-sight (LOS) phase space observations of cluster PSBs at $z \sim 1$ \citep{2014ApJ...796...65M}, we environmentally selected PSBs in the same halo mass range ($10^{14} < M_{\mathrm{200,crit}} / \mathrm{M_{\odot}} < 2 \cdot 10^{15}$) and found close agreement with observations.
    In particular, cluster PSBs are preferentially located in a narrow region of phase space with projected cluster-centric radii $R \sim (0.15-0.5) \, \mathrm{R_{200,crit}}$. The fact that both simulated and observed cluster PSBs are found in the same preferential region of phase space suggests a shared environmentally driven mechanism relevant to the formation of PSBs, which is specific to galaxy clusters, such as ram-pressure stripping.
    When evaluating cluster PSBs at different redshifts and cluster masses, we find that cluster PSBs at $z \lesssim 0.5$ are overwhelmingly infall dominated, especially towards higher cluster masses. 
    This further supports the idea that, different to the PSBs in the field, ram-pressure stripping shuts down star formation of previously active galaxies, thus leading to the identification of cluster PSBs within a distinct region of phase space. 
    
    \item Cluster PSBs further show no signs of significantly increased AGN or SNe feedback at recent look-back-times. In other words, we find no evidence suggesting that AGN feedback is triggered via ram-pressure stripping during cluster infall for PSBs. We also find no evidence that the AGN is responsible for quenching cluster PSBs. 
    This is further supported by the similarity in AGN feedback between the PSB and SF sample: While the former population is quenched at $t_{\mathrm{lbt}}=0\,$Gyr, the latter is not. Hence, we conclude that cluster PSBs are primarily shut down via environmental quenching, likely ram-pressure stripping.

\end{itemize}

To summarise, PSBs with stellar mass $M_* \geq 5 \cdot 10^{10} \, \mathrm{M_{\odot}}$ at $z \sim 0$ typically evolve as follows: First, PSB progenitors, which at $t_{\mathrm{lbt}}=2.5\,$Gyr are predominantly found in halos with more than one galaxy, experience a merger event. Specifically, $89\%$ of PSBs experience at least one merger within the last $t_{\mathrm{lbt}}=2.5\,$Gyr, with $65\%$ undergoing at least one major merger. 
Second, the merger provides additional gas and/or facilitates the inflow of gas onto the PSB progenitor, often triggering a starburst phase. After the merger, the BH accretion, and thereby the AGN power output ($P_{\mathrm{AGN,PSB}} \geq 10^{56}\,$erg/Myr), typically increases significantly, especially at higher stellar masses. A quick shutdown in star formation follows, which is often accompanied by a dispersal and heating of (previously star-forming) gas.
Lastly, a PSB galaxy remains, i.e. a galaxy with a young stellar population and quiescent levels of star formation. 

Strikingly, this evolution is different for PSBs found in galaxy clusters: While cluster PSBs also experience an abundance of mergers, leading to star formation enhancement, they are found in a distinct region of phase space, implying a shared environmentally driven quenching mechanisms.
Moreover, cluster PSBs are usually experiencing their first infall, especially in higher mass clusters, favouring a density dependent quenching mechanism such as ram-pressure stripping. In other words, although the merger abundance, associated with an increased SFR in cluster PSB progenitors prior to their infall, is similar to our global sample, the reason for the shutdown in star formation is not.

To conclude, we find that PSBs experience starbursts due to merger events, independent of their environment, but the quenching mechanisms strongly depend on environment: While AGN feedback is the dominant quenching mechanism for field PSBs, PSBs in galaxy clusters are quenched by ram-pressure stripping due to the hot cluster environment. Thus, for field galaxies their cold gas fraction prior to quenching from the AGN is important to whether they stay star-forming or become PSBs, while for cluster PSBs their infall orbit is the most important factor for quenching, as discussed already by \citep{2019MNRAS.488.5370L}. This likely leads to very different fundamental properties of PSBs in the field and clusters, but to study this in detail remains to be done in a future study.

\section*{Acknowledgements}
We thank Felix Schulze, Ulrich Steinwandel, Ludwig B\"{o}ss, and Tadziu Hoffmann for helpful discussions.
The \textit{Magneticum Pathfinder} simulations were partially performed at the Leibniz-Rechenzentrum with CPU time assigned to the Project "pr86re". 
This work was supported by the Deutsche Forschungsgemeinschaft (DFG, German  Research  Foundation)  under  Germany’s  Excellence Strategy – EXC-2094 – 390783311.
Information on the \textit{Magneticum Pathfinder} project is available at \url{http://www.magneticum.org}.

\section*{Data Availability}

The data underlying this article will be shared on reasonable request to the corresponding author.



\bibliographystyle{mnras}

\bibliography{research}



\bsp	
\label{lastpage}
\end{document}